\documentclass[oldversion]{aa}

\usepackage{amsmath}
\usepackage{natbib}
\usepackage{epsfig}
\usepackage{txfonts}

\newcommand{\nue}{\nu_{\rm e}} 
\newcommand{\nuebar}{{\bar \nu}_{\rm e}}

\newcommand{\Msol}{M_{\odot}}

\newcommand{\dr}{{{\rm d}r}}

\newcommand{\dS}{{{\rm d}S}}

\newcommand{\foe}{{10^{51}\,\mathrm{erg}}}

\newcommand{\eexp}{{E_{\rm exp}}}

\newcommand{\texp}{{t_{\rm exp}}}

\newcommand{\Mns}{{M_{\rm ns}}}

\newcommand{\dMg}{\Delta M_{\rm gain}}
  
\newcommand{\Rs}{{R_{\rm s}}}  
\newcommand{\Rg}{{R_{\rm g}}}  
\newcommand{\Rc}{{R_{\rm c}}}  
\newcommand{\vns}{{v_{\rm ns}}}

\newcommand{\Lib}{{L_{\rm ib}}}

\newcommand{\rib}{{R_{\rm ib}}} 
\newcommand{\Rib}{{R_{\rm ib}}} 
\newcommand{\tib}{{t_{\rm ib}}}

\newcommand{\deltas}{{\delta_{\rm shock}}} 
 
\newcommand{\deltag}{{\delta_{\rm gain}}} 
\newcommand{\deltac}{{\delta_{\rm crit}}}

\newcommand{\drhodSl}{ \left( \frac{\partial \rho}{\partial S}   \right)_{P,Y_l}}

\newcommand{\tadv}{\tau_{\rm adv}^{\rm g}}

\newcommand{\tconv}{\tau_{\rm conv}}

\newcommand{\vgainav}{\langle |v_r| \rangle_{\rm gain}}

\newcommand{\Qaac}{Q_{\rm aac}}

\bibpunct{(}{)}{;}{a}{}{,}

\begin{document}

\bibliographystyle{aa}

\title{Multidimensional supernova simulations with
       approximative neutrino transport}
\subtitle{II. Convection and the advective-acoustic cycle in the supernova core}

\author{
        L. Scheck\inst{1}     \and
        H.-Th. Janka\inst{1}  \and
        T. Foglizzo\inst{2}   \and
        K. Kifonidis\inst{1}
       }
        
\offprints{H.-Th.~Janka} 
\mail{thj@mpa-garching.mpg.de}

\date{received; accepted}      
    
\institute{Max-Planck-Institut f\"ur Astrophysik,
           Karl-Schwarzschild-Stra{\ss}e 1, 
           D-85741 Garching, Germany \and
           Service d'Astrophysique, DSM/DAPNIA, CEA-Saclay,
           91191 Gif-sur-Yvette, France}

\abstract{
Performing two-dimensional hydrodynamic simulations including a detailed
treatment of the equation of state of the stellar plasma and for the
neutrino transport and interactions, we investigate here the interplay
between different
kinds of non-radial hydrodynamic instabilities that can play a role during
the postbounce accretion phase of collapsing stellar cores. The
convective mode of instability, which is driven by the 
negative entropy gradients caused by neutrino heating or by variations in
the shock strength in transient phases of shock expansion and contraction,
can be identified clearly by the development of typical Rayleigh-Taylor
mushrooms. 
However, in those cases where the gas in the postshock region is rapidly 
advected towards the gain radius, the growth of such a buoyancy instability 
can be suppressed. 
In this situation the shock and postshock flow can nevertheless develop 
non-radial asymmetry with an oscillatory growth in the amplitude. This
phenomenon has been termed ``standing (or spherical) accretion 
shock instability'' (SASI).
% by Blondin et al.\ (2003). It is shown here that
%   \cite{Blondin+03}. 
It is shown here that the SASI oscillations can trigger
convective instability, and like the latter, they lead to an increase in the
average shock radius and in the mass of the gain layer.
Both hydrodynamic instabilities in combination stretch the
advection time of matter accreted through the neutrino-heating layer and
thus enhance the neutrino energy deposition in support of the
neutrino-driven explosion mechanism. A rapidly contracting and more compact
nascent neutron star turns out to be favorable for explosions, because
the accretion luminosity and neutrino heating are greater and the growth
rate of the SASI is higher. Moreover, we show that
the oscillation period of the SASI observed in our simulations
agrees with the one estimated for the advective-acoustic cycle (AAC),
in which perturbations are carried by the accretion flow from the shock
to the neutron star and pressure waves close an amplifying
global feedback loop. 
A variety of other features in our models, as well as differences in their
behavior, can also be understood on the basis of the AAC hypothesis.
The interpretation of the SASI in our simulations as a purely acoustic 
phenomenon, however, appears difficult.} 
% (Blondin \& Mezzacappa 2006)


\authorrunning{Scheck et al.}
\titlerunning{Multidimensional supernova simulations}

\maketitle

%=====================================================================
\section{Introduction}
\label{sec:intro}

Hydrodynamic instabilities play an important role in core-collapse
supernovae, because on the one hand they may be crucial for starting 
the explosion and on the other hand they may
provide a possible explanation for the observed anisotropy of
supernovae.  There is a growing consensus that the neutrino-driven
explosion mechanism of core-collapse supernovae does not work in
spherical symmetry for progenitors more massive than about
$10\,\Msol$. None of the recent simulations with one-dimensional (1D)
hydrodynamics and a state-of-the-art description of the neutrino
transport develops an explosion
\citep{RJ02,Liebendoerfer+01,Liebendoerfer+05,Thompson+03,Buras+03,Buras+06,Buras+06b}.
However, multi-dimensional effects were recognised to be helpful. In
particular it was shown that convection is able to develop below
the stalled supernova shock and that it can increase the efficiency of
neutrino heating significantly \citep{HBFC94,BHF95,JM95,JM96}. Current 
two-dimensional (2D) simulations are thus considerably closer to the 
explosion threshold than 1D models
\citep{Buras+03,Buras+06,Buras+06b}, and shock revival and
the onset of an explosion has
been reported recently for a 2D calculation with an $11.2\,\Msol$ progenitor
\citep{Buras+06b}.  In earlier 2D simulations, in which the angular
size of the numerical grid was constrained to less than $180^{\circ}$
and in simulations in which
the approximative description of the neutrino transport
resulted in a fast onset of the explosion, convection was dominated
by rather small angular scales of several ten degrees
\citep{JM94,JM96}.  However, in recent 2D calculations of
\cite{Buras+06,Buras+06b,Scheck+04,Burrows06,Burrows07}, and
\citeauthor{Scheck+06} (\citeyear{Scheck+06}, henceforth Paper~I),
a slower development of the explosion and the use of a full 
$180^{\circ}$ grid
allowed for the formation of pronounced global (dipolar and 
quadrupolar) modes of asymmetry.

The anisotropy in these models is of particular interest, as it 
might provide the explanation for
two results from observations: Firstly, spectropolarimetry
(\citealt{Wang+01,Wang+03,Leonard+06}, and references therein)
revealed that a non-spherical ejecta distribution is a common feature
of many core-collapse supernovae and is probably caused by the
explosion mechanism itself, since the anisotropy increases if deeper
layers of the ejecta are probed. In the case of Supernova 1987A this
non-spherical distribution of the ejecta can even be directly imaged
with the Hubble Space Telescope \citep{Wang+02}. Secondly, neutron
stars move through interstellar space with velocities much higher than
those of their progenitors  
\citep[e.g.,][]{Cordes+93,LL94,HP97,Arzoumanian+02,Zou+05,Chatterjee+05,Hobbs+05}, 
in some cases with more than $1000\,\mathrm{km/s}$. It was suggested by
\cite{Herant95} and demonstrated with hydrodynamic simulations by
\cite{Scheck+04,Scheck+06} that neutron star velocities of this
magnitude can result from strongly anisotropic (in the most extreme
cases ``one-sided'' i.e., dipole-dominated) explosions, in which the
total linear momentum of the ejecta must be balanced by a
correspondingly high recoil momentum of the neutron star.

In multi-dimensional simulations convective motions break the initial
global sphericity and support the explosion (or bring the model
closer to the explosion threshold) by transporting cool matter from
the shock to the gain radius where neutrino heating is strongest
and by allowing hot matter to rise and to increase the pressure 
behind the stalled shock. 
However, it is not clear whether convection can
also be responsible for the development of low modes in the postshock
accretion
flow, as suggested by \cite{Herant95} and \cite{Thompson00}.  The
$l=1$ pattern studied by \cite{Herant95} was motivated by a perturbation
analysis of volume-filling convection in a fluid sphere by
\cite{Chandra61}, who found the dipole ($l=1$) mode to be the most 
unstable one.
In fact, \cite{Woodward+03} and \cite{Kuhlen+03} demonstrated with
three-dimensional simulations that the $l=1$ mode dominates the
convection in red-giant and main-sequence stars. \cite{Blondin+03},
however, investigating an idealized setup in 2D hydrodynamic
simulations, discovered that an adiabatic
accretion flow below a standing shock develops a non-radial, oscillatory
instability, which they termed ``standing accretion shock
instability'' or SASI, and which is dominated by the $l=1$ or $l=2$
modes. This suggests that the low-mode asymmetries found to 
develop in supernova cores in multi-dimensional models 
may be caused by global instabilities different from convection.
\cite{Foglizzo+06} performed a linear stability analysis for 
a problem that resembles the stalled shock situation
in supernovae, taking into
account the limited radial size of the convectively unstable layer
below the shock and the finite advection of matter through this region. 
The latter process turns out to have a stabilising effect and can hamper
the growth of convection significantly. In particular, the lowest
modes are convectively unstable only if the ratio of the convective
growth timescale to the advection time through the unstable layer is
small enough. \cite{Foglizzo+06} estimate that this may not be
the case in general and support the suggestion that instabilities 
different from
convection may be responsible for the occurrence of low-order modes
of asymmetry in the postshock accretion flow.

The ``advective-acoustic cycle'', in short AAC
\citep{Foglizzo_Tagger00,Foglizzo01,Foglizzo02}, is a promising
candidate for explaining such a (SASI) instability. 
It is based on the acoustic
feedback produced by the advection of entropy and vorticity
perturbations from the shock to the forming neutron star. By means of
linear stability analysis, \cite{Galletti_Foglizzo05} showed that due
to the AAC the flow in the stalled accretion shock phase of
core-collapse supernovae is unstable with respect to non-radial
perturbations, and that the highest growth rates are found for the
lowest degree modes (in particular for the $l=1$ mode).  

The situation studied by \cite{Blondin+03} and
\cite{Galletti_Foglizzo05} was, however, strongly simplified compared
to real supernovae.  \cite{Blondin+03} observed the growth of
non-radial perturbations in a flow between an accretion shock and an
inner boundary, which was located at a fixed radius. The boundary
conditions were taken from a stationary flow solution.  Furthermore,
neither a realistic description of the equation of state of the gas
nor the effects of neutrinos were taken into account by
\cite{Blondin+03}. Improving on this, \cite{Blondin_Mezzacappa06}
adopted an analytic neutrino cooling function
\citep{Houck_Chevalier92}, and \cite{Ohnishi+06} in addition took
into account neutrino heating and used the more realistic equation of
state from \cite{Shen+98}. Both groups concur in that low-mode
instabilities develop also in these more refined simulations.  The
nature of the instability mechanism is, however, still a matter of
debate.  While \cite{Ohnishi+06} consider the AAC as the cause of the
low-mode oscillations, \cite{Blondin_Mezzacappa06} argue that a
different kind of instability, which is purely acoustic and does not
involve advection, is at work in their simulations. Yet, the
eigenmodes found in the latter simulations were also reproduced in a
linear study of \cite{Foglizzo+06b}, who demonstrated that at least
for higher harmonics the instability is the consequence of an
advective-acoustic cycle. \cite{Laming+07}, finally, suggested the
possibility that feedback processes of both kinds can occur and
differ in dependence of the ratio of the accretion shock radius to
the inner boundary of the shocked flow near the neutron star
surface.

The work by \cite{Blondin_Mezzacappa06}, \cite{Ohnishi+06}, and
\cite{Foglizzo+06b} shows that non-radial SASI instability of the 
flow below a standing accretion shock occurs also when neutrinos 
(which could have a damping influence) are taken into account.
This is in agreement with a linear stability analysis of the
stationary accretion flow by \cite{Yamasaki_Yamada+07}, who
included neutrino heating and cooling, and studied the influence
of varied neutrino luminosities from the proto-neutron star. They
found that for relatively low neutrino luminosities the growth of
an oscillatory non-radial instability is favored, with the most
unstable spherical harmonic mode being a function of the 
luminosity, whereas for sufficiently high neutrino luminosity a
non-oscillatory instability grows. They attributed the former to
the AAC and the latter to convection.

All these studies concentrated on steady-state accretion flows,
made radical approximations to the employed neutrino physics,
and considered idealized numerical setups with special boundary
conditions chosen at the inner and outer radii of the considered
volume. Because of these simplifications such studies are
not really able to assess the importance of the different kinds
of hydrodynamic instabilities for supernova explosions. 
The growth rates of these
instabilities depend on the properties of the flow, and are thus
constant for stationary flows. In real supernovae, however, the flow
changes continuously, because the shock adapts to the varying mass
accretion rate, the neutrino heating below the shock changes, and the
proto-neutron star contracts. Therefore the growth rates also vary,
and a priori it is not clear whether they will be high enough for
a long enough time to allow a growth of some instability to the
nonlinear phase on a timescale comparable to the explosion timescale
(which itself can be influenced by the instability and is a
priori also unknown).

The aim of this work is therefore to go some steps further in the
direction of realism and to abandon the assumption of a stationary
background flow. To this end we study here the growth of
hydrodynamic instabilities in a ``real'' supernova core, i.e., we 
follow in 2D simulations the post-bounce evolution of the infalling
core of a progenitor star as provided by stellar evolution calculations,
including a physical equation of state for the stellar plasma and a
more detailed treatment of the neutrino physics than employed in 
the previous works. The considered
models were computed through the early phase of collapse until 
shortly after bounce by using state-of-the-art multi-group neutrino
transport \citep{Buras+03,Buras+06,Buras+06b}.
In the long-time post-bounce simulations performed by us, we then
used an approximative description of the neutrino transport based on 
a gray (but non-equilibrium) integration of the neutrino number and 
energy equations along characteristics 
(for details of the neutrino treatment, see \citealt{Scheck+06} [Paper~I]).
Compared to supernova simulations with a state-of-the-art energy-dependent
description of the neutrino transport (in spherical symmetry see, e.g.,
\citealt{RJ02,Liebendoerfer+01,Liebendoerfer+05,Thompson+03}, and for
multi-group transport also in 2D, see, e.g., 
\citealt{Buras+03,Buras+06,Buras+06b}) the models presented here thus
still employ significant simplifications. Such an approximative neutrino
treatment must therefore be expected to yield results that can differ 
quantitatively from
those of more sophisticated transport schemes. Nevertheless our approach
is able to capture the qualitative features of the better treatments.
It is certainly significantly more
elaborate (and ``realistic'') than the schematic neutrino source terms
employed by \cite{Foglizzo+06b} and \cite{Blondin_Mezzacappa06}, and
the local neutrino source description (without transport) adopted by 
\cite{Ohnishi+06} and \cite{Yamasaki_Yamada+07}. 
We consider our approximation as
good enough for a project that does not intend to establish the
viability of the neutrino-heating mechanism but which is interested mostly
in studying fundamental aspects of the growth of non-radial hydrodynamic
instabilities in the environment of supernova cores including the influence
that neutrino cooling and heating have in this context.

We made use of one more approximation that reduces the complexity of
our simulations compared to full-scale supernova models, namely, we
did not include the neutron star core but replaced it by a Lagrangian
(i.e., comoving with the matter) inner grid boundary 
that contracts with time to smaller radii, mimicking the shrinking of
the cooling nascent neutron star. 
At this moving boundary the neutrino luminosities
produced by the neutron star core were imposed as boundary conditions.
This had the advantage that we could regulate the readiness of a 
model to explode or not explode, depending on the size of the chosen
core luminosities and the speed of the boundary contraction. 
The inner boundary of our computational grid is impenetrable for 
the infalling accretion flow, but the accreted matter settles into
the surface
layer of the forming neutron star, similar to what happens outside
of the rigid core of the compact remnant at the center of a 
supernova explosion.
This is different from the various kinds of ``outflow boundaries''
employed in the literature\footnote{\cite{Blondin+03} used a 
``leaky boundary'' and \cite{Ohnishi+06} a ``free outflow
boundary'', both assuming non-zero radial velocity at the grid
boundary. In contrast, \cite{Blondin_Mezzacappa06} adopted a 
``hard reflecting boundary''. Although in this case the radial velocity
at the boundary is taken to be zero, there is still a non-zero mass
flux as the density near the boundary goes to infinity. The difference
can be important for the spurious generation of acoustic feedback by
vorticity perturbations.}, although 
\cite{Blondin+03} and \cite{Blondin_Shaw07} reported about tests
with several different 
prescriptions for the boundary treatment without finding any
significant influence on the growth of the SASI. Our modeling  
approach therefore follows \cite{Scheck+04} and Paper~I, where indeed 
the development of low-mode flow (with dominant $l=1$ and $l=2$ modes)
between shock and neutron star was found. In these previous papers
we, however, did not attempt to identify the mechanism(s) that were
causal for the observed phenomenon and just mentioned that convection
and the acoustically-driven or AAC-driven SASI may yield an 
explanation for the large global asymmetries seen to develop during
the neutrino-heating phase of the stalled shock. There was no 
analysis which mechanism was active and why it had favorable
conditions for growth.

In the present work we return to these questions. 
In particular we aim here at exploring the following points:
\begin{itemize}
\item What is the timescale for a non-radial instability of the 
  stalled accretion shock (SASI) to develop in a supernova core?
  How is it influenced by neutrino effects?
\item Can the instability be identified as consequence of an amplifying
  advective-acoustic cycle, of a growing standing pressure wave
  \citep{Blondin_Mezzacappa06} or of something else?
\item What determines its growth rate? For which conditions does the
  instability grow faster than convection and which influence may
  this have on the subsequent (nonlinear) evolution?
\item What is the relationship between convection and the instability
  in the nonlinear phase?
\item What is the possible supportive role of the SASI in the context
  of neutrino-driven explosions and in creating the low-mode ejecta
  asymmetry identified as cause of large neutron star kicks by 
  \cite{Scheck+04} and \cite{Scheck+06}?
% \item What is the influence of the instability on the explosion
%   energy and the neutron star recoil?
\end{itemize}

In order to address these questions we will first summarise the most
important properties of convection and of the AAC in the gain layer in
Sect.~\ref{sec:instabilities}. This will serve us as basis for the
later analysis of our set of two-dimensional simulations.
In Sect.~\ref{sec:numsetup} we will describe
the computational methods and numerical setup we use for these
simulations and will motivate our choice of parameter values
for the considered models. We will present the
simulation results in Sect.~\ref{sec:results} and will discuss them in
detail in Sects.~\ref{sec:linear} and \ref{sec:nonlinear}.
Section~\ref{sec:conclusions}, finally, contains our conclusions.

%=====================================================================
\section{Hydrodynamic instabilities}
\label{sec:instabilities}

%---------------------------------------------------------------------
\subsection{Linear and nonlinear convective growth of perturbations}
\label{sec:convection}

In a hydrostatic, inviscid atmosphere, regions with negative entropy
gradients (disregarding possible effects of composition gradients)
are convectively unstable for all wavelengths. Short wavelength
perturbations grow fastest, with a local growth rate
$\omega_{\rm buoy}>0$ equal to the
imaginary part of the complex Brunt-V{\"a}is{\"a}la frequency:
\begin{equation}
  \omega_{\rm buoy} \equiv \sqrt{-a_{\rm grav} \,
{\cal C} \, / \, \rho}.\label{omegabuoy}
\end{equation}
Here $a_{\rm grav}<0$ is the local gravitational acceleration, $\rho$ is the
density and
\begin{equation}
  {\cal C} \equiv \drhodSl \cdot \frac{\dS}{\dr},
%+ \drhodYl \cdot \frac{\dYl}{\dr}
\end{equation}
where $S$ is the entropy, $P$ is the pressure, and $Y_{l}$ is the total
lepton number per nucleon.  Note that ${\cal C}>0$ is the
instability condition for Schwarzschild convection.

\cite{Foglizzo+06} pointed out that in the stalled shock phase, the
convective growth timescale $\omega_{\rm buoy}^{-1}$ in the unstable
layer below the shock is of the same order as the timescale for
advection from the shock to the gain radius,
\begin{equation}
  \tadv \equiv \int_{\Rg}^{\Rs} \frac{\dr}{|v_r(r)|},
\label{eq:def_tadv}
\end{equation}
where $\Rg$ is the gain radius, $\Rs$ the shock radius and $v_r$ the
radial velocity. Advection is stabilising because it gives
perturbations only a finite time to grow
in the gain region, before they are advected
into the stable layer below the gain radius.
Considering the local growth rate
$\omega_{\rm buoy}(r)$ given by Eq.~(\ref{omegabuoy})
in a reference frame advected
with the flow, the amplitude $\delta$ of a small-wavelength
perturbation may grow during its advection from the shock to the gain
radius, at best by a factor $\exp(\chi)$,
\begin{equation}
  \delta_{\rm gain} = \delta_{\rm shock} \, \exp(\chi),
\label{eq:deltas_deltag}
\end{equation}
where the quantity
\begin{equation}
  \chi \equiv \int_{\Rg}^{\Rs} \omega_{\rm buoy}(r) \,
\frac{\dr}{|v_r(r)|} = \tadv / \tconv
\label{eq:def_chi}
\end{equation}
can be interpreted as the ratio of the advection timescale to
the average local growth timescale the perturbation experiences,
$\tconv \equiv \langle \omega_{\rm buoy}^{-1} \rangle$ (the latter
quantity is implicitly defined by Eq.~\ref{eq:def_chi}). Thus it
would appear that in order to reach a given perturbation amplitude
at the gain radius, a certain seed perturbation amplitude of the
matter crossing the shock would be necessary.

%  ----- linear case -----

However, a linear stability analysis reveals that the stationary 
accretion flow below the shock is globally unstable and perturbations
can grow from {\em arbitrarily} small initial seeds, if sufficient time
is available \citep{Foglizzo+06}. 
According to \cite{Foglizzo+06} this is the case for
a limited range $[l_{\rm min},l_{\rm max}]$ of modes for which $\chi$
exceeds a critical value $\chi_0$,
\begin{equation}
 \chi > \chi_0\ , \quad\mathrm{where}\quad \chi_0 \approx 3.
\label{eq:cond_chi}
\end{equation}
For $\chi<\chi_0$ the flow remains linearly stable, even though
a negative entropy gradient is present.

% ----- nonlinear case -----

The analysis of \cite{Foglizzo+06} applies only for the linear phase
of the instability, i.e. for small perturbation amplitudes. However,
it is possible that the situation has to be considered as nonlinear
right from the beginning, i.e. that the seed perturbations grow to
large amplitudes already during their advection to the gain radius. In
this context ``large'' can be defined by considering the buoyant
acceleration of the perturbations.

For a small bubble, in which the density $\rho$ is lower than the one
of the surrounding medium, $\rho_{\rm surr}$, the convective growth
during the advection to the gain radius may lead to an increase of the
relative density deviation $\delta \equiv |\rho-\rho_{\rm
  surr}|/\rho_{\rm surr}$ (which can be considered as the perturbation
amplitude) as given by Eq.~\eqref{eq:deltas_deltag}.  The bubble
experiences a buoyant acceleration $|a_{\rm grav}| \, \delta$ towards
the shock, which is proportional to the local gravitational
acceleration $a_{\rm grav}$.  The time integral of the buoyant
acceleration becomes comparable to the advection velocity, when the
perturbation amplitude reaches a critical value
\begin{eqnarray}
  \deltac &\equiv& \frac{ \vgainav }{ \langle a_{\rm grav}
\rangle_{\rm gain} \, \tadv }\\
  & \sim& \frac{\vgainav^2}{\Rs \, \langle a_{\rm grav}
\rangle_{\rm gain} }\frac{\Rs}{\Rg-\Rs}\sim {\cal O}(1\%)\,,
\label{eq:delta_crit}
\end{eqnarray}
where $\vgainav$ and $\langle a_{\rm grav}\rangle_{\rm gain}$ are the
average values of the radial velocity and the gravitational
acceleration in the gain layer, respectively.  For $\deltag > \deltac$
a small-scale perturbation is able to rise against the accretion flow.
If the whole flow is perturbed, the buoyant motions on small scales
affect the situation globally and could allow for the onset of
convective overturn also on larger scales. Note that in contrast to
the linear growth of the instability this process does not require
$\chi>\chi_0$ but it does require large enough seed
perturbations,
\begin{equation}
  \deltas > \frac{\deltac}{\exp(\chi)}\,.
\label{eq:cond_deltas}
\end{equation}
A sufficient condition for the suppression of convection is therefore
that neither Eq.~\eqref{eq:cond_chi} nor Eq.~\eqref{eq:cond_deltas}
are fulfilled.

%----------------------------------------------------------------------
\subsection{The advective-acoustic cycle}
\label{sec:aac}

\begin{figure}
\includegraphics[width=8.5cm]{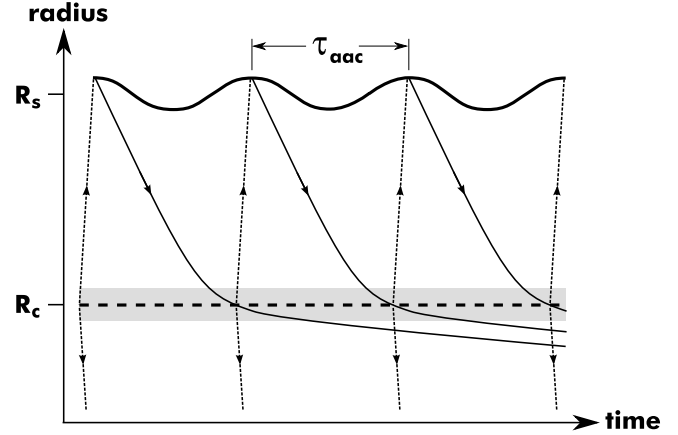}
\caption{Schematic view of the advective-acoustic cycle between the
  shock at $\Rs$ (thick solid line) and the coupling radius, $\Rc$
  (thick dashed line), in the linear regime, shown for the case
  where the oscillation period of the shock ($\tau_{\mathrm{osc}}$)
  equals the cycle duration, $\tau_{\mathrm{aac}}$. Flow lines carrying
  vorticity perturbations downwards are drawn as solid lines, and the
  pressure feedback corresponds to dotted lines with arrows. In the
  gray shaded area around $\Rc$ the flow is decelerated strongly.
  }
\label{fig:scheme}
\end{figure}

A second hydrodynamic instability has recently been recognised to be
of potential importance in the stalled shock phase.
Blondin et al. (2003) noticed that the stalled accretion shock becomes
unstable to non-radial deformations even in the absence of entropy gradients,
a phenomenon termed SASI. It can be interpreted as the result of an
``advective-acoustic cycle'' (in short AAC), as first discussed by
\cite{Foglizzo_Tagger00} in the context of accretion onto black holes,
and later studied for supernovae by \cite{Galletti_Foglizzo05} and
\cite{Foglizzo+06b} by means of linear stability analysis. The
explanation of these oscillations is based on the linear coupling
between advected and acoustic perturbations due to flow gradients.

Although this linear coupling occurs continuously throughout the
accretion flow from the shock to the neutron star surface, some
regions may contribute more than others to produce a pressure feedback
towards the shock and establish a global feedback loop. The analysis
of the linear phase of the instability in Sect.~\ref{sec:linear}
reveals the importance of a small region at a radius $\Rc$ above the
neutron star surface, where the flow is strongly decelerated. The
feedback loop can be described schematically as follows: small
perturbations of the supernova shock cause entropy and vorticity
fluctuations, which are advected downwards. When the flow is
decelerated and compressed above the neutron star surface, the
advected perturbations trigger a pressure feedback.
This pressure feedback perturbs the
shock, causing new vorticity and entropy perturbations.  Instability
corresponds to the amplification of perturbations by a factor $|\Qaac|
>1$ through each cycle.

The duration $\tau_{\rm aac}^{\mathrm{f}}$ of each cycle is a fundamental
timescale. It corresponds to the time needed for the advection of
vortical perturbations from the shock to the coupling radius $\Rc$,
where the pressure feedback is generated, plus the time required by
the pressure feedback to travel from this region back to the shock.

\begin{table*}[tpbh!]
\begin{center}
  \caption{Important quantities for the simulations discussed in this
    paper.} 
\label{tab:restab_limcas}
\begin{tabular}{lrrrrrrrrrrr}
\hline
\hline
Name$^\dagger$ & $\Lib$  & $R_{\rm ib}^{\rm f}$ & $\tib$ & $\delta_{\rm i}$    & $L_{\nu_{\mathrm{e}}+\bar\nu_{\mathrm{e}}}$ & $t_{\rm nl}$ & $\texp$ & $\dMg$
       & $\eexp$ & $\Mns$   & $v_{\mathrm{ns}}$ \\
       & [B/s] & [km]     & [s]    &       & [B/s]   & [s]       & [s]     & [$\Msol$] & [B]  & [$\Msol$] & [km/s] \\
\hline
W00FA  &    -- &   8.0 & 0.5  & $10^{-3} \, v_r$ &    0.0   &   --      &   --     &     -- &   --    &     --    &  --   \\
\hline
W00F   &   0.2 &   8.0 & 0.5  & $10^{-3} \, v_r$ &   99.0   & 0.154     & 0.194    &  0.004 &  0.37   &    1.50   &  200  \\
W00    &   0.2 &  15.0 & 1.0  & $10^{-3} \, v_r$ &   51.7   & 0.346     &   --     &     -- &   --    &    1.63   &   $-$3  \\
W00S   &   0.2 &  15.0 & 4.0  & $10^{-3} \, v_r$ &   29.6   &  --       &   --     &     -- &   --    &    --     &  --   \\
W05S   &   7.4 &  15.0 & 4.0  & $10^6$cm/s       &   38.3   &  --       &   --     &     -- &   --    &    --     &  --   \\
W05V   &   7.4 &  15.0 & 10.0 & $3\times 10^7$cm/s & 33.5   &  --       &   --     &     -- &   --    &    --     &  --   \\
\hline
W12F   &  29.7 &  10.5 & 0.25 & $10^{-3} \, v_r$ &  109.3   & 0.144     & 0.164    &  0.010 &  0.87   &    1.44   & $-$558  \\
W12F-c &  29.7 &  10.5 & 0.25 & ${\cal{O}}(10^{-2}) \, v_r$ & 112.5  & 0.090     & 0.117    &  0.015 &  0.94   &    1.41   &  612  \\
\hline
\end{tabular}
\end{center}

$^{\dagger}$
    The constant boundary luminosity is denoted by $\Lib$,
    $R_{\rm ib}^{\rm f}$ is the asymptotic inner boundary radius (see
    Eq.~\ref{eq:ribtimdep}),
    $\tib$ the contraction timescale of the inner boundary,
    $\delta_{\rm i}$ the amplitude of the initial velocity perturbations
    (which was chosen to be proportional to the local radial velocity
    $v_r$ in most cases --- see text for details),
    $L_{\nu_{\mathrm{e}}+\bar\nu_{\mathrm{e}}}$ is the luminosity
    of electron neutrinos and antineutrinos at a radius of
    500$\,$km at 150$\,$ms after the start of the simulations,
    $t_{\rm nl}$ the
    time at which the average lateral velocity in the gain
    layer exceeds $10^9\,\mathrm{cm/s}$, $\dMg$ the mass in the
    gain layer at the time $\texp$ when the explosion starts,
    $\eexp$ the explosion energy, $\Mns$ the neutron star mass, and
    $\vns$ the neutron star velocity as computed from momentum
    conservation. The last three quantities are given at $t=1\,$s, for
    Model W00F at $t=750\,$ms, and $t_{\mathrm{exp}}$ is defined
    as the moment when the energy of the matter in the gain layer
    with positive specific energy
    exceeds $10^{49}\,$erg. Model W00FA is a hydrodynamic simulation
    without including neutrino effects. Only Models W00F, W12F, and
    W12F-c developed explosions and values for the corresponding
    explosion and neutron star parameters are given. The neutrino
    luminosities imposed at the inner boundary are kept constant
    during a time $t_L = 1\,$s. The energy unit $1\,\mathrm{B} = \foe$
    is used and all times are measured from the start of the simulation,
    i.e.  $t=0\,$s means $16\,$ms after core bounce.
\end{table*}

The oscillatory exponential growth resulting from the AAC can be 
described by a
complex eigenfrequency $\omega = \omega_r + i \omega_i$ satisfying the
following equation:
\begin{equation}
\exp(-i\omega  \tau_{\rm aac}^{\mathrm{f}})=\Qaac,\label{eq:dispersion}
\end{equation}
where the real part $\omega_r$ is the oscillation frequency and the
imaginary part $\omega_i$ is the growth rate of the AAC.  Note that
Eq.~(\ref{eq:dispersion}) is a simplified form of Eq.~(49) of
\cite{Foglizzo02}. For the sake of simplicity, it neglects the
marginal influence of the purely acoustic cycle of pressure waves
trapped between the shock and the accretor. This hypothesis is
motivated by the recent estimate of the efficiencies $\Qaac$,
${\cal R}_{\rm ac}$ of the advective-acoustic and of the purely acoustic
cycles, respectively, obtained by \cite{Foglizzo+06b} in a simpler set-up
when the frequency is high enough to allow for a WKB approximation.
According to their Figs.~8 and 9, the purely acoustic cycle is always stable
($|{\cal R}_{\rm ac}|<1$), whereas the advective-acoustic cycle is unstable
($|\Qaac|>1$)\footnote{This discussion applies for the growth behavior
of the advective-acoustic cycle and of the purely acoustic cycle in the
presence of a pressure feedback produced at the coupling radius and
reaching the shock. The calculation of ${\cal R}_{\rm ac}$ and
$\Qaac$ by \cite{Foglizzo+06b} does {\em not} assume a purely radial
acoustic feedback but fully takes into account the azimuthally
traveling sound waves as well as evanescent pressure waves (pseudosound),
which do not propagate. Nevertheless, it is currently a controversial
issue whether the analysis by \cite{Foglizzo+06b} allows one to draw
conclusions on the kind of instability proposed by \cite{Blondin_Mezzacappa06}
and \cite{Blondin_Shaw07}, who advocate a growth mechanism driven by
sound waves traveling solely in the angular direction. 
The possibility of understanding the development of SASI modes in our
hydrodynamic supernova simulations by an acoustic cycle with non-radial
sound wave propagation will be discussed in Sect.~\ref{sec:linear}.}.
According to Eq.~(\ref{eq:dispersion}), the oscillation period
$\tau_{\rm  osc}\equiv 2\pi/\omega_r$ of the AAC depends both on the
duration $\tau_{\rm aac}^{\mathrm{f}}$ of the cycle and on the phase 
$\varphi$ of $\Qaac$:
\begin{equation}
\omega_r\tau_{\rm aac}^{\mathrm{f}}+\varphi=2n\pi,\label{phaseQ}
\end{equation}
where $n$ is an integer labelling the different harmonics. 
In the particular flow studied by
\cite{Foglizzo+06b}, the oscillation period of the fundamental mode is
a good estimate of the duration of the cycle
($\tau_{\rm  osc}\sim\tau_{\rm aac}^{\mathrm{f}}$).
This simple relationship is not obvious a priori. For example,
if $\Qaac$ were real and negative (i.e., $\varphi = \pi$), 
the oscillation period would scale like
$\tau_{\rm  osc}\sim 2 \tau_{\rm aac}^{\mathrm{f}}$ (because enhancing
feedback requires a phase coherence between the amplifying mechanism
and the shock oscillation).

% The fundamental mode of the AAC is thus
% identical to the duration $ \tau_{\rm aac}^{\rm f}$ of the cycle only
% in the particular case when $\Qaac$ is real and positive. However,
% according to \cite{Foglizzo+06b} $|\varphi|$ is small and therefore
% $\tau_{\rm osc} \approx \tau^{\rm f}_{\rm aac}$ is a reasonable
% approximation for the fundamental oscillation period, which we will
% use in the following.

The amplitude of perturbations in the AAC increases like
$\exp(\omega_i \, t)$, with a growth rate $\omega_i$ deduced from
Eq.~(\ref{eq:dispersion}):
\begin{equation}
  \omega_i \equiv \frac{ \ln( |\Qaac| ) }{ \tau_{\rm aac}^{\mathrm{f}} }.
\label{eq:def_sigmaaac}
\end{equation}
Comparing Eqs.~\eqref{eq:def_chi} and \eqref{eq:def_sigmaaac} it
is interesting to note that a small advection timescale suppresses 
the growth of entropy-driven convection whereas it
leads to higher growth rates for the AAC (neglecting the
logarithmic dependence on $\Qaac$). Thus the AAC may operate under
conditions which are not favourable for convection and vice versa.
Investigating this further is one of the goals of this work.

%=====================================================================
%\section{Computational methods and numerical setup}
\section{Numerical setup and models}
\label{sec:numsetup}

In order to investigate the importance of instabilities like the ones
discussed in the previous section during the post-bounce evolution of
core-collapse supernovae, we performed a series of two-dimensional (2D)
hydrodynamic simulations. For this purpose
we used the same numerical setup as in Paper~I. We employed the version
of the hydrodynamics code that was described by \cite{Kifonidis+03}.
It is based on the piecewise parabolic method (PPM) of \cite{CW84},
assuming axisymmetry and adopting spherical
coordinates $(r,\theta)$. The calculations were performed on a
polar grid that had typically 800 zones in radial
direction and 360 zones in lateral direction (extending from polar angle
$\theta=0$ to $\theta=\pi$). The lateral grid was equidistant while the
radial grid had logarithmic spacing with a ratio of radial zone size
to radius that did not exceed 1\%. For the neutrino number and energy
transport we applied a gray, characteristics-based transport scheme
that was able to efficiently approximate the transport in the 
transparent and semi-transparent regimes up to optical
depths of several 100. Only transport in the radial direction
was taken into account, but we allowed for lateral variations of the 
neutrino flux by solving one-dimensional transport equations 
independently for all discrete polar angles of the $r$-$\theta$ grid. 
A detailed description of the transport method is given in Paper~I.

\begin{figure}[tpb!]
\centering
\resizebox{\hsize}{!}{\includegraphics{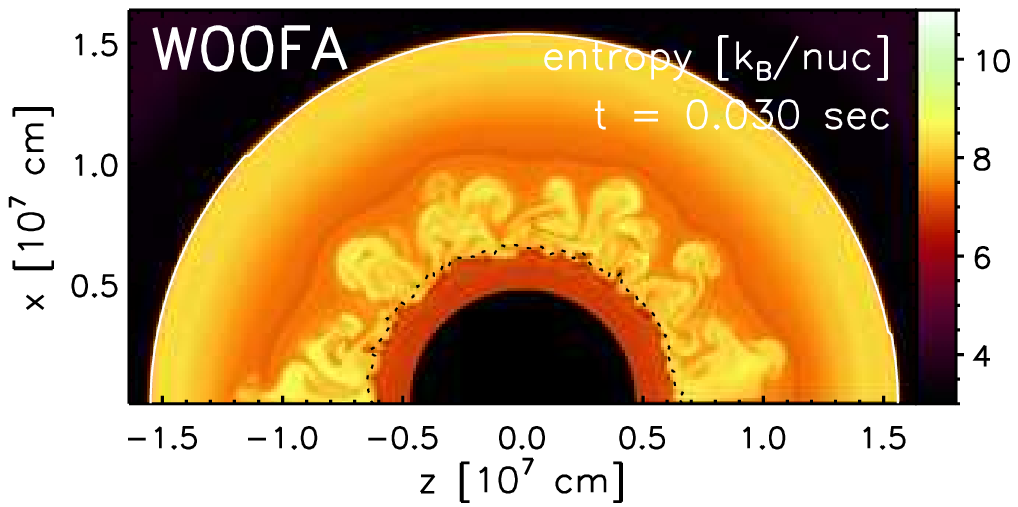}}
\resizebox{\hsize}{!}{\includegraphics{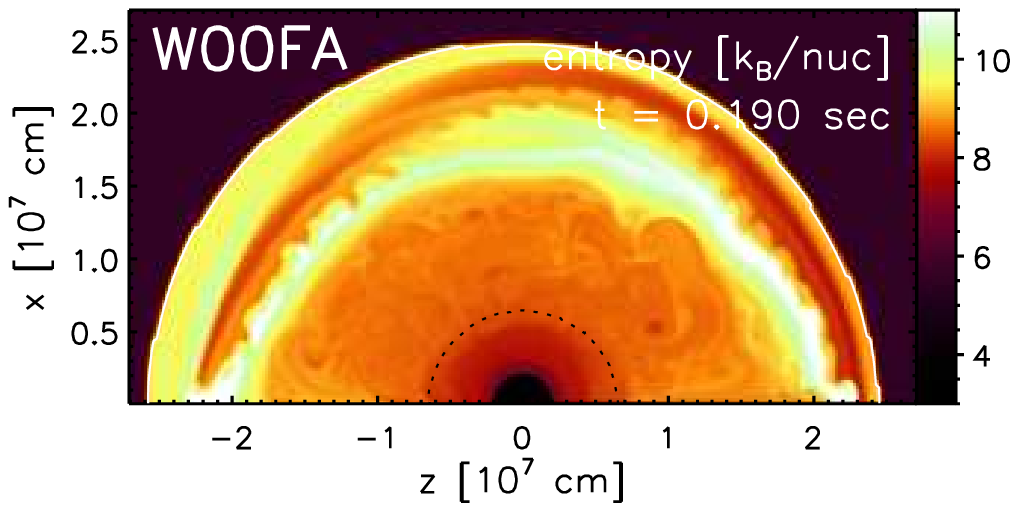}}
\caption{Entropy distribution of model W00FA $30\,$ms and $190\,$ms after
  the start of the simulation.  The initial entropy profile and
  postshock entropy gradients caused by shock motions
  give rise to weak convection. A low-amplitude $l=1$ oscillation
  develops. 
 {\em (Color figures are available in the online version of our paper.)}}
\label{fig:limcas_stot_w00fa}
\end{figure}

\begin{figure}[tpb!]
\centering
\resizebox{\hsize}{!}{\includegraphics{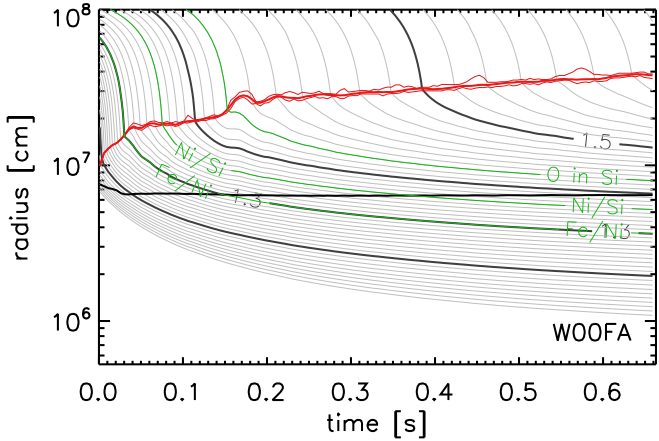}}
\caption{Mass-shell trajectories for model W00FA. The spacing of the
  thin lines is $0.01\,\Msol$. Green lines mark the mass shells at
  which the composition of the progenitor changes. The red lines are
  the minimum, average, and maximum shock radii, the black line marks
  the radius, at which the average density is
  $10^{11}\,\mathrm{g/cm^3}$. The difference between minimum and
  maximum shock radius is caused by bipolar shock oscillations (see
  Fig.~\ref{fig:limcas_stot_w00fa}).
 {\em (A color figure is available in the online version of our paper.)}}
\label{fig:limcas_mshell_w00fa}
\end{figure}

\subsection{Initial and boundary conditions}

We used an initial model (the `W' model from Paper~I) that was
obtained by evolving the $15\,\Msol$ supernova progenitor s15s7b2 
of \cite{WW95} through collapse and core bounce until shock 
stagnation in a simulation 
with a detailed, energy-dependent treatment of neutrino transport
(\citealt{Buras+03}; see their Model s15).
We started our runs at a time of $16\,$ms after core
bounce. In order to enable the growth of hydrodynamic instabilities we
perturbed the initial model, unless noted otherwise, by adding random,
zone-to-zone velocity perturbations of $0.1\%$ amplitude.

The neutron star core (i.e. typically the innermost $1.1\,\Msol$) was 
not included in our simulations but was replaced by a contracting inner
boundary of the computational grid. Boundary conditions
were imposed there for the hydrodynamics and the neutrino
transport, and a point-mass potential of the excised core was adopted
to account for the gravitational influence of this region. Although
the treatment of gravity is not of primary relevance for the 
fundamental questions studied in this paper, we mention here that 
the description of the gravitational potential took into account
the self-gravity of the gas on the grid with its two-dimensional
distribution, as well as an approximative treatment of general 
relativistic effects (for details, see Paper~I).

The inner grid boundary was placed at a Lagrangian shell with 
enclosed mass of $M=1.1\,\Msol$ at which we imposed conditions
describing hydrostatic equilibrium. Its radius was assumed to evolve  
according to
\begin{equation}
  \Rib(t) =  \frac{R_{\rm ib}^{\rm i}}{1 \, + \, (1-\exp(-t/\tib)) \,
  (R_{\rm ib}^{\rm i}/R_{\rm ib}^{\rm f}-1)} \ ,
\label{eq:ribtimdep}
\end{equation}
where $R_{\rm ib}^{\rm f}$ is the final (asymptotic) boundary radius,
$\tib$ is the contraction timescale and $R_{\rm ib}^{\rm i} \approx
65\,$km is the initial radius, which is given by the initial model.
The neutrino luminosities from the neutron star core, which we imposed 
at the inner boundary, were assumed to be constant during the 
first second after core bounce. This simple choice can be justified
by the results of core-collapse simulations with sophisticated neutrino
transport (see Paper~I). 

With this approach we parametrized the cooling and shrinking of
the core of the nascent neutron star
and its neutrino emission, which all depend on the incompletely
known properties of the nuclear equation of state. Different choices
of the boundary motion and strength of the neutrino emission allowed
us to vary the properties of the supernova explosion and of
the developing hydrodynamic instabilities in the region between
neutron star and stalled shock. It is very important to note that
the stagnation radius of the stalled shock reacts sensitively not
only to the mass infall rate from the collapsing progenitor star
and to the rate of neutrino heating in the gain layer,
but also to the contraction behavior of the neutron star. A faster
contraction usually leads to a retraction of the shock, whereas a 
less rapid shrinking of the neutron star allows the shock to 
expand and stagnate at a larger radius. This, of course, causes
important differences of the postshock flow and thus affects the
growth of non-radial hydrodynamic instabilities.

In some of the simulations discussed here, the rapid contraction
of the forming neutron star caused the density and sound
speed at the inner boundary to become so high 
that the hydrodynamic timestep was severely limited by the 
Courant-Friedrich-Lewy (CFL) condition. Moreover, when the optical
depth in this region increased to more than several hundred, numerical
problems with our neutrino transport method occurred unless very fine
radial zoning was chosen, making the timestep even smaller. In such
cases we moved the inner grid boundary to a larger radius and bigger 
enclosed mass (i.e., we increased the excised neutron star core).
Hereby we attempted to change the contraction behavior of the  
nascent neutron star as little as possible.
The new inner boundary was placed at a radius $\rib{\!\!\! '}\,$ where the 
optical depth for electron neutrinos was typically around $100$. 
When doing this, the gravity-producing mass of the inner core was 
adjusted appropriately \citep[see][]{Arcones+06} and the boundary neutrino
luminosities were set to the values present at $\rib{\!\!\! '}\,$ at
the time of the boundary shifting, thus 
making sure that the gravitational acceleration, the neutrino flux, and
neutrino heating and cooling above $\rib{\!\!\! '}\,$ followed a continuous 
evolution. The parameters in Eq.~(\ref{eq:ribtimdep}) were adjusted 
from the old values $R_{\rm ib}^{\rm i}$ and $R_{\rm ib}^{\rm f}$ to
new values $\widetilde{R}_{\rm ib}^{\rm \,i}$ and 
$\widetilde{R}_{\rm ib}^{\rm \,f}$, respectively, in the following way:
\begin{align}
\widetilde{R}_{\rm ib}^{\rm \,i} &\;=\; R_{\rm ib}^{\rm i} \times
                                      (\rib{\!\!\! '}\,/\rib)\,, \nonumber\\
\widetilde{R}_{\rm ib}^{\rm \,f} &\;=\; R_{\rm ib}^{\rm f} \times
                                      (\rib{\!\!\! '}\,/\rib)\,. 
\label{eqn:bndry_shift}
\end{align}
This simple rescaling had the consequence that during the subsequent 
evolution small differences of the contraction velocity of the new inner
boundary and therefore of the settling neutron star appeared, which led
also to minor changes of the decay of the neutrino luminosities with
time. Nevertheless, no significant impact on the simulations
was observed, e.g., Model W00F developed an explosion at the same
time, independent of whether or not the boundary was shifted according 
to our recipe.

%----------------------------------------------------------------------

\begin{figure}[tpbh!]
\centering
%\begin{tabular}{cc}
\includegraphics[angle=0,width=8.5cm]{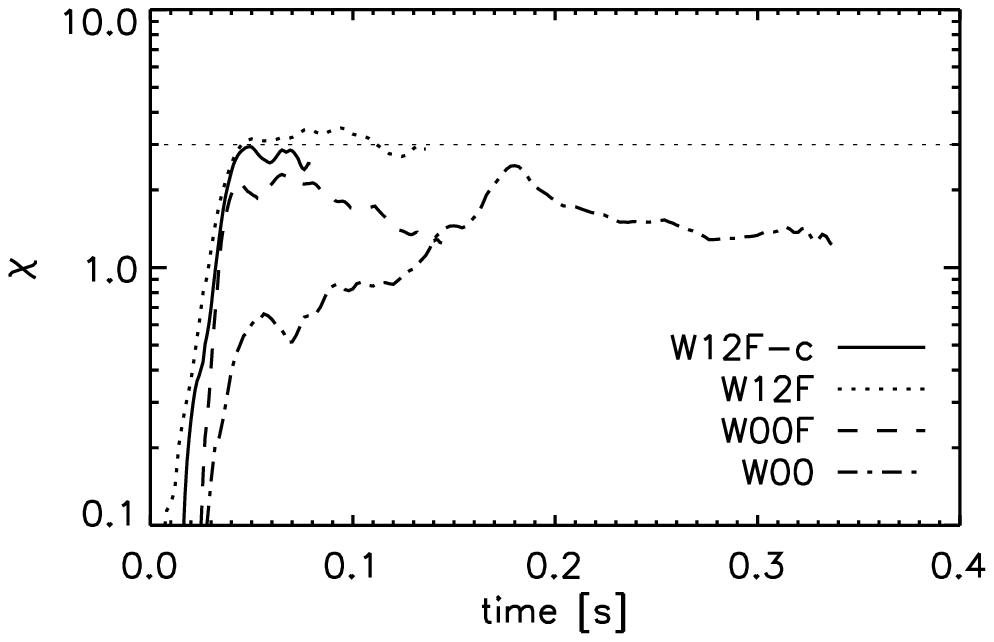}       %&
\includegraphics[angle=0,width=8.5cm]{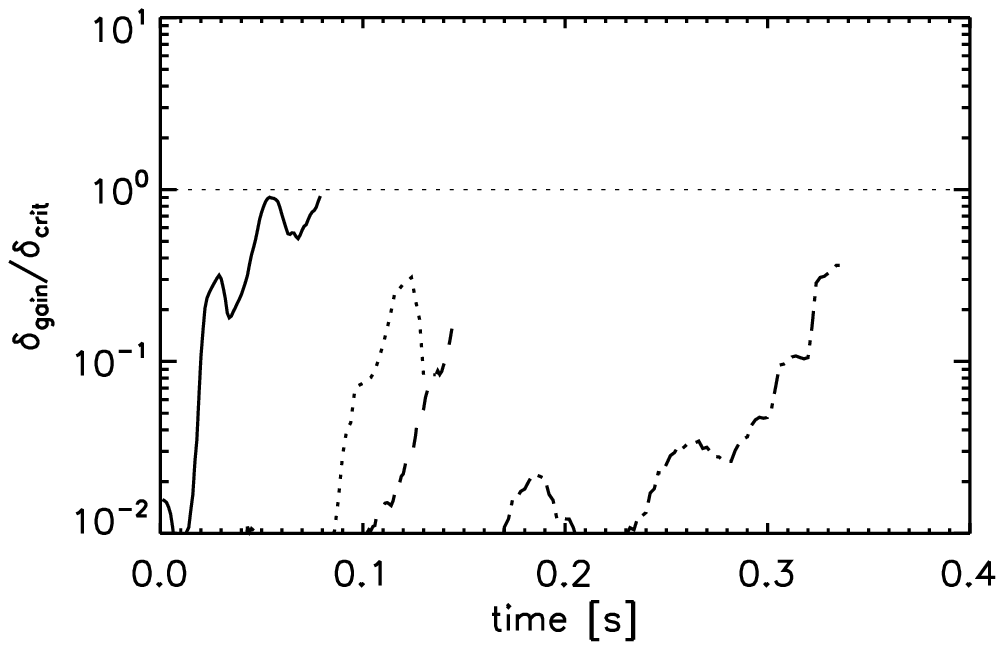}
%\end{tabular}
\caption{Evolution of the quantity $\chi$ (upper panel, see
  Eq.~\ref{eq:def_chi}) and of the ratio $\deltag/\deltac$ (lower panel,
  see Eqs.~\ref{eq:deltas_deltag}, \ref{eq:delta_crit}) for Models
  W12F-c, W12F, W00F and W00. All lines end 10$\,$ms before the
  nonlinear phase begins at
  time $t_{\rm nl}$, when the average lateral velocity in the
  gain layer exceeds $10^9\,\mathrm{cm/s}$. At later times $\chi$ and
  $\deltac$ cannot be measured reliably any longer.  In all models,
  $\chi \lesssim \chi_0 \approx 3$ at $t<t_{\rm nl}$. Only in Model
  W12F-c the ratio $\deltag/\deltac$ gets very close to unity before
  $t=t_{\rm nl}$,
  which means that only in this model convection is able to set in
  faster than the SASI.}
\label{fig:chi_denspert}
\end{figure}

%----------------------------------------------------------------------

\begin{figure}[tpbh!]
\resizebox{\hsize}{!}{\includegraphics{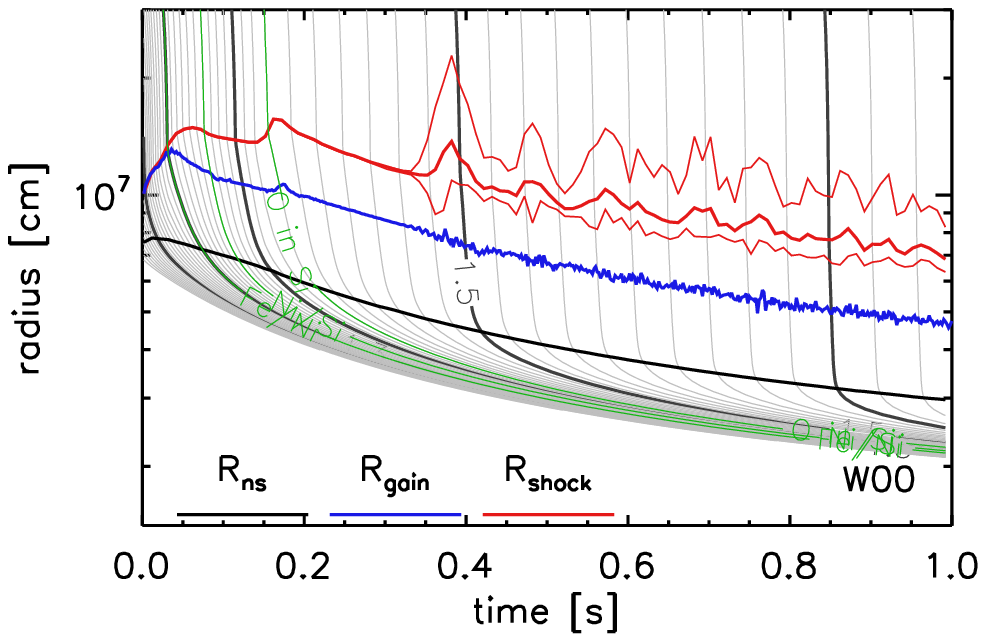}}
\resizebox{\hsize}{!}{\includegraphics{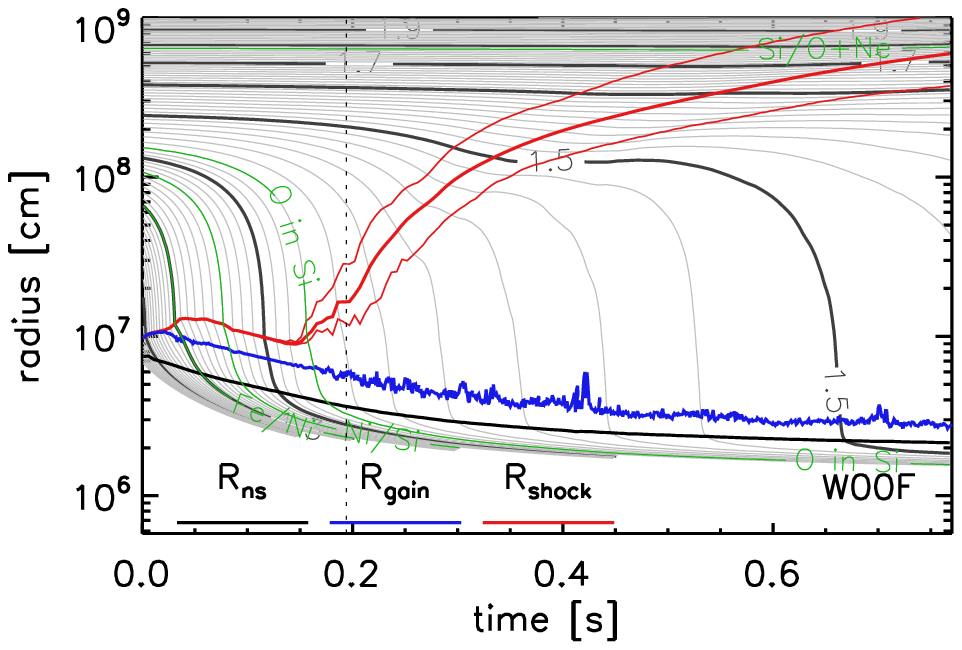}}
\caption{Same as Fig.~\ref{fig:limcas_mshell_w00fa}, but for Models W00
  (upper panel) and W00F (lower panel). The blue line marks the position
  of the gain radius. Up to $t \approx 300\,$ms
  Model W00 remains nearly spherical and evolves like the corresponding
  one-dimensional simulation. However, an
  initially very weak $l=1$ oscillation mode in the postshock flow
  grows in this phase and finally becomes nonlinear, causing strong
  shock oscillations. Yet, this model does not explode. Although the
  shock expands transiently in a quasi-periodic manner, the average shock
  radius decreases and all matter remains bound. In Model W00F an
  $l=2$ mode develops and starts to affect the shape of the shock at $t
  \approx 150\,$ms, much earlier than in Model W00. The oscillations
  become nonlinear, and at $t=194\,$ms (marked by a dotted vertical line)
  the model explodes.
  {\em (Color figures are available in the online version of our paper.)}}
\label{fig:limcas_w00_mshells}
\end{figure}

\subsection{Model parameters}
\label{sec:modelpars}

The characteristic parameters and some important quantities of the
eight models investigated here are listed in Table~\ref{tab:restab_limcas}.
The models differ concerning the included physics, assumed boundary
conditions, and the initial perturbations used to seed the
growth of hydrodynamic instabilities.

The most simplified case we considered, Model W00FA, is a purely 
hydrodynamic simulation without including neutrino effects. This 
choice follows \cite{Blondin+03}, who also ignored neutrinos.
In comparison with the other models we computed, it allows
us to study the influence of neutrino cooling and heating.
\cite{Blondin+03} also placed an inner boundary at a fixed radius and 
applied outflow conditions there to allow for a steady-state accretion
flow (alternatively, they also tested reflecting conditions with a 
cooling term to keep the shock at a steady radius). 
In contrast, in Model W00FA accretion is enabled by the 
retraction of the inner boundary of the computational grid, which 
mimics the Lagrangian motion of a mass shell in a contracting
neutron star. Another difference from \cite{Blondin+03} is the fact that 
in our models the accretion rate shrinks when infalling matter from 
the less dense layers at increasingly larger radii reaches the shock. 
Thus the development of hydrodynamic instabilities occurs
in a situation that is generically non-stationary. 

In five other simulations we included neutrinos and chose boundary
conditions such that the growth of convection was suppressed. This allowed
us to identify and study other instabilities like the SASI more easily.
The suppression of convection could be achieved by prescribing vanishing
or negligibly low core luminosities. In such cases only the
luminosity produced between the inner boundary and the gain radius
causes neutrino energy deposition in the gain layer. Therefore the neutrino
heating remains weak, resulting in a shallow entropy gradient and consequently
in a large growth timescale for convection. This implies that for low core
luminosities the ratio of the advection to the buoyancy timescale, $\chi$
(Eq.~\ref{eq:def_chi}), remains below the critical value and therefore
in spite of a negative entropy gradient in the neutrino heating region, the 
postshock layer remains convectively stable due to the rapid advection 
of the gas down to the gain radius (see Sect.~\ref{sec:convection}).
The five simulations where this is the case are Models W00F, W00, W00S, 
W05S, and W05V. These models differ in the prescribed contraction of the
inner boundary. Models W00 and W00F employ the ``standard'' and
``rapid'' boundary contraction, respectively, of Paper~I. In
order to cover a wider range of advection timescales --- which will 
help us to gain deeper insight into the mechanism that causes the
low-mode instability found in our simulations (see Sect.~\ref{sec:linear})
--- we performed three simulations with slower boundary contraction,
namely Models W00S, W05S, and W05V (Table~\ref{tab:restab_limcas}). 
In the last two models the core neutrino luminosity has a 
non-negligible (but still fairly low) value. The correspondingly 
enhanced neutrino heating leads to larger shock radii and thus longer
advection timescales. Models W00F, W00, and W00S were computed with
our standard initial perturbations (0.1\% random noise on the
velocity). For Models W05S and W05V an $l=1$ velocity perturbation
was applied. This allowed us to suppress high-mode noise and to 
measure the oscillation period of the low-mode instability despite
the low growth rates in these models.

For Models W12F and W12F-c, finally, we adopted boundary conditions
that were guided by core-collapse simulations with sophisticated
multi-group neutrino transport. The contraction
of the inner boundary was chosen to match the motion of the
corresponding mass shell in such simulations for the same
progenitor \citep{Buras+03}. The boundary luminosity we imposed 
led to typical explosion energies of about $\foe$. Despite the
non-negligible core luminosity convection in these models was
suppressed because of the rapid boundary contraction. The latter caused
the radius of the stalled shock to become rather small, and consequently
the accretion velocities in the postshock layer were very large.
Therefore the advection timescale was short and
the parameter $\chi$ did not exceed the critical value
of about 3. In such a situation the amplitude of the progenitor
perturbations can decide about whether convection sets in (starting
in the nonlinear regime as discussed in Sect.~\ref{sec:convection}) 
or not. Since the properties of the perturbations in the progenitor
star are not well known, we decided to explore two cases, 
one (Model W12F) with small initial perturbations (our standard 
0.1\% velocity perturbation) such that the growth of convection
was suppressed, and another case where the initial perturbations 
were large enough so that convection could develop. 
For the latter model, W12F-c, we used the same perturbations as 
for Models W12-c and W18-c of Paper~I with amplitudes of up 
to several percent and a spatial variation as given by the velocity
fluctuations that had grown during a 2D core-collapse simulation
of a 15$\,M_\odot$ star (Model s15r of \citealt{Buras+03} and
\citealt{Buras+06b}).

\begin{figure*}[tbph!]
\centering
\begin{tabular}{cc}
\includegraphics[angle=0,width=8.0cm]{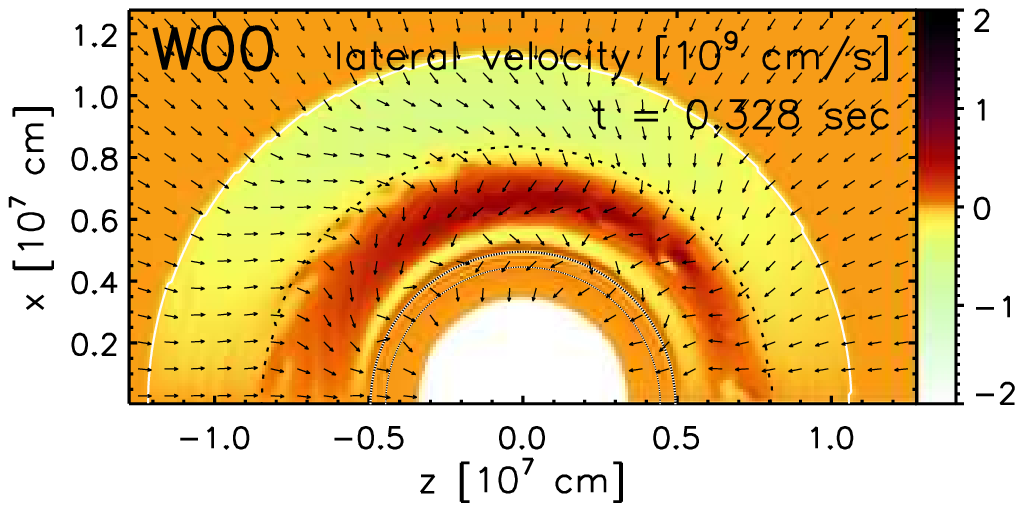} &
\includegraphics[angle=0,width=8.0cm]{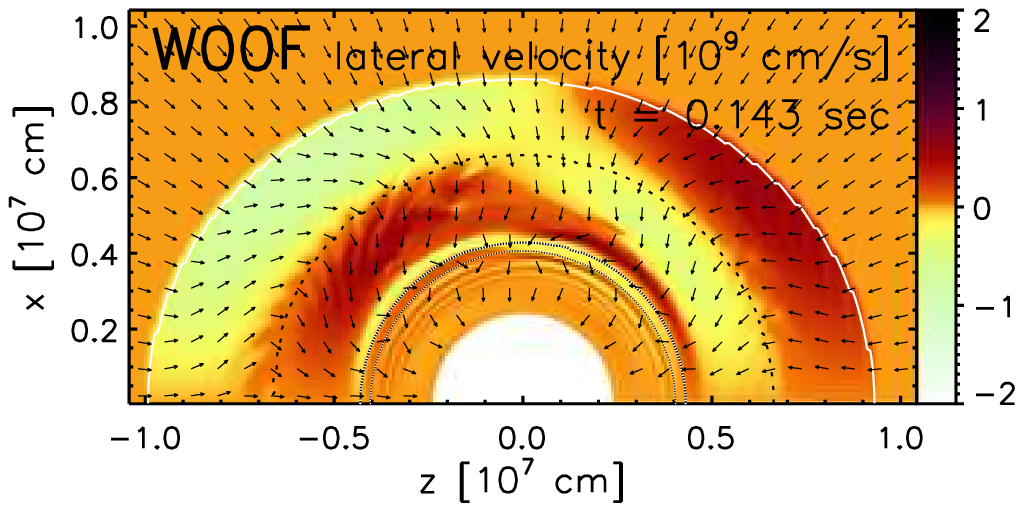} \\
\includegraphics[angle=0,width=8.0cm]{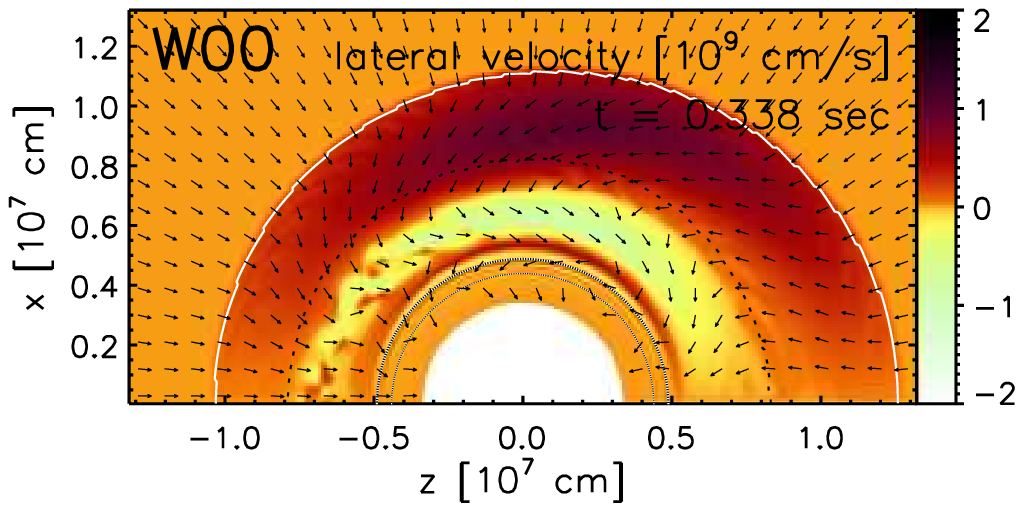} &
\includegraphics[angle=0,width=8.0cm]{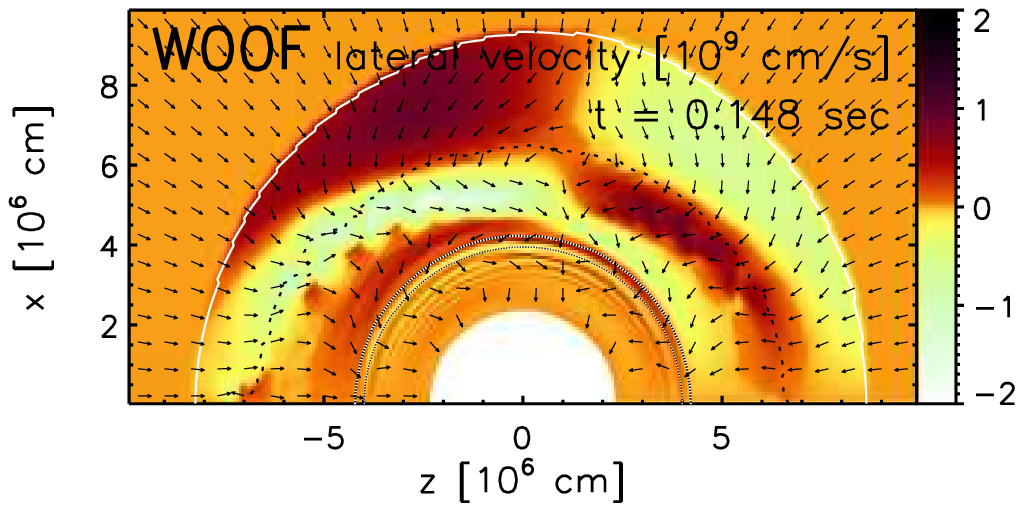} \\
\includegraphics[angle=0,width=8.0cm]{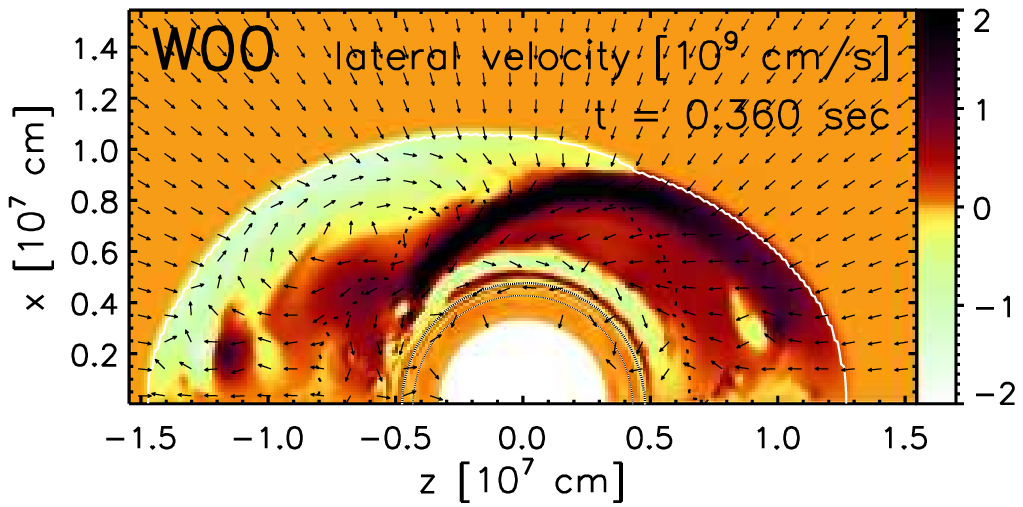} &
\includegraphics[angle=0,width=8.0cm]{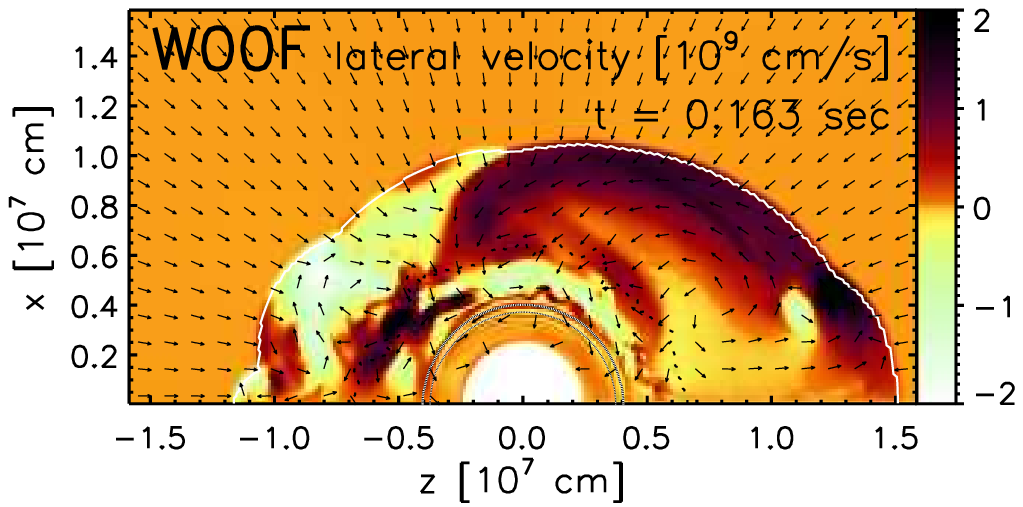} \\ %153
\end{tabular}
\caption{Lateral velocities (color coded; superimposed are the vectors
  of the velocity field, which
  indicate the direction of the flow) for Models W00 and W00F.  The
  white lines mark the shock, the black dotted lines the gain radius.
  For both models we show the situation at two times near $t=t_{\rm
    nl}$, at which the oscillations are in opposite phases (i.e. the
  times differ by half an oscillation period), and at a third time, when
  the oscillations have run out of phase (see
  Sect.~\ref{sec:trigger_conv}). In Model W00 an $l=1$ SASI mode develops,
  i.e., the still nearly spherical shock moves back and forth along
  the $z$-axis, whereas in Model W00F an $l=2$ SASI mode becomes dominant,
  i.e., the
  shock oscillates between a prolate and an oblate deformation. The
  postshock matter attains high lateral velocities, because the radial
  preshock flow hits the shock at an oblique angle when the shock is
  nonspherical or when its center is displaced from the grid center.
  {\em (Color figures are available in the online version of our paper.)}}
\label{fig:w00_evo_vely}
\end{figure*}

\begin{figure*}[tbph!]
\centering
\begin{tabular}{cc}
\includegraphics[angle=0,width=7.7cm]{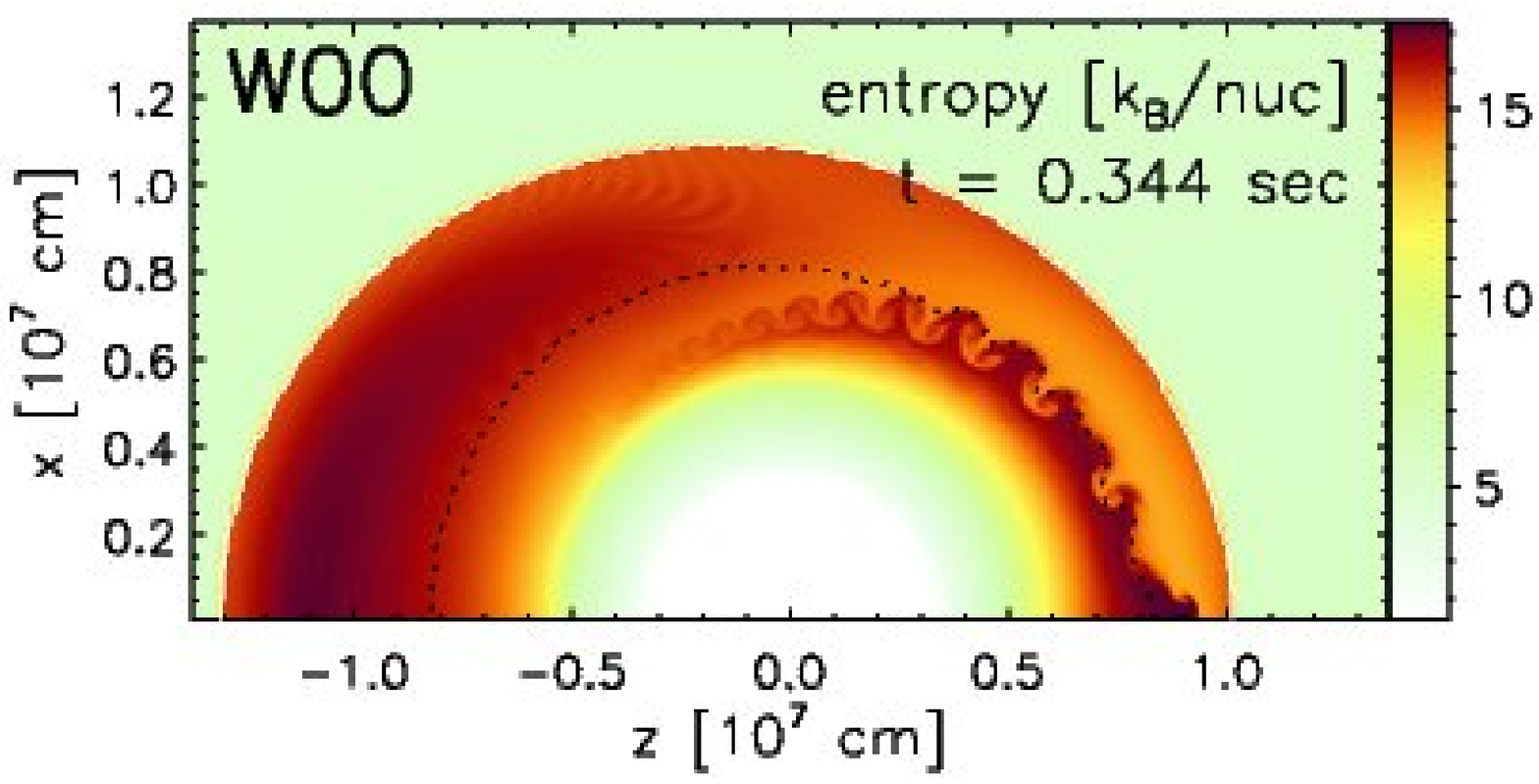} &
\includegraphics[angle=0,width=7.7cm]{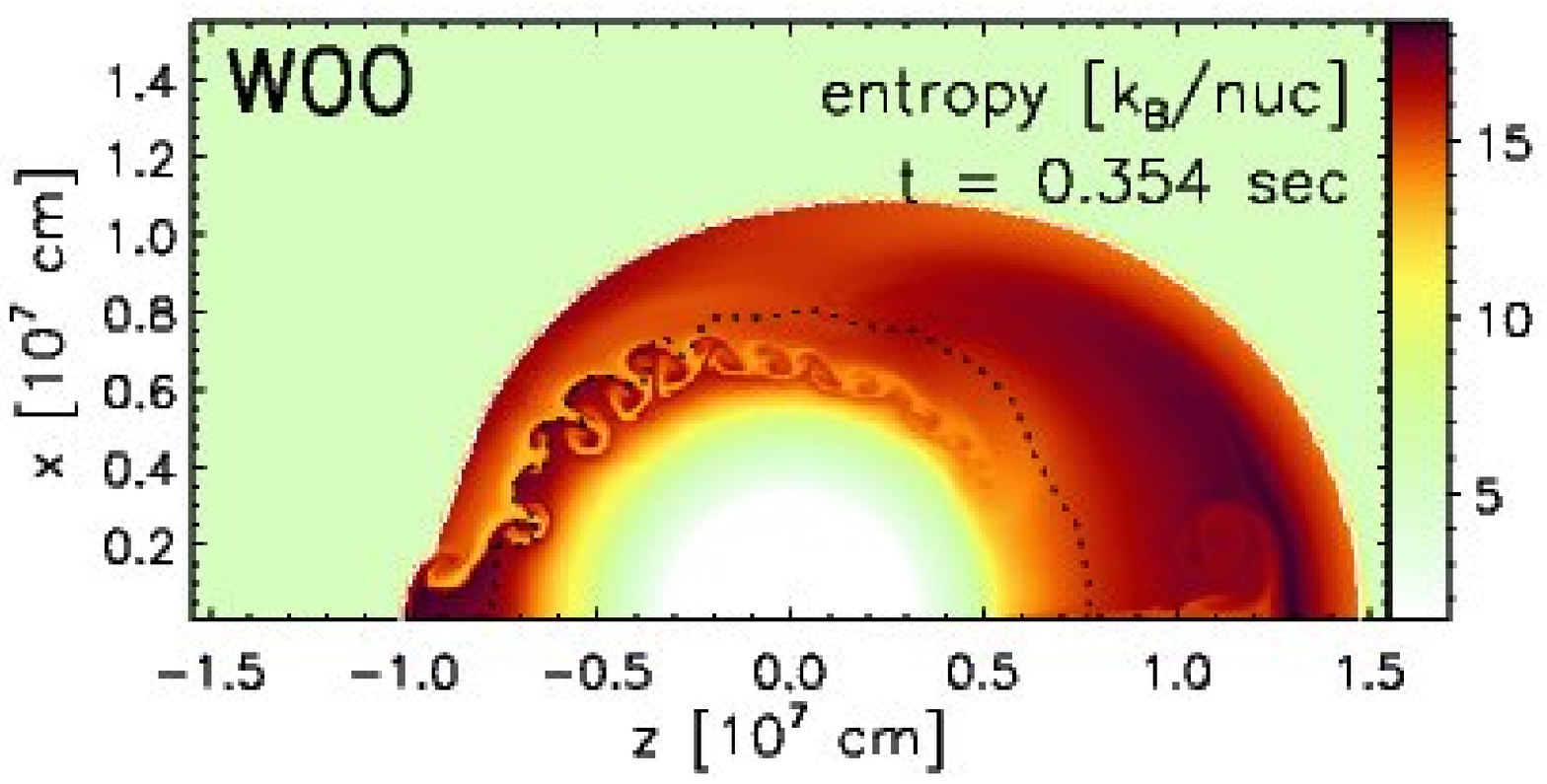} \\
\includegraphics[angle=0,width=7.7cm]{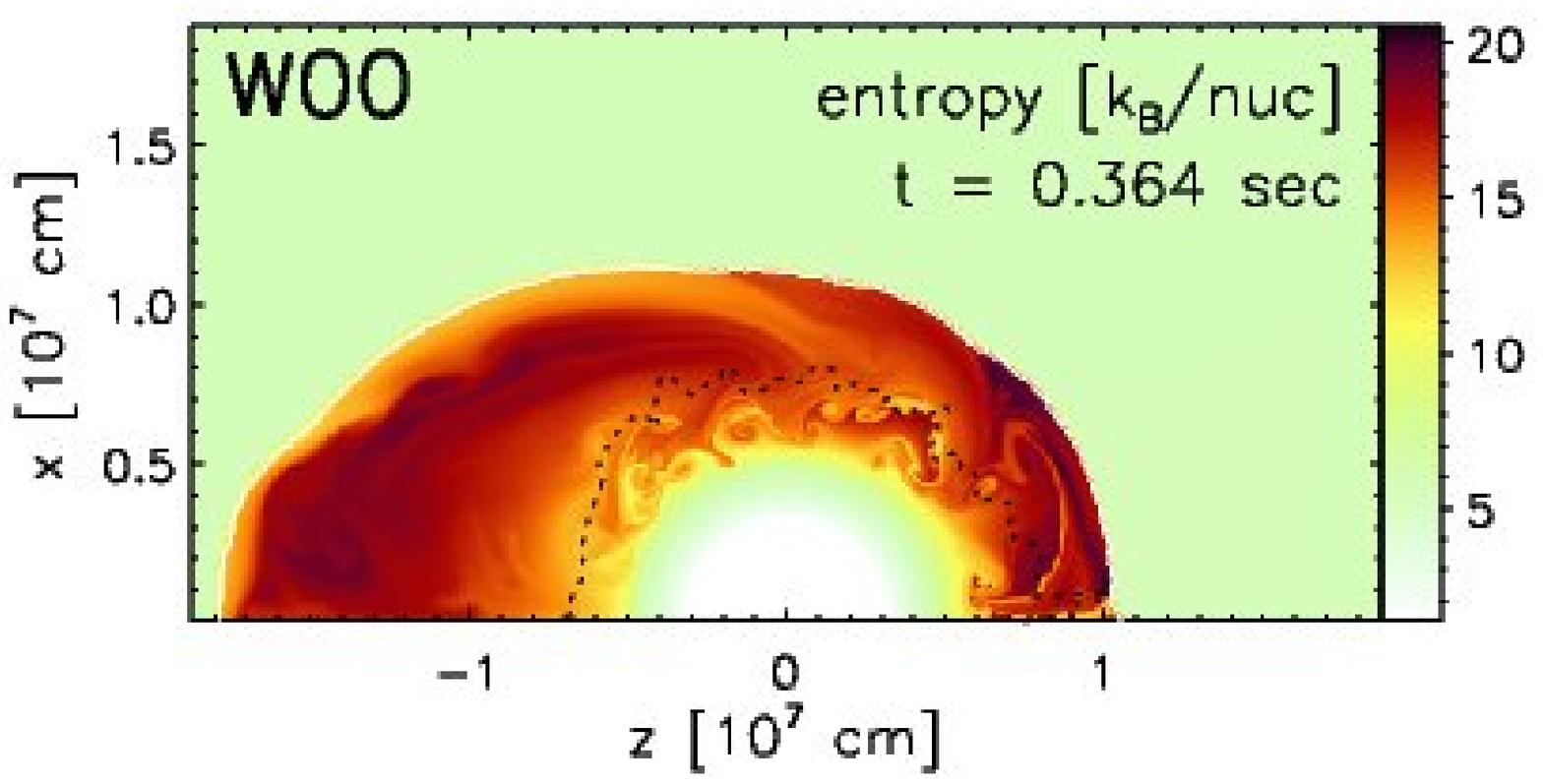} &
\includegraphics[angle=0,width=7.7cm]{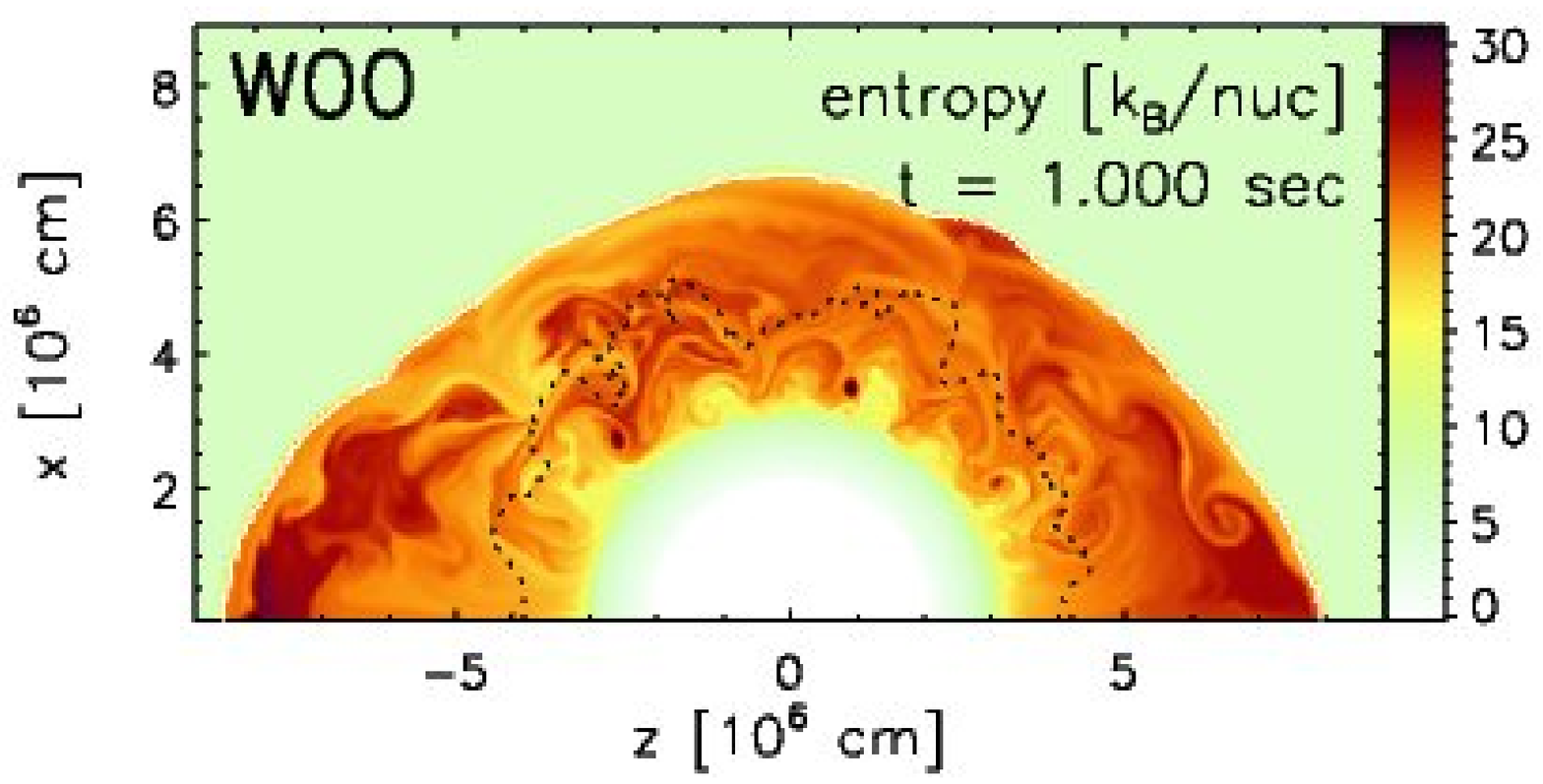}
\end{tabular}
\caption{Entropy distribution of Model W00 for several moments
  near the beginning of the nonlinear phase (the displayed times
  have a separation of
  half an oscillation period), and at $t=1\,$s.  Within each SASI
  oscillation cycle the postshock entropies vary strongly and steep,
  unstable entropy gradients develop in the postshock flow. Finally,
  the Rayleigh-Taylor growth timescale becomes smaller than the
  oscillation period and the characteristic mushroom structures
  are able to grow. In the
  subsequent evolution the low-mode oscillations saturate and the
  model does not develop an explosion.
  {\em (Color figures are available in the online version of our paper.)}}
\label{fig:w00_evo_stot}
\end{figure*}

%[30,60,90,120,156,180,250,500,1000]
\begin{figure*}[tbph!]
\centering
\begin{tabular}{cc}
\includegraphics[angle=0,width=7.7cm]{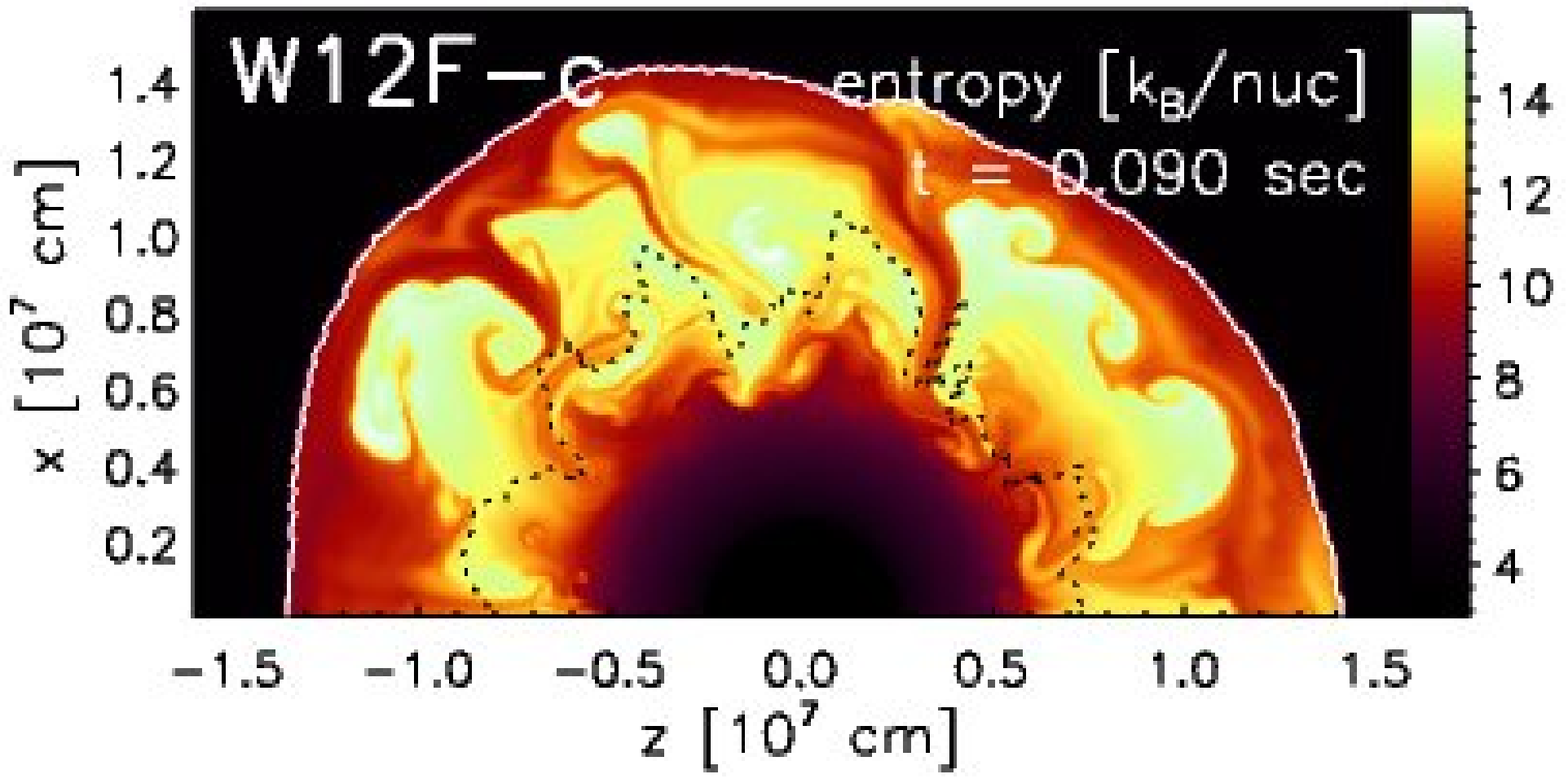}  &
\includegraphics[angle=0,width=7.7cm]{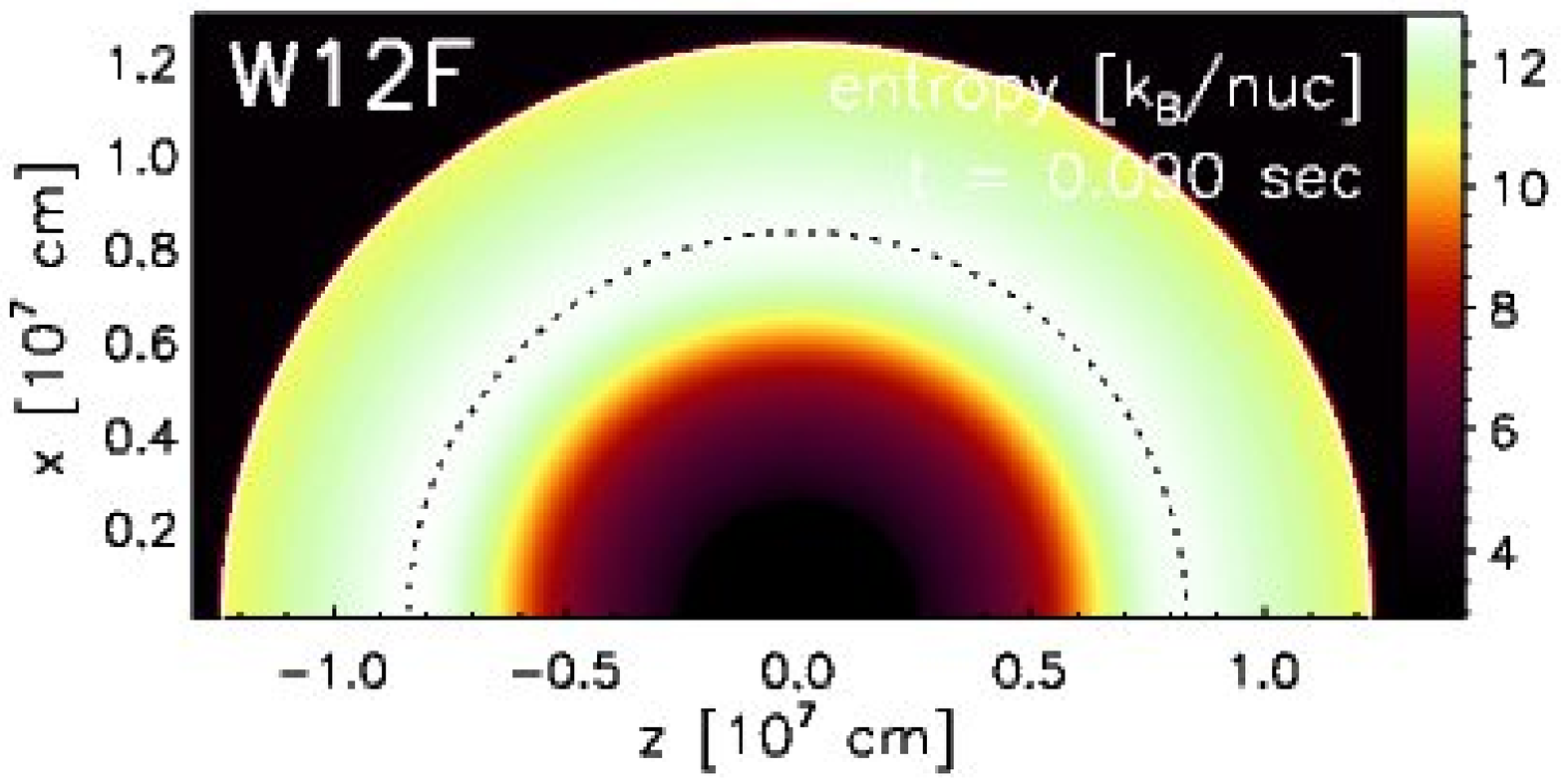}  \\
\includegraphics[angle=0,width=7.7cm]{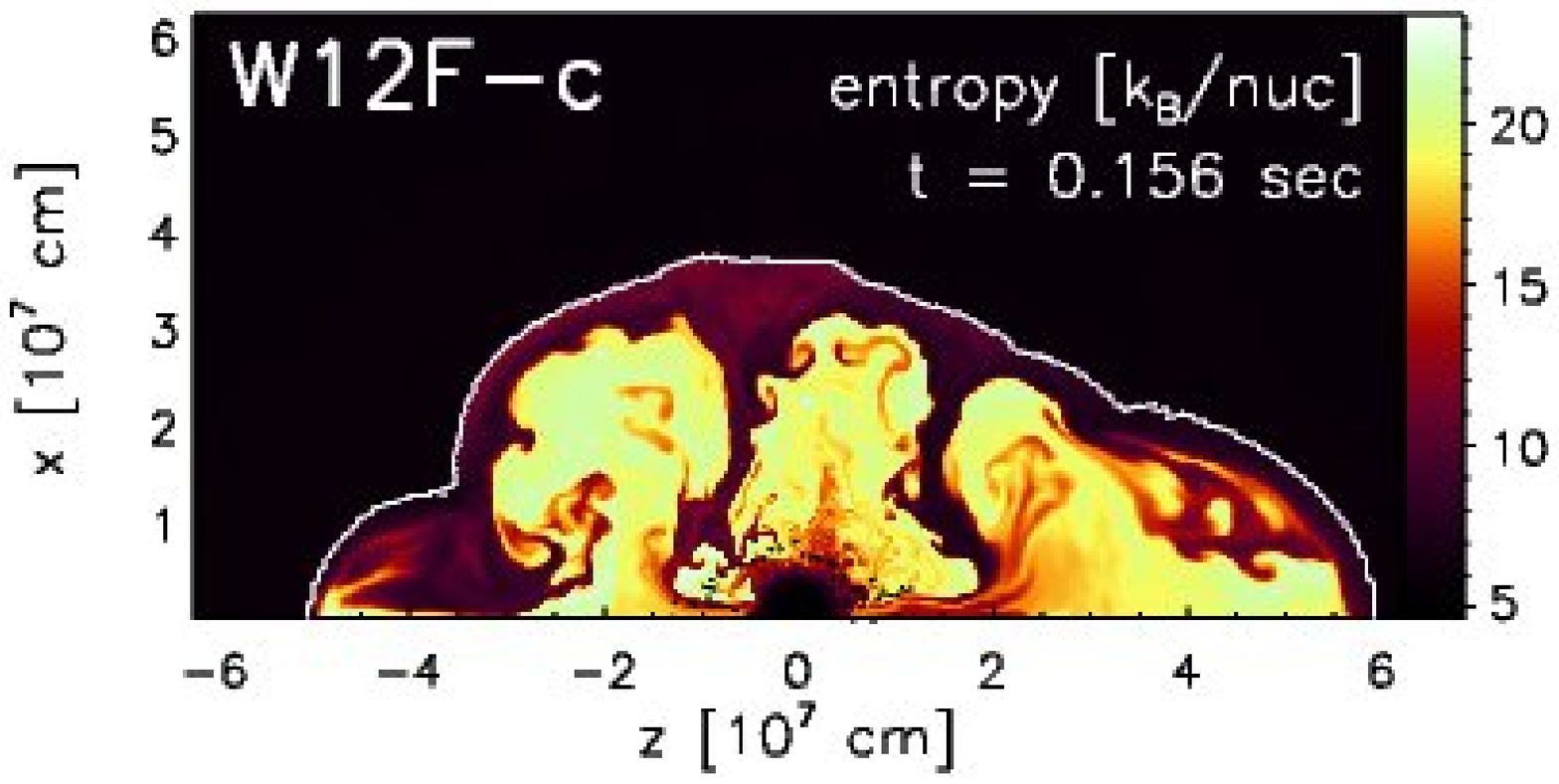} &
\includegraphics[angle=0,width=7.7cm]{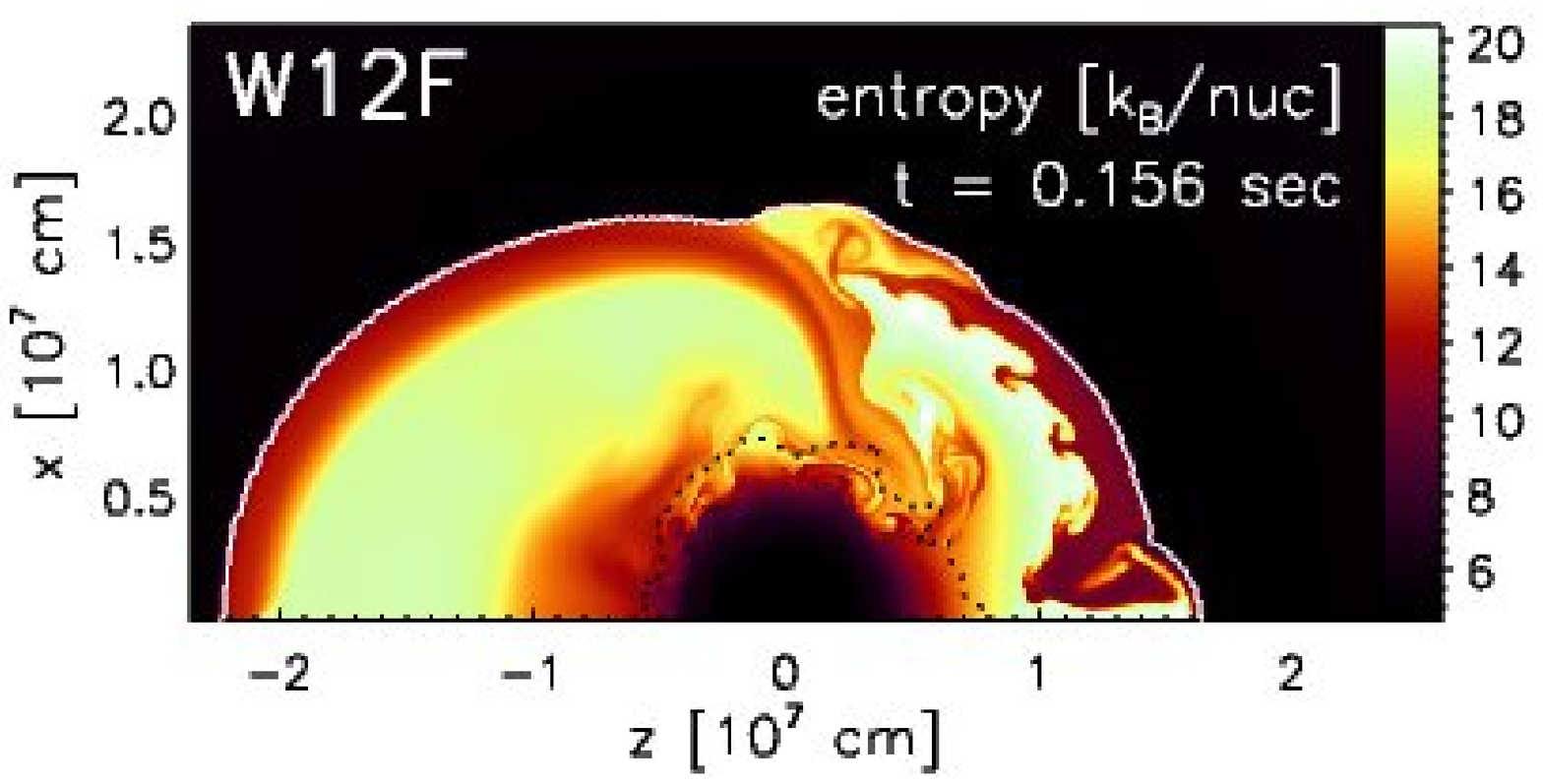} \\
\includegraphics[angle=0,width=7.7cm]{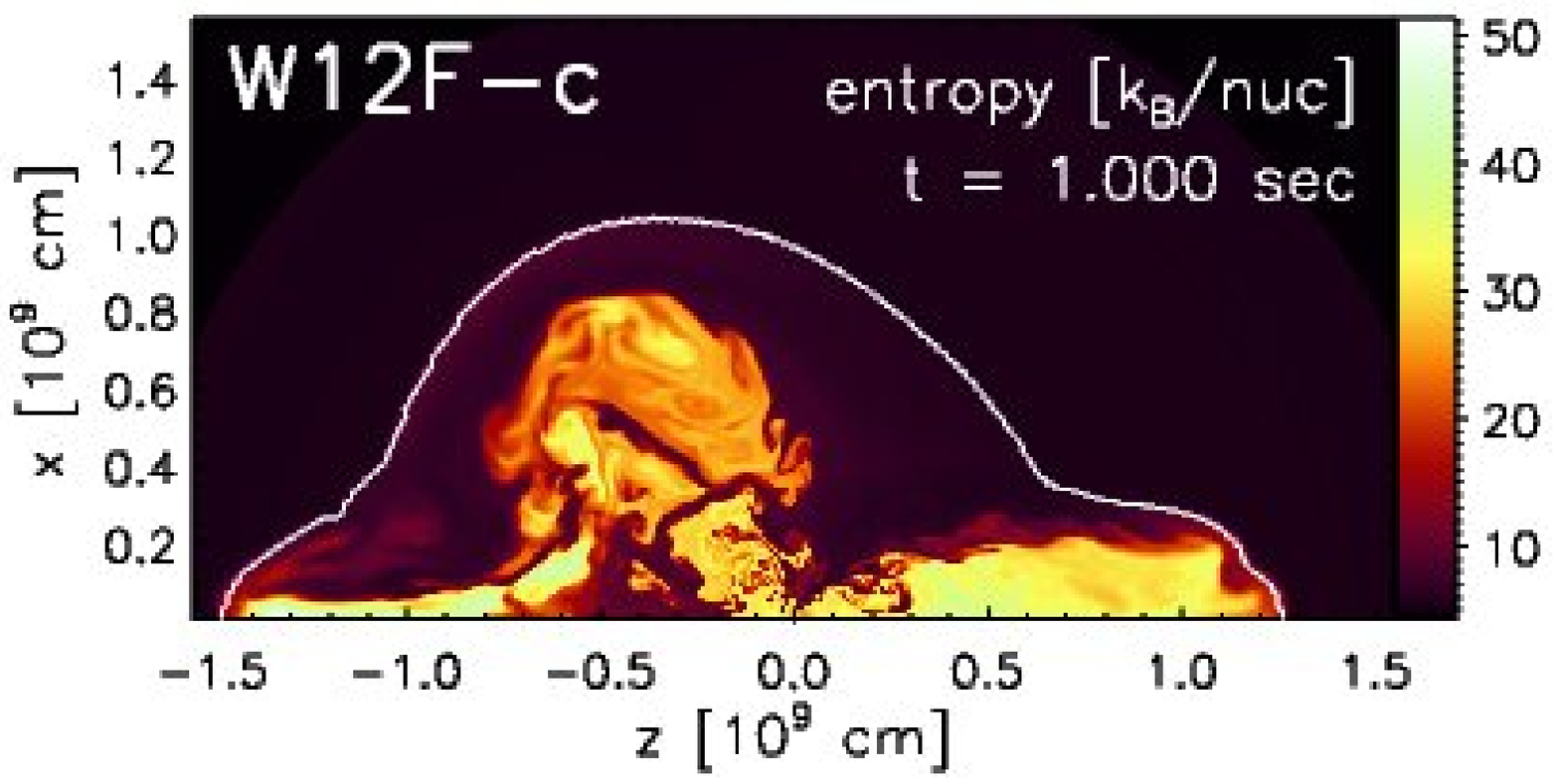}  &
\includegraphics[angle=0,width=7.7cm]{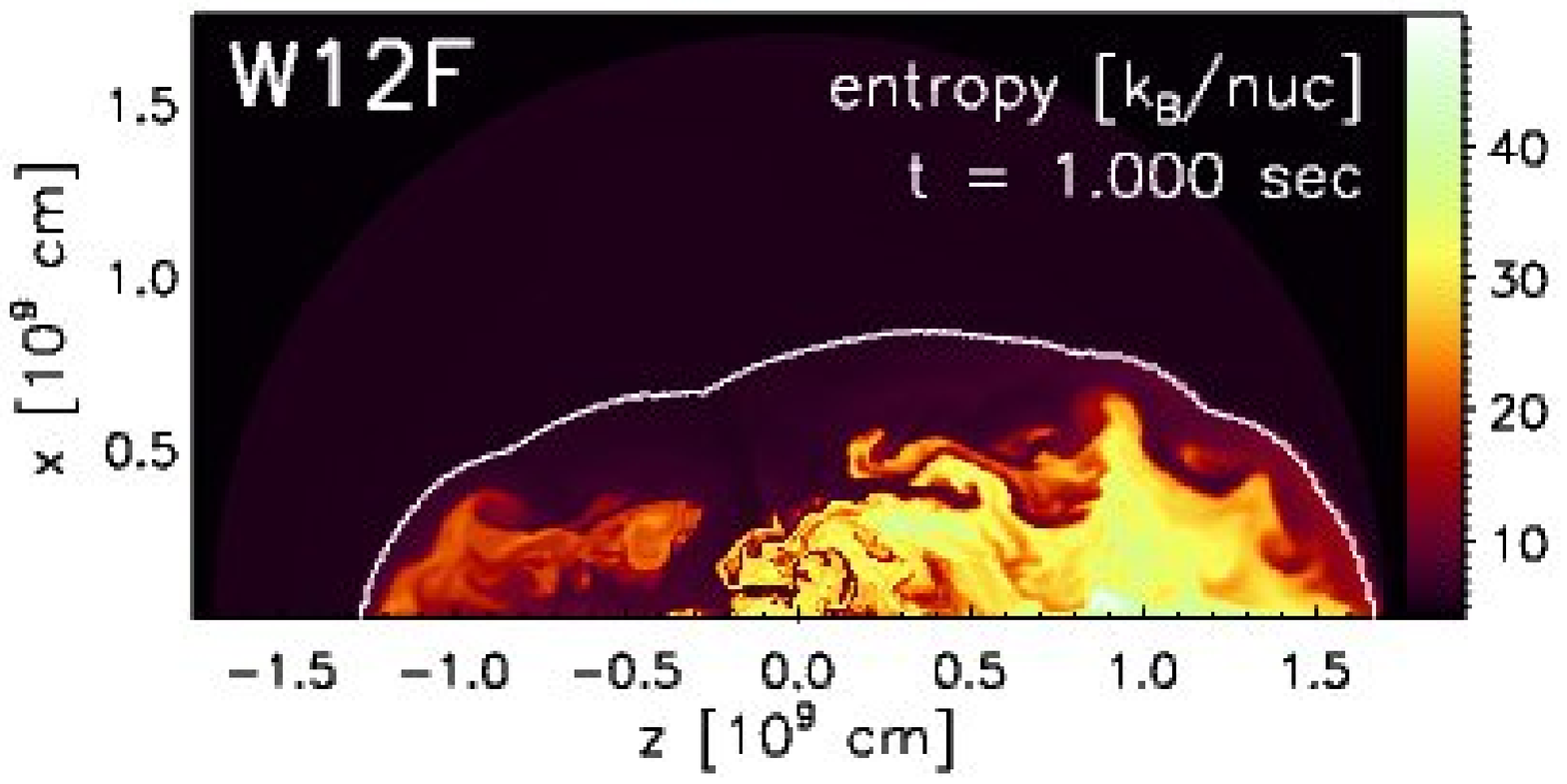}
\end{tabular}
\caption{Entropy distribution of Models W12F-c (left
  column) and W12F (right column) for several times.
  Model W12F-c quickly develops anisotropies
  because of the onset of convection, whereas in Model W12F
  convection is initially suppressed and low-mode SASI oscillations become
  visible after about $100\,$ms. After these oscillations have grown
  to large amplitude and have begun to trigger convection also in Model W12F,
  the two models explode in a qualitatively very similar way, although the
  detailed structure and asymmetry of the postshock flow and supernova shock
  are clearly different.
  {\em (Color figures are available in the online version of our paper.)}}
\label{fig:w12f_evo_stot}
\end{figure*}

%=====================================================================
\section{Results}
\label{sec:results}

In this section we will give an overview of the simulation results,
whose interpretation will be given in more detail in Sects.~\ref{sec:linear}
and \ref{sec:nonlinear}.

%----------------------------------------------------------------------
\subsection{A model without neutrinos}
\label{sec:w00fa}

Although Model W00FA does not include neutrino heating, convective
fluid motions develop in this case because a
convectively unstable region with a negative entropy gradient
is present at $r \approx 70\,$km already in the initial conditions
of our simulations. This feature is a consequence of the decreasing
shock strength before shock stagnation. Soon after we start model
model run W00FA,
buoyant bubbles form in the unstable region and rise towards the shock
(Fig.~\ref{fig:limcas_stot_w00fa}). Convective action continues
during the whole simulation because neutrino cooling, which could
damp convection, is disregarded, and because convectively unstable
entropy gradients are created by shock motions that are caused
by variations of the preshock accretion rate and by bipolar shock
oscillations due to SASI modes (see Sect.~\ref{sec:trigger_conv}). 
However, without
neutrino heating the convective overturn does not become as strong and
dynamical as in the simulations of Paper~I, where neutrino effects
were included. Also the bipolar shock oscillations are rather weak
(the shock deformation amplitude does not exceed 15\%) and occur
quasi-periodically with a period of 20--50\,ms
(Fig.~\ref{fig:limcas_mshell_w00fa}).

These multi-dimensional processes do not affect the overall
evolution of the model and the shock position as a function of time
is almost identical to the one found in a corresponding
one-dimensional simulation. In spite of the contraction of the 
inner grid boundary, the shock expands slowly and continuously
(Fig.~\ref{fig:limcas_mshell_w00fa}). A transient
faster expansion occurs at $t \approx 150\,$ms, when a composition
interface of the progenitor star falls through the shock and the
mass accretion rate drops abruptly. After $660\,$ms we stopped the
simulation. At this time the shock had reached a radius of $~400\,$km.

Although the shock expands slowly, this does not lead to an
explosion because without neutrino heating the specific energy of
the matter behind the shock remains negative. The shock expansion
takes place because matter piles up in the postshock region and
forms an extended atmosphere around the neutron star. This slowly
pushes the shock further out in response to the adjustment of
hydrostatic equilibrium by the accumulation of mass in the downstream
region. Since in the absence of cooling processes the matter cannot
lose its entropy, it is not able to settle down onto the neutron
star quickly. Therefore the behavior of Model W00FA is
destinctively different from the situation obtained in supernova
simulations with neutrino transport, and it also differs from
the stationary flow that was considered by \cite{Blondin+03}.
The postshock velocity in Model W00FA is much lower and the
shock radius becomes larger.

Neutrino cooling is therefore essential 
to obtain a quasi-steady state accretion flow when simulations
are performed in which the central neutron star is included
(in our models it is partly excised and replaced by an impenetrable
inner grid boundary). Only when neutrinos remove energy and
reduce the entropy of the gas can the matter be integrated into 
the dense surface layers of the compact object. The rapid flow
of the gas from the shock to the neutron star implies short
advection timescales, which are crucial for
the growth of the SASI (see the discussion in Sec.~\ref{sec:aac}). 
Although the accretion flow that develops in our supernova simulations 
is similar to the one assumed by \cite{Blondin+03} and 
\cite{Ohnishi+06}, there are still potentially important
differences. Because of the contraction of the neutron star
and due to the density gradient in the collapsing star, the mass
accretion rate varies (usually decreases) with time and the accretion
between shock and neutron star surface never becomes perfectly
stationary. Our simulations also differ from those of 
\cite{Ohnishi+06} and \cite{Blondin_Mezzacappa06} by our
more detailed treatment of the neutrino effects. Altogether
this allows us to assess the questions how non-radial hydrodynamic
instabilities develop at more realistic model conditions for the
supernova core than considered in previous studies, and how such
instabilities may influence the onset of the supernova explosion.

\begin{figure}[tbph!]
\centering
\resizebox{\hsize}{!}{\includegraphics{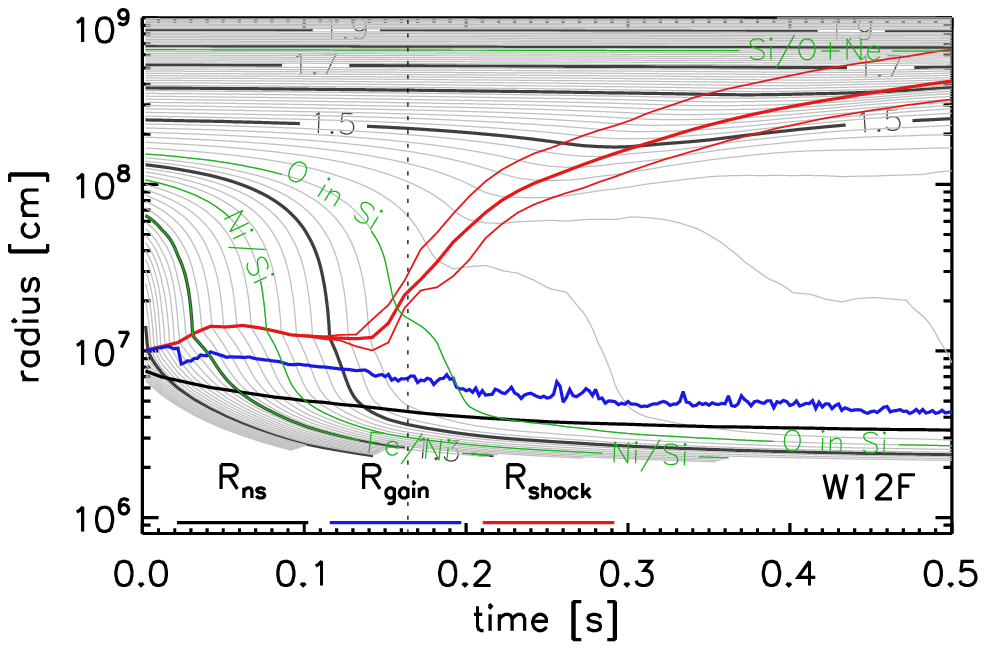}}
\resizebox{\hsize}{!}{\includegraphics{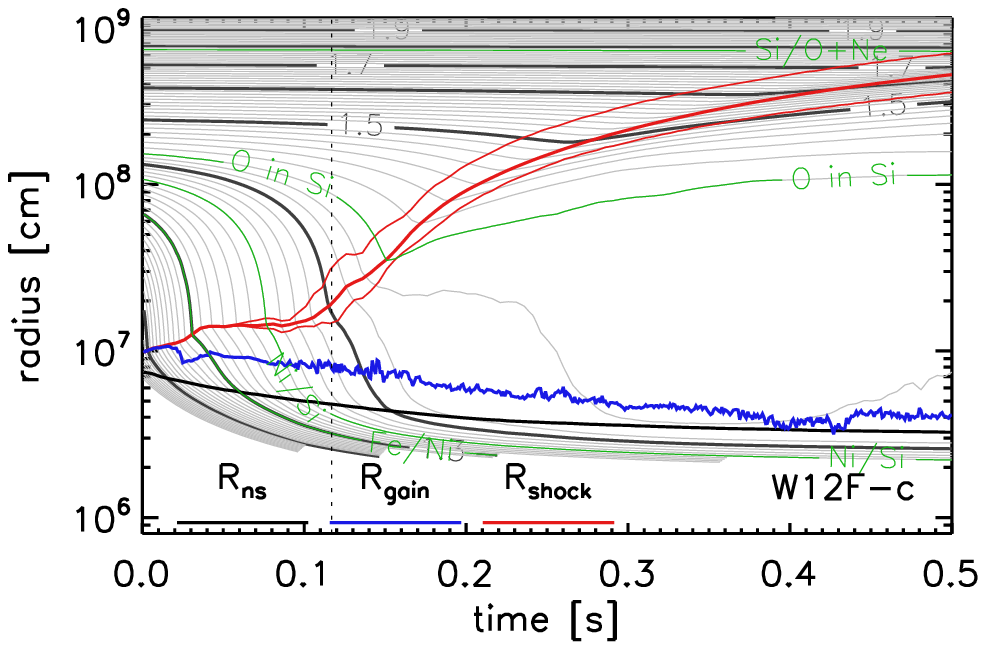}}
\caption{{\it Upper panel:} Same as Fig.~\ref{fig:limcas_w00_mshells}, but
  for Model W12F. After an initial phase, in which the model remains
  nearly spherically symmetric, the SASI becomes strong enough to deform
  the shock and to trigger convection.  This model explodes at
  $t\approx160\,$ms (marked by the vertical dotted line), after the
  oxygen-enriched silicon shell is has fallen through the shock.  {\it
    Lower panel:} Mass-shell plot for Model
  W12F-c. Convective activity starts to deform the shock in this model
  at $t\approx 60\,$ms, and the explosion occurs already at $t\approx 120\,$ms,
  before the oxygen-enriched silicon layer (whose inner boundary is indicated
  by the line labelled with ``O in Si'') has fallen through the shock.
  {\em (Color figures are available in the online version of our paper.)}}
\label{fig:limcas_w12f_mshells}
\end{figure}

\begin{figure}[tbph!]
\centering
\resizebox{\hsize}{!}{\includegraphics[angle=0]{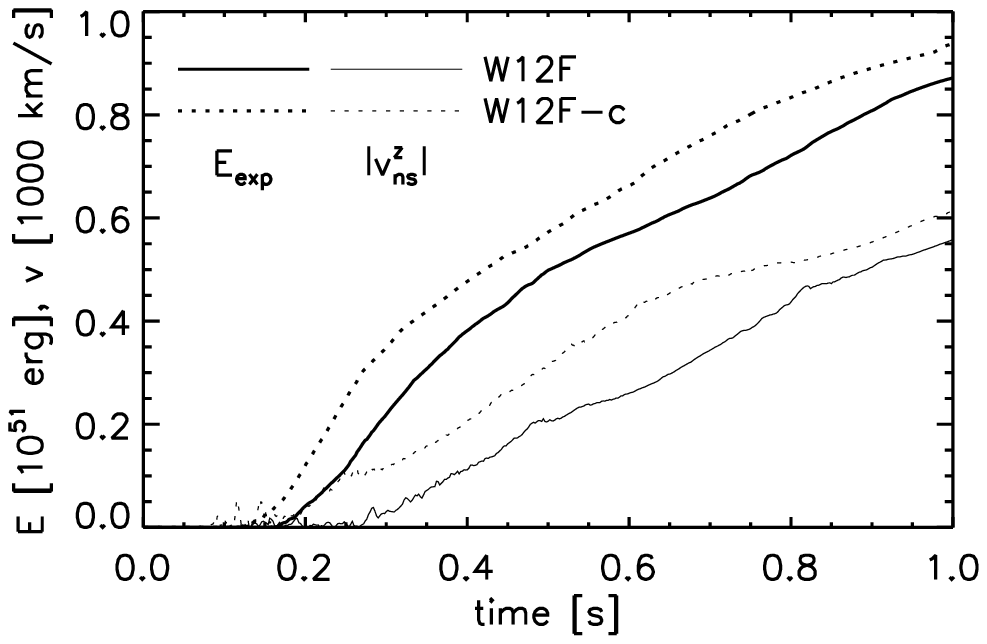}}
\caption{Evolution of the explosion energy (thick) and the neutron
  star velocity (thin) for Models W12F (solid) and W12F-c (dotted).}
\label{fig:w12f_eexp_vns}
\end{figure}

\subsection{Models with suppressed convection}
\label{sec:w00}

In the models including neutrino transport the accreted matter 
loses energy and entropy by
neutrino cooling and thus is able to settle down onto the neutron
star, following the contraction of the inner boundary. Comparing the
mass shell trajectories of the neutrinoless Model~W00FA and of 
Model~W00 (shown in
Figs.~\ref{fig:limcas_mshell_w00fa} and \ref{fig:limcas_w00_mshells},
respectively) this difference becomes evident. Since the accreted matter
does not pile up, also the shock turns around after an initial expansion
phase and recedes continuously during the later evolution (except for
a short, transient expansion phase a $t\approx150\,$ms, which is 
initiated when a composition interface of the progenitor star
crosses the shock). Due to the
miniscule boundary luminosity the neutrino heating remains weak and
the parameter $\chi$ of Eq.~(\ref{eq:def_chi}) stays below the critical
value (Fig.~\ref{fig:chi_denspert}). Consequently, there is no evidence 
of convection in the gain layer and Model~W00 evolves nearly
spherically symmetrically in the first $300\,$ms.

However, already several ten milliseconds after the start of the
simulation a lateral velocity component (which changes direction with
a period of about $30\,$ms) is observable in the flow between shock
and neutron star surface. % (Fig.~\ref{fig:limcas_rtimg_w00f}).
The amplitude of this $l=1$ oscillation mode starts to increase
continuously after $t \approx 100\,$ms and grows by a factor of about
two per period. However, the amplitude is not large enough to affect
the shape of the shock before $t \approx 250\,$ms because the 
finite resolution of the numerical grid prevents the shock from
being pushed out by less than one radial zone and thus it remains
perfectly spherical for low oscillation amplitudes (lateral variation
is already visible in the postshock flow, though).

In the subsequent evolution the shock radius is initially still slowly
decreasing and the shock shape remains approximately spherical, but
the shock surface moves back and forth along the axis of 
symmetry assumed in our two-dimensional simulations. The
direction of the postshock flow changes periodically and the flow 
transports matter between the southern and the northern hemispheres
(Fig.~\ref{fig:w00_evo_vely}).  This situation is quite
similar to the bipolar oscillations encountered in some models
discussed in Paper~I and also in full-scale supernova simulations
with sophisticated neutrino transport
\citep{Buras+06b}.  However, because convection is absent, the flow
pattern and the shape of the shock are much less structured
in Model W00.

At $t \approx 360\,$ms the amplitude of the shock oscillations has
become very large, the shock radii at the poles differ by up to
$50\,$km, whereas the average shock radius is only about $100\,$km. In
this phase the entropy behind the shock starts to vary strongly with
time and angle (Fig.~\ref{fig:w00_evo_stot}). Steep negative entropy
gradients ($\dS/\dr = {\cal O}(1\,k_{\mathrm{b}}/{\rm km})$) develop and
Rayleigh-Taylor instabilities start to grow at the boundaries between low-
and high-entropy matter. The postshock flow reaches lateral velocities
of several $10^9\,$cm/s and supersonic downflows towards the neutron
star form (see Sect.~\ref{sec:trigger_conv} for a discussion of these
processes).  Within a few oscillation cycles the whole postshock flow
becomes very similar to the nonlinear convective overturn present at
the onset of the explosion in those models of Paper~I where the explosion
energy was rather low. 

However, in contrast to these simulations of Paper~I, Model W00 does 
not explode. At $t
\approx 390\,$ms the bipolar oscillations reach their maximum
amplitude. In the further evolution they become weaker and on average
the shock radius decreases (Fig.~\ref{fig:limcas_w00_mshells}).  The
slow decay of activity is interrupted by several short phases of
stronger shock expansion and bipolar oscillation, which occur
quasi-periodically every $50$--$100\,$ms. When we stop the simulation
at $t=1\,$s the shock has retreated to a radius of only $~70\,$km on
average.

Models W00S, W05S, and W00V, in which a slowly contracting neutron
star was assumed, evolve qualitatively
very similar to Model W00. However, with increasing contraction timescale
the oscillation period becomes longer (up to $100\,$ms) and the
growth rate of the low-mode instability decreases. All these models
are dominated by an $l=1$ SASI mode and none of them is able to explode.

Also Model W00F with its rapidly contracting inner boundary evolves
initially quite similar to Model W00 (Fig.~\ref{fig:limcas_w00_mshells}). 
However, all timescales are
shorter: The oscillation amplitude starts to grow already after
$50\,$ms, the shock becomes non-spherical at $t \approx 130\,$ms and
convection sets in at $t \approx 160\,$ms. Furthermore, 
Fig.~\ref{fig:w00_evo_vely} shows that in this model
the $l=2$ mode (i.e. oscillation between prolate and oblate states) is
initially more strongly excited than the $l=1$ mode, which starts to
dominate only just before the onset of the explosion.

\begin{figure}[tbph!]
\centering
\includegraphics[angle=0,width=8.5cm]{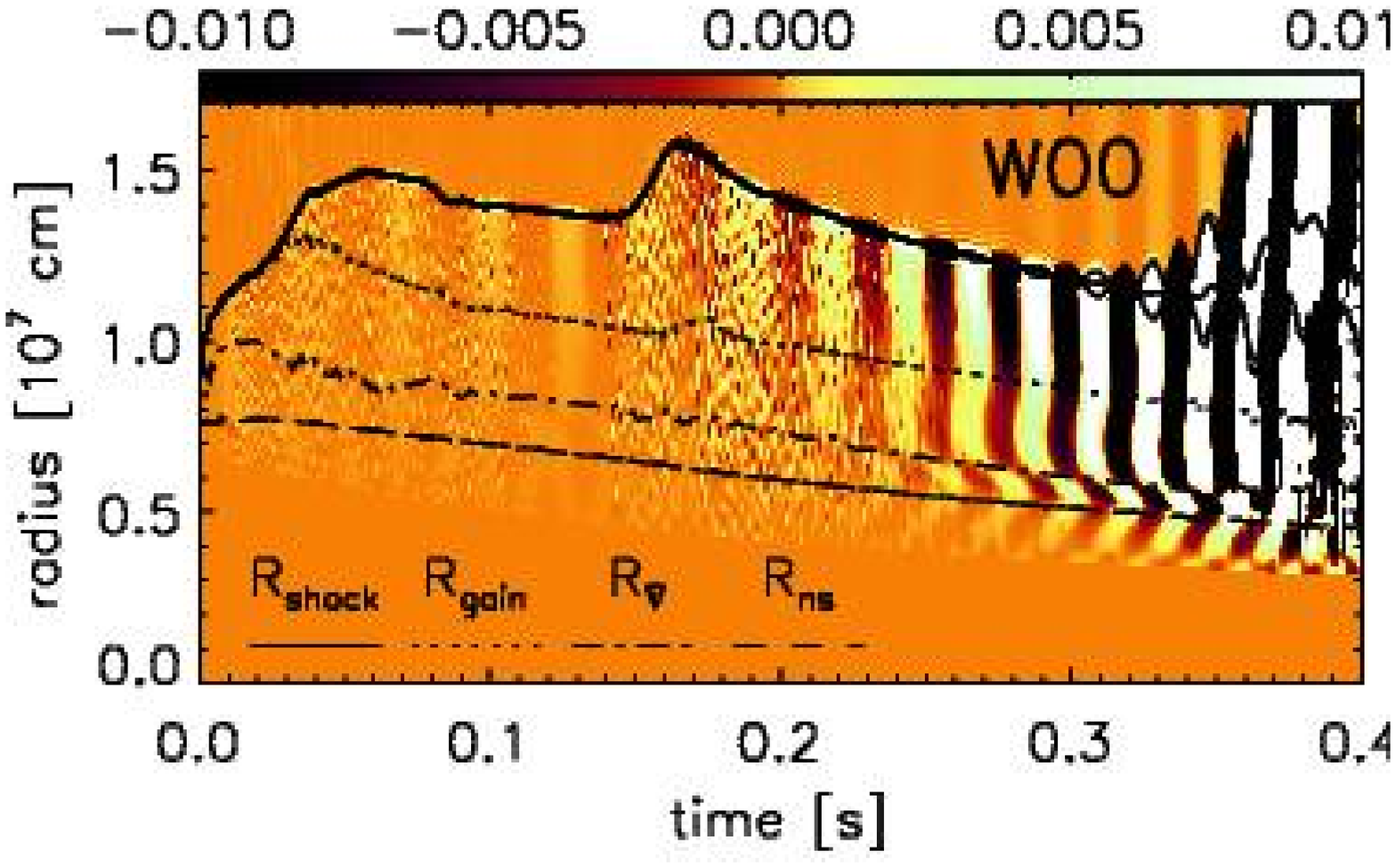}  
\includegraphics[angle=0,width=8.5cm]{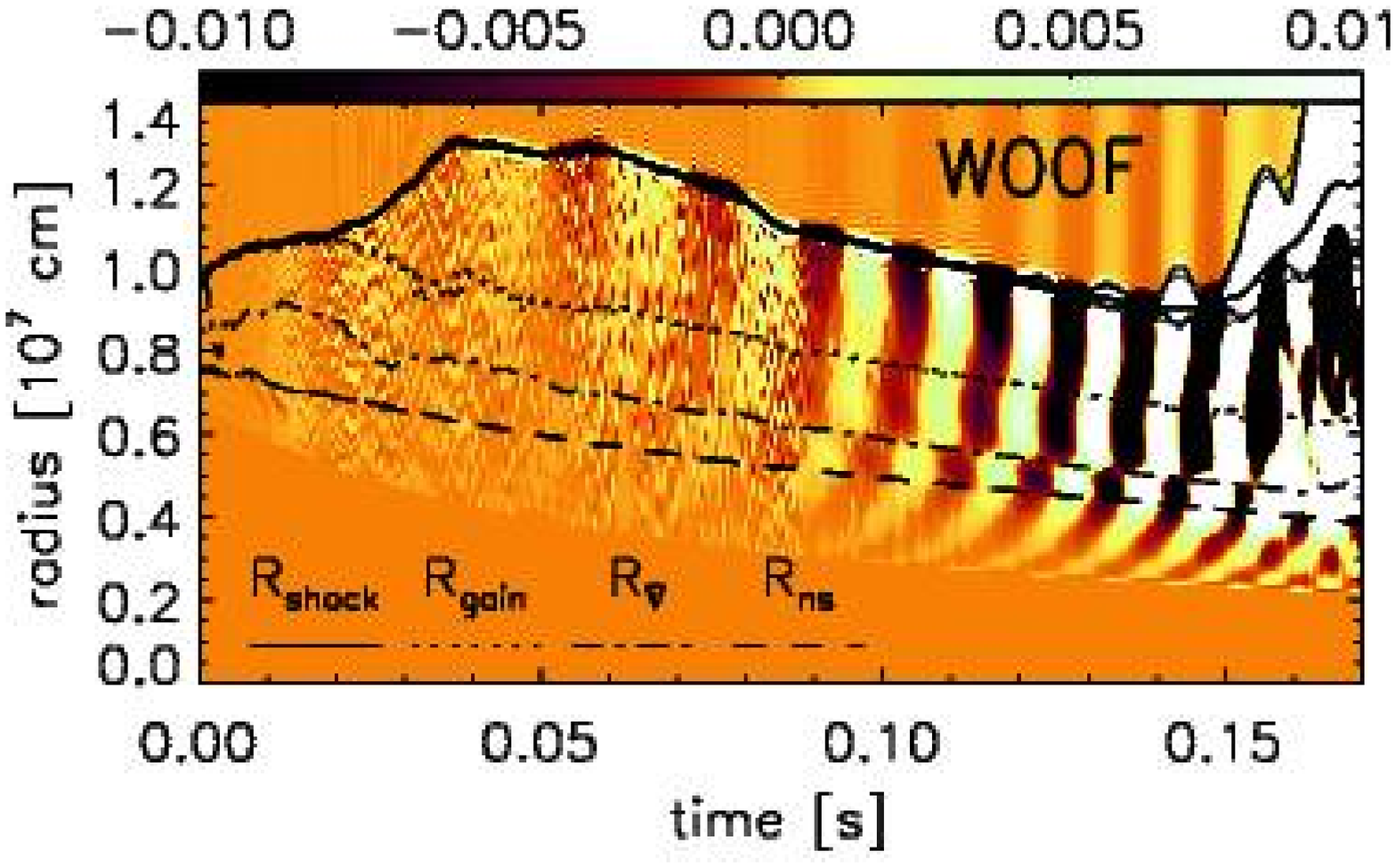}  
\includegraphics[angle=0,width=8.5cm]{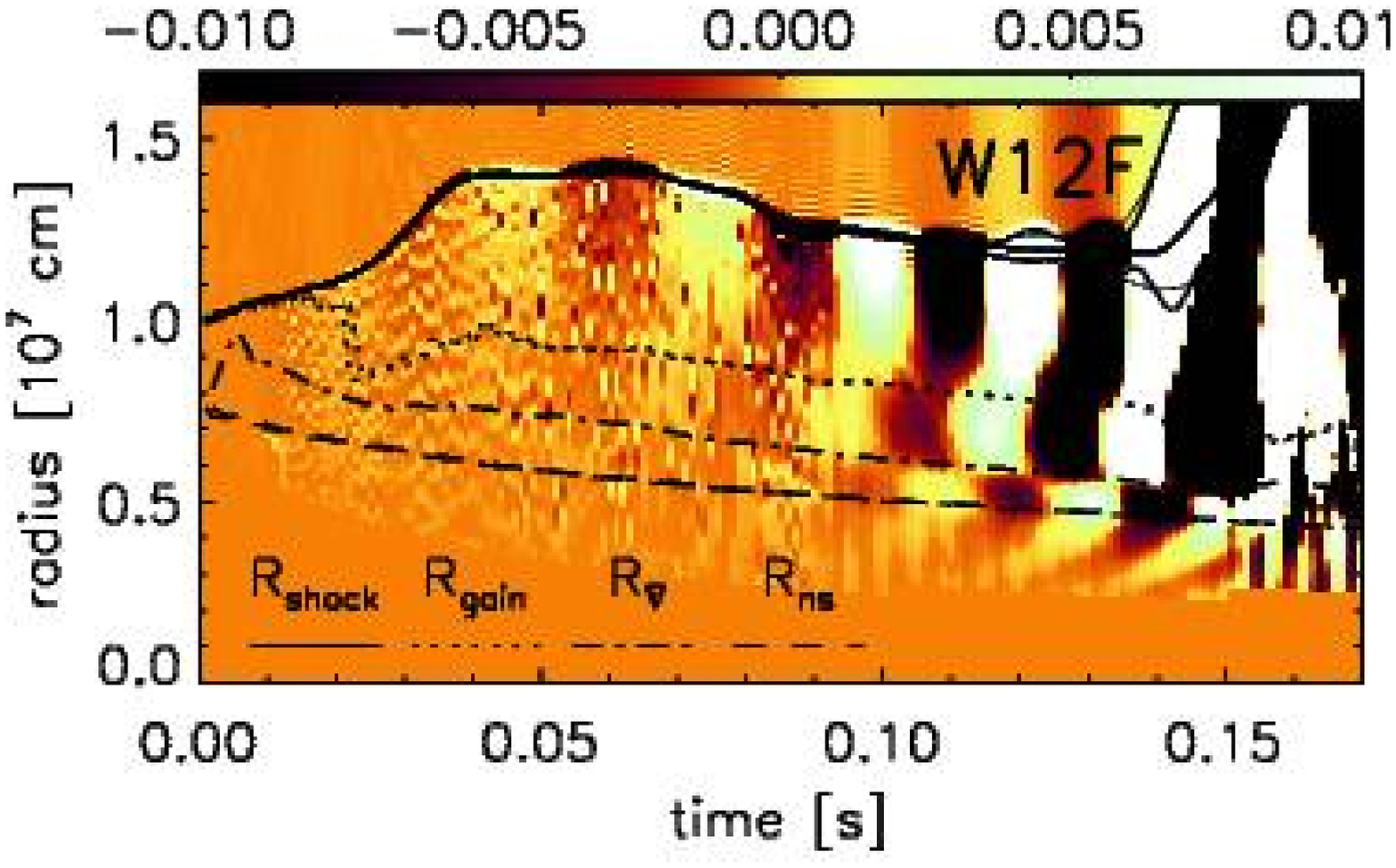}
\caption{Time evolution of the amplitude of the dominant spherical
  harmonics mode of the pressure, normalized by the amplitude of the
  $l = 0$ mode, as function of radius for Models W00, W00F and W12F.
  The solid lines are the minimum,
  average, and maximum shock radius, the dotted line is the gain
  radius, the dashed line is the neutron star surface (defined as the
  location where the density is $10^{11}\,$g$\,$cm$^{-3}$), and the
  dash-dotted line marks the position, $R_{\nabla}(t)$, of the largest 
  velocity gradient. A low-mode oscillation develops in the
  postshock flow.  A pronounced phase shift is visible at a radius
  $R_{\varphi}(t)$ that agrees well with the position of the largest
  velocity gradient. The ``noise'' (short-wavelength sound waves)
  visible in the early phase after bounce is caused by the shock
  propagation and is not related to the advective-acoustic cycle.
  {\em (Color figures are available in the online version of our paper.)}}
\label{fig:rtimg_press}
\end{figure}

\begin{figure}[tbph!]
\centering
\includegraphics[angle=0,width=8.5cm]{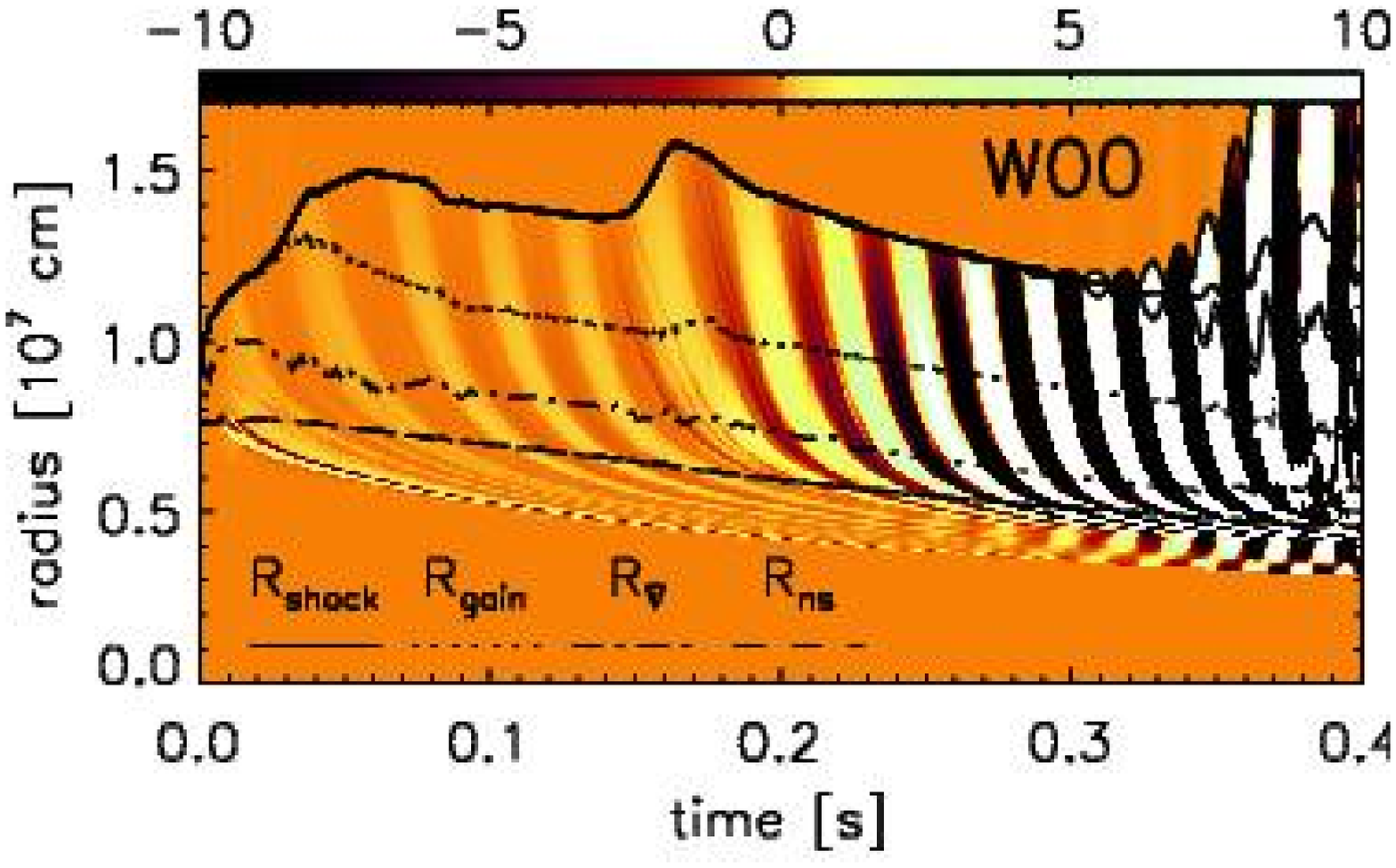}
\includegraphics[angle=0,width=8.5cm]{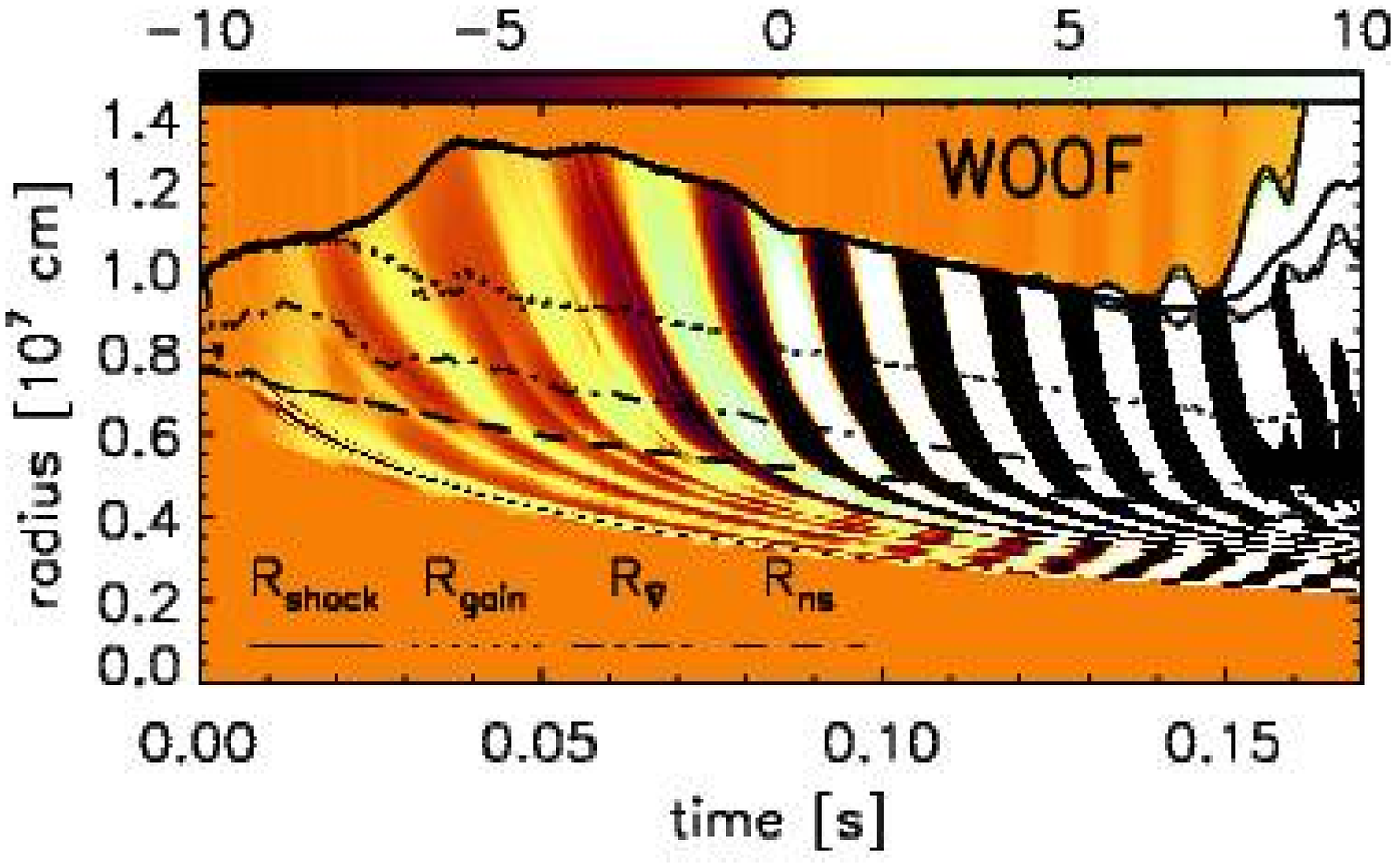}
\includegraphics[angle=0,width=8.5cm]{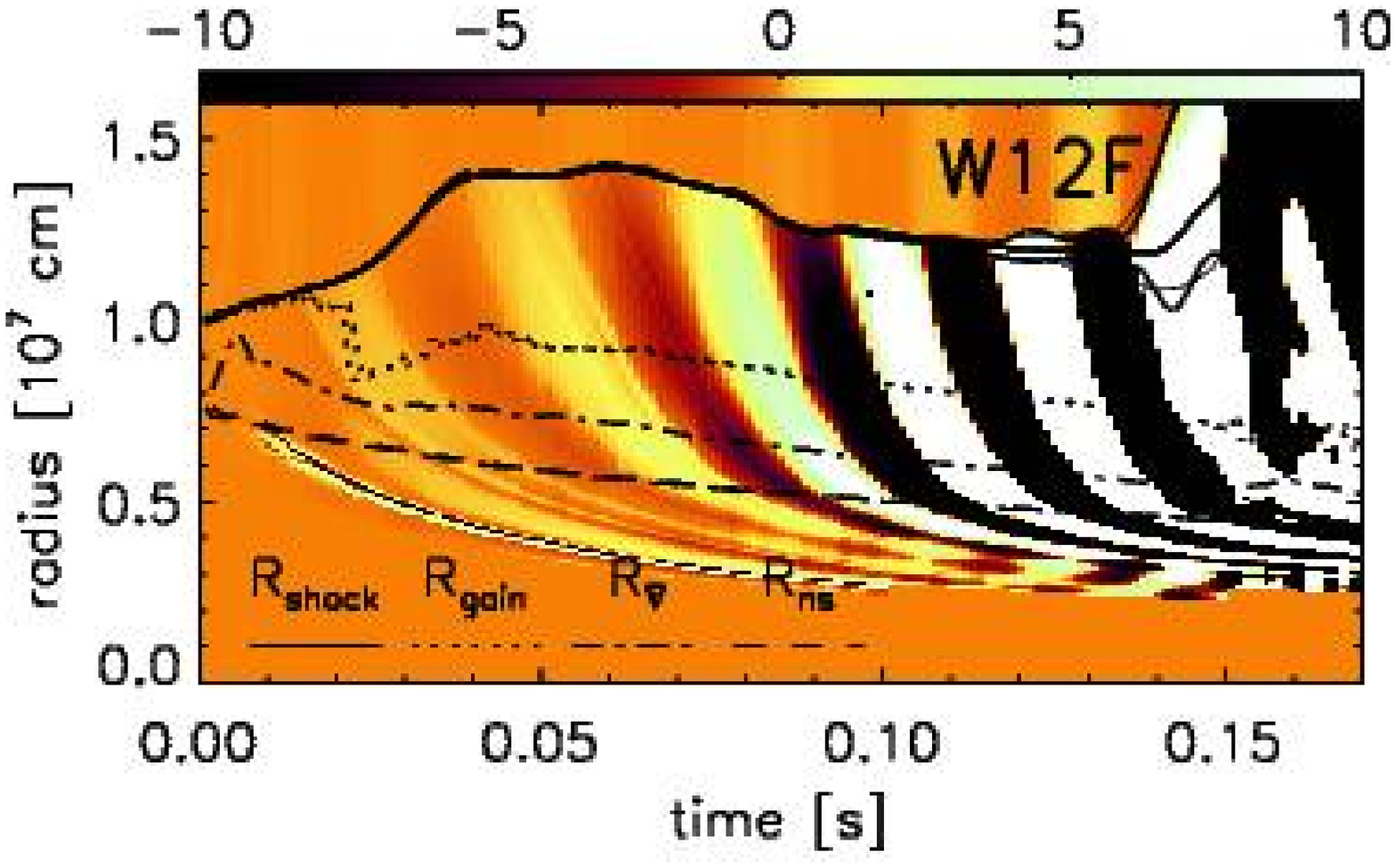}
\caption{
  Time evolution of the amplitude of the dominant spherical harmonics
  mode of the quantity $A(r,t,\theta)$ of Eq.~(\ref{eq:def_A}),
  displayed as function of radius for Models W00, W00F and W12F.
  The lines have the same meaning as in Fig.~\ref{fig:rtimg_press}.
  As in the latter figure, a zebra-like pattern becomes visible here
  already several ten milliseconds after bounce a zebra-like,
  indicating that matter with a
  nonvanishing lateral velocity component is advected from the
  shock towards the neutron star.
  {\em (Color figures are available in the online version of our paper.)}}
\label{fig:rtimg_vort}
\end{figure}

In contrast to the models with slower boundary contraction, the
continuous neutrino heating in Model W00F is strong enough to
trigger an explosion at $\texp=194\,$ms. This difference is caused by
the fact that the faster contraction leads to gravitational energy 
release (the accreted matter heats up by compression) and thus to higher
neutrino luminosities (see Sec.~\ref{sec:explosion} for further
discussion). The anisotropic gas distribution caused by the low-mode
oscillations becomes frozen in when the shock accelerates outward. The
shock develops a prolate deformation and a single accretion funnel
forms in the northern hemisphere. Since the explosion attains a 
large-scale asymmetry, the anisotropic distribution of the ejecta exerts 
a strong gravitational force that causes an acceleration of the newly 
formed neutron star (see Paper~I for details about this process and
the procedure of evaluating (postprocessing) our simulations for the 
resulting kick velocity of the neutron star\footnote{Due to the fact
that the neutron star core is replaced in our simulations
by an inner grid boundary and thus anchored at the grid center,
our conclusions on the neutron star kick velocities need to be
confirmed by independent hydrodynamic models without such a numerical
constraint. \cite{Burrows06,Burrows07} seem to be unable to 
reproduce our findings 
with their recent simulations, using a code setup that
allows the neutron star to move and employing a strictly momentum
conserving implementation of the gravitational effects in the
fluid equation of motion.
However, the numerical consequences of their new treatment of the 
gravity source term, and in particular 
the supposed superiority compared to other (standard) treatments,
must be demonstrated by detailed numerical tests, which 
\cite{Burrows06,Burrows07} have so far not presented. The exact reasons
for the potentially discrepant results are therefore unclear to
us and may be manifold. 
We strongly emphasize here that the neutron star kicks reported by
\cite{Scheck+06} were calculated by two independent methods of 
postprocessing analysis. First, making use of total linear
momentum conservation, the neutron star recoil was estimated
from the negative value of the momentum of all gas on the 
computational grid at the end of the simulations
(note that using an inner grid boundary at some radius 
$r > 0$ leads to momentum transfer to the excised inner core so
that the gas in the simulation domain does not retain zero 
$z$-momentum). Second, the different time-dependent
forces that can contribute to the neutron star acceleration, 
i.e., the gravitational force excerted by the anisotropically
distributed matter around the neutron star as well as the momentum
transfer associated with anisotropically accreted or outflowing
gas of the neutron star, were
added up and then integrated in time. Both of these completely 
independent approaches led to estimates for the neutron star
kick velocities in very nice agreement with each other.}) 
Due to the miniscule boundary luminosity the energy of the explosion
remains rather low ($0.37\times\foe$ at 750$\,$ms after bounce,
see Table~\ref{tab:restab_limcas}, and $0.5\times\foe$ for the extrapolated
value at $1\,$s), but 
the neutron star attains a fairly high kick velocity ($\vns \approx 
200\,$km/s at 750$\,$ms post bounce and estimated 350$\,$km/s 
for $t=1\,$s).

%--------------------------------------------------------------------
\subsection{Models with typical explosion energies}
\label{sec:w12f}

While the simulations discussed so far demonstrate clearly the
existence of a non-radial instability that is not convection, they 
were based on the assumption that the core neutrino luminosities are
negligibly small. In contrast, in W12F and
W12F-c boundary luminosities were assumed such that the
explosion energies reached values close to those considered to be 
typical of core-collapse supernovae. 
An overview of the evolution these models can be obtained from 
Figures \ref{fig:w12f_evo_stot} and \ref{fig:limcas_w12f_mshells},
where we show entropy distributions at several times and the
mass-shell plots, respectively.

In Model W12F-c, in which large initial seed perturbations were assumed
(cf.\ Table~\ref{tab:restab_limcas}), the first convective bubbles form
at $t\approx 60\,$ms, and at $t\approx 90\,$ms the whole gain layer has
become convective (see Fig.~\ref{fig:w12f_evo_stot}). From this time
on the total energy in the gain layer rises continuously and already
at $\texp \approx 120\,$ms the first zones acquire positive total energy and
the model explodes. The 
initially weakly perturbed Model W12F behaves differently in the first
$200\,$ms. There is no sign of convection and for the first
$100\,$ms the shock radius evolves as in a corresponding one-dimensional 
model. However, as in Model W00 a weak $l=1$ oscillation
mode is present in the postshock flow already at early times
($t\approx30\,$ms) and grows exponentially to large amplitudes. At
about $t_{\rm nl} \approx 150\,$ms steep convectively unstable
entropy gradients are generated behind the oscillating shock and within
two cycle periods a situation develops that strongly resembles model
W12F-c at the onset of the explosion. Also Model W12F explodes, though
a bit later than Model W12F-c, at $t=164\,$ms.

Although the pre-explosion evolution and the explosion timescales
of the two models are different, the models behave quite similar
after the explosion has set in. The convective structures merge and
downflows form at the interface between expanding, neutrino-heated gas
and the matter with lower entropy just behind the shock.
The number of downflows decreases with time and from $t \approx
200\,$ms on a single downdraft dominates the anisotropic gas
distribution. Its position
differs in the two models, as does the shape of the shock, but the
explosion energies and even the neutron star velocities grow nearly
in the same way after $t\sim 0.3\,$s and reach essentially the same
values at the end of our simulations (Fig.~\ref{fig:w12f_eexp_vns}).

\begin{figure}
\centering
\includegraphics[angle=0,width=8.5cm]{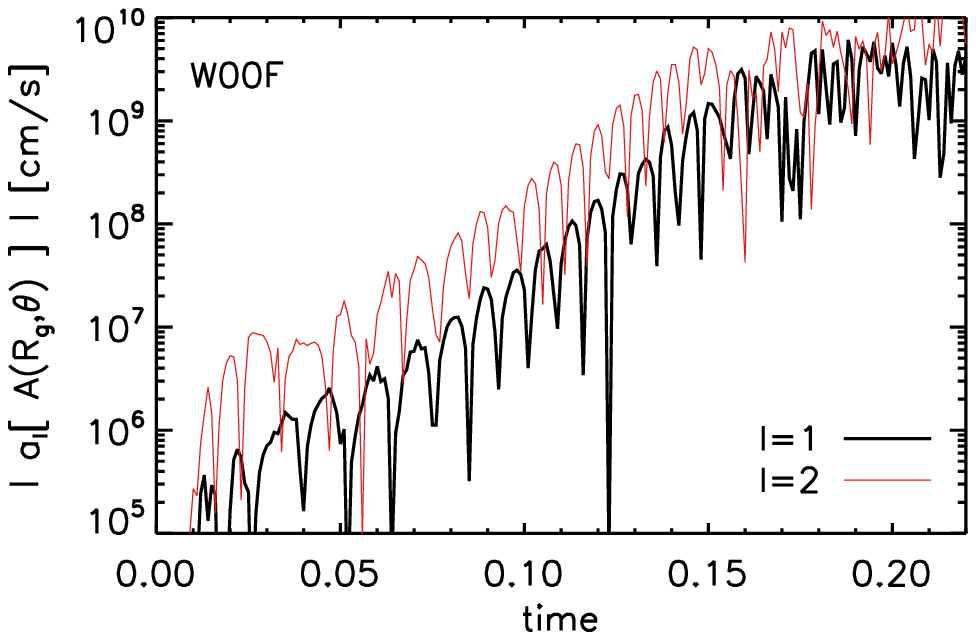}
\caption{Oscillatory growth of the amplitudes $a_1$ and $a_2$ of the
  $l =1,\,2$ spherical harmonics components of the quantity $A(r,t,\theta)$
  of Eq.~(\ref{eq:def_A}) at the gain radius of Model W00F.
  {\em (A color figure is available in the online version of our paper.)}}
\label{fig:coefficients}
\end{figure}

%=====================================================================

%\section{Interpretation of the linear phase: identification of the AAC}
\section{The linear phase of the instability: identification of the AAC}
\label{sec:linear}

We now turn to a detailed investigation of the question which physical
mechanism is responsible for the SASI that we have
seen in the models discussed in the previous section.

\subsection{Motivation and method}

In a number of studies \citep[e.g.,][]{Galletti_Foglizzo05,Burrows06,Ohnishi+06,Scheck+06,Foglizzo+06b,Yamasaki_Yamada+07}
the advective-acoustic cycle was identified
or invoked as the cause of the SASI oscillations that were found in these
studies to occur as observed by \cite{Blondin+03}.
This interpretation is
currently challenged by \cite{Blondin_Mezzacappa06}, who advocate 
as an explanation of the SASI modes a purely acoustic process, which is
driven by sound waves traveling solely in non-radial direction
\citep{Blondin_Shaw07}. One difficulty of deciding about the correct
interpretation is 
due to the fact that the oscillation timescale of the SASI can either be
understood
as the acoustic timescale along a well chosen transverse path, or the
advection time down to a suitably chosen coupling radius. 
From the physics point
of view, however, the foundations of the advective-acoustic mechanism are
well documented (see the papers cited in Sect.~\ref{sec:aac} and the 
references therein), whereas the purely acoustic mechanism is still 
incompletely understood \citep{Laming+07}.
In particular \cite{Blondin_Mezzacappa06} argued that the existence of a
different gradient of the momentum flux on both sides of the shock is
responsible for the instability.
This argument, however, is so inconclusive that it was used by
\citeauthor{Nobuta+94} (\citeyear{Nobuta+94}, Fig.~10)
in order to reach the exactly
opposite conclusion, namely the stability of a stationary shock
in an accretion disk.

Independent of any timescale consideration, \cite{Foglizzo+06b} were able
to directly measure the efficiencies of both advective-acoustic and purely
acoustic cycles using a WKB approximation, i.e. for perturbations whose
wavelength is shorter than the size of the flow gradients near the shock. For
every unstable eigenmode for which this quantitative estimate was possible, it
showed the stability of the purely acoustic cycle and the instability of the
advective-acoustic one.  The WKB approximation is unfortunately unable to
treat accurately the lowest frequency modes, whose wavelength is comparable
to the radius of the shock. This argument in principle leaves room for
alternative explanations of the instability of the lowest frequency
modes. This is
why we do not discard the possibility of a purely acoustic, unstable cycle a
priori, despite its unsatisfactory theoretical foundation. 

The quantities and results shown in
Figs.~\ref{fig:rtimg_press}--\ref{fig:tscaling}
in the present paper are supposed to characterize the development of the
SASI in a time-dependent environment and to serve comparison of the SASI 
properties with the
expectations of either an advective-acoustic or a purely acoustic process.
Using a projection of perturbations on spherical harmonics, the time evolution
of the radial structure of the most unstable eigenmode is visualized, and the
oscillation frequency $\omega_r$ and growth rate $\omega_i$ can be measured.
The oscillation timescale is then compared to some reference timescales
associated with advection and acoustic waves. The acoustic timescales chosen
for this comparison are $\tau_{\rm sound}^{\rm rad}$, computed along a radial
path crossing the shock diameter and back, and $\tau_{\rm sound}^{\rm lat}$,
computed along the circumference at the shock radius (i.e., immediately 
behind the shock position):
\begin{eqnarray}
\tau_{\rm sound}^{\rm rad}&\equiv
&2\int_{R_{\rm ib}}^{R_{\rm sh}}\frac{{\rm d}r}{c-v}
+2\int_{R_{\rm ib}}^{R_{\rm sh}}\frac{{\rm d}r}{c+v}
+\frac{4R_{\rm ib}}{c_{\rm s,ib}}\,,\\
\tau_{\rm sound}^{\rm lat}&\equiv& \frac{2\pi R_{\rm sh}}{c_{\rm sh}}\,.
\end{eqnarray}
Following \cite{Foglizzo+06b}, the reference timescales chosen for the
advective-acoustic cycle are the advection time $\tau_{\rm adv}^\nabla$ from
the shock to the radius $R_\nabla$ of maximum deceleration, and an estimate
of the full cycle timescale $\tau_{\rm aac}^\nabla$ based on a radial
approximation for simplicity:
\begin{eqnarray}
\tau_{\rm adv}^\nabla&\equiv &\int_{R_\nabla}^{R_{\rm sh}}\frac{{\rm d}r}{|v|}\,,
\label{eq:tadv}\\
\tau_{\rm aac}^\nabla&\equiv& \int_{R_\nabla}^{R_{\rm sh}}\frac{{\rm d}r}{|v|}+
\int_{R_\nabla}^{R_{\rm sh}}\frac{{\rm d}r}{c-|v|}\,.
\label{eq:taac}
\end{eqnarray}
The consistency of the advective-acoustic interpretation is further tested by
comparing the timescale of deceleration $|{\rm d}v/{\rm  d}r|^{-1}$ with the
oscillation time of the instability.
If velocity gradients are indeed responsible for
the acoustic feedback, unstable flows should correspond to abrupt
deceleration while smoothly decelerated flows should be stable. Moreover, the
amplification factor $Q$ during one oscillation is compared to the value
measured in the simpler setups studied by \cite{Blondin_Mezzacappa06} and
\cite{Foglizzo+06b}.

\subsection{Extracting eigenfrequencies from the simulations}

In Fig.~\ref{fig:rtimg_vort}, advected perturbations
are displayed by the amplitudes of the largest modes of the spherical
harmonics of a quantity $A(r,t,\theta)$, which turns out to be 
particularly useful
for a quantitative analysis of the SASI. It is defined as
\begin{equation}
  A(t,r,\theta) \equiv \frac{1}{\sin \theta} \,
  \frac{\partial}{\partial\theta}
  \left( v_{\theta}(t,r,\theta)\,\sin\theta \right)\,,
\label{eq:def_A}
\end{equation}
with $r^{-1}A$ being the divergence of the lateral velocity component,
i.e., $A \equiv r \mathrm{div}(v_{\theta}\,\vec{e}_{\theta})$, which
scales with the size of the lateral velocity of the fluid motion.
At the gain radius, its expansion in spherical harmonics
$Y_{l,m}(\theta,\phi)$ is written
as
\begin{equation}
  A(t,R_{\rm g}(t),\theta) = \sum_{l=0}^{\infty} a_l(t) \, Y_{l,0}(\theta,0)\,,
\label{eq:ylm_decomp}
\end{equation}
where due to the assumption of axisymmetry only $m=0$ has to be
considered.

For $l>0$, the spherical harmonics coefficients $a_l$ of this quantity
are proportional to the ones of the shock displacement (see
\citealt{Foglizzo+06}, Appendix F), so $A(t,\Rg,\theta)$ contains
basically the same information as $\Rs(t,\theta)$.  As
\cite{Blondin_Mezzacappa06}, we prefer to consider a local quantity
$A(t,R_{\rm g}(t),\theta)$ in the postshock layer here
rather than the shock displacement $\delta
R = \Rs(\theta)-\langle\Rs\rangle_{\theta}$ (used in
\citealt{Blondin+03} and \citealt{Ohnishi+06}), because $A$ is much
less affected by noise ($A(t)=0$ for a non-stationary spherical flow,
whereas $\Rs(t)$ is varying) and allows one to measure the oscillation
period and the growth rate much more sensitively than it is possible
by using $\Rs$. Tests showed that for our models, in which relatively
large seed perturbations were imposed on the infalling stellar matter
ahead of the shock, $A$ as defined in Eq.~(\ref{eq:def_A}) yields a
cleaner measure of the SASI even for very low amplitudes than the
perturbed entropy or pressure considered by
\cite{Blondin_Mezzacappa06}. As an example, the absolute values of the
coefficients $a_1$ and $a_2$ are shown as functions of time for
Model W00F in Fig.~\ref{fig:coefficients}.

For a given mode $l$ the oscillation period $\tau_{\mathrm{osc},l}$
can be determined
from the minima of $|a_l(t)|$, which occur at times
$t=t^n_{\mathrm{min}, l}$ when $n$ is a counter for the minima.
The detection of the minima works reliably only when the amplitude
is large enough (it therefore fails in the first
$10$--$20\,$ms) and is also not feasible when convective
instabilities involve a broad range of frequencies in the
nonlinear phase. During one cycle of mode $l$ the corresponding
coefficient $a_l(t)$ becomes zero twice, therefore the cycle period can
be measured as
\begin{equation}
  \tau_{\mathrm{osc},l}(t^n_{\mathrm{min}, l}) =
  t^{n+1}_{\mathrm{min}, l} - t^{n-1}_{\mathrm{min}, l}\,.
\label{eq:def_Taac_meas}
\end{equation}
The evolution of the period of the $l=1$ modes, $\tau_{\mathrm{osc},1}$,
evaluated with Eq.~(\ref{eq:def_Taac_meas}), is displayed for three of
our models in Fig.~\ref{fig:toscevo}.

In order to measure the cycle efficiency, $Q$, we use again the
coefficients $a_l$ defined in Eq.~(\ref{eq:ylm_decomp}). We detect the
positions of the maxima of $|a_l(t)|$, which occur at times
$t=t^n_{\mathrm{max}, l}$ ($n$ now being the counter for the maxima):
if the oscillations of mode $l$ are dominated by the $(k+1)$-th harmonic,
%(i.e. $\tau_{\rm aac}^k = \tau^{\rm f}_{\rm aac}/k$, where $\tau^{\rm f}_{\rm
%aac}$ is the period of the fundamental mode)
$|a_l(t)|$ has $2(k+1)$ maxima during one fundamental cycle period $\tau_
{\rm aac}^{\rm f}$, so the amplification per {\em  fundamental} cycle can be
measured as
\begin{equation}
 Q(t^n_{\mathrm{max}, l}) \equiv \exp\left\lbrack2\pi(k+1)
 \frac{\omega_i}{\omega_r}
\right\rbrack\sim \left\lbrack\frac{ a_l(t^{n+1}_{\mathrm{max}, l}) }
{ a_l(t^{n-1}_{\mathrm{max}, l}) }\right\rbrack^{k+1}.
\label{eq:def_Qaac_meas}
\end{equation}
This method fails, if several of the harmonics are excited with
similar strength. However, these phases can be identified and typically
one of the harmonics dominates clearly (mostly the fundamental mode,
$k=0$). The efficiencies measured by using Eq.~\eqref{eq:def_Qaac_meas}
are shown in Fig.~\ref{fig:Q_aac}.

\subsection{Interpretation of the results}

\begin{figure}[tbph!]
\centering
\includegraphics[angle=0,width=8.5cm]{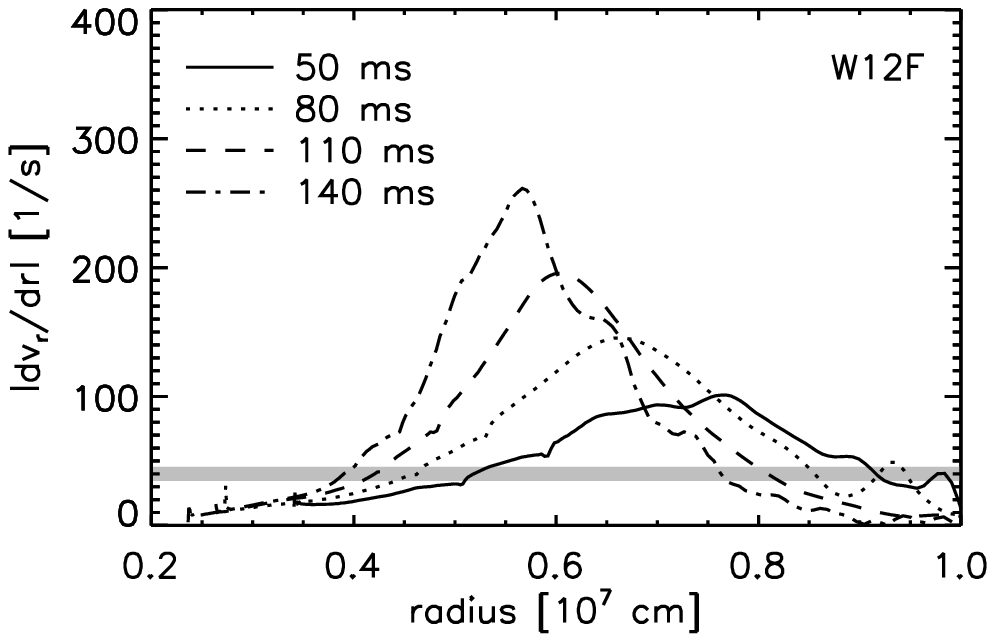}
\includegraphics[angle=0,width=8.5cm]{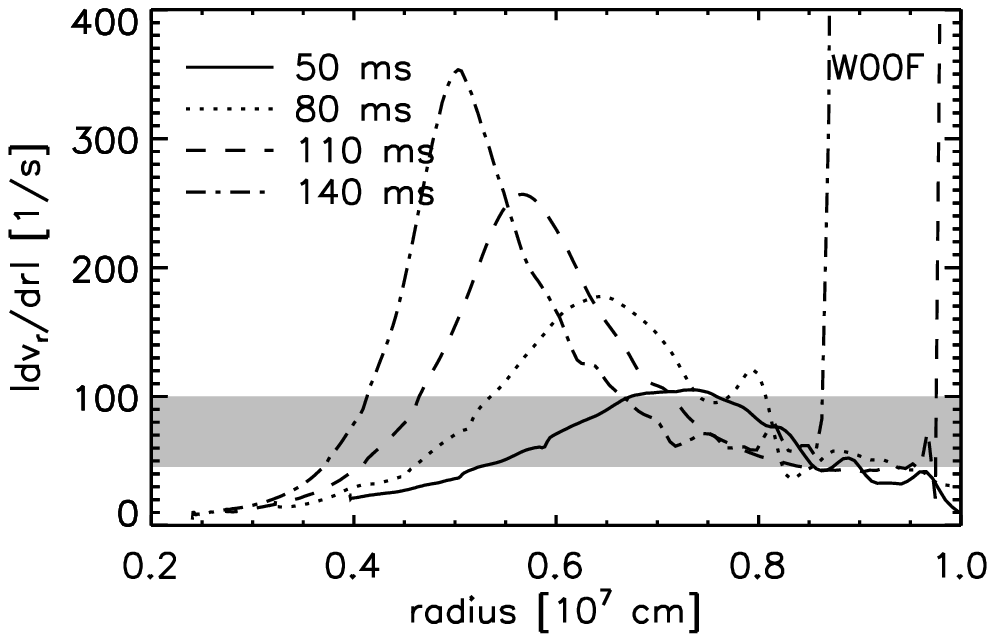}
\includegraphics[angle=0,width=8.5cm]{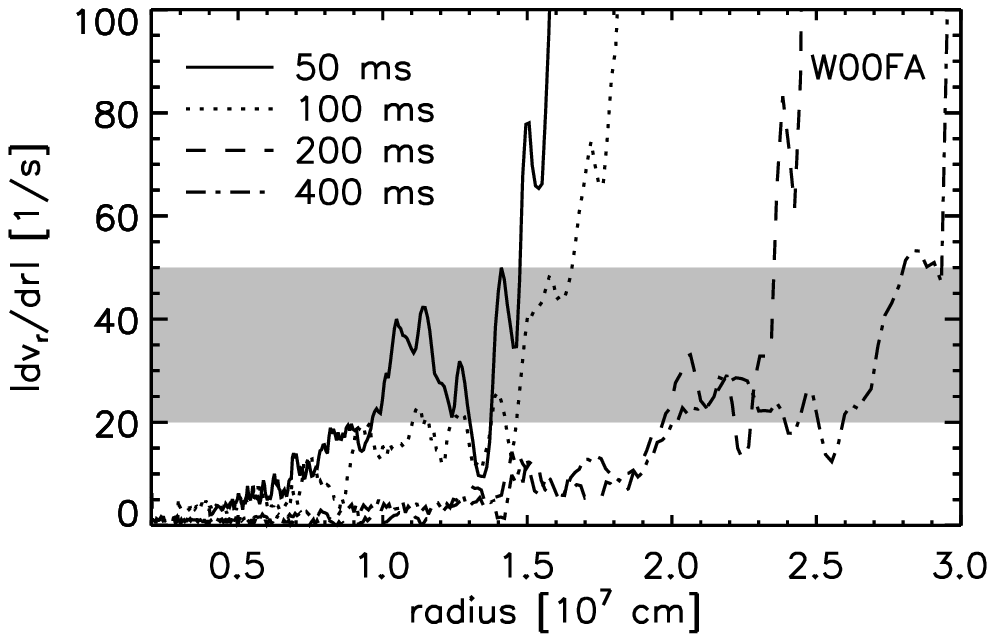}
\caption{Absolute values of the radial derivative of the radial
  velocity component as functions of radius for Models W12F, W00F,
  and W00FA at several times. The gray-shaded
  area indicates the range of values of $\tau_{\rm osc}^{-1}$
  during these times. In the models including neutrinos a
  pronounced ``deceleration peak'' forms with a maximum value
  significantly higher than $\tau_{\rm osc}^{-1}$, whereas such
  a feature is absent in Model W00FA.}
\label{fig:dvelx_dr_w00fa}
\end{figure}

The projection of acoustic and advected perturbations on spherical
harmonics reveals that the shock oscillations are associated
with coherent pressure fluctuations and with the downward advection
of perturbations produced at the shock. This association is
visible in all simulated cases, and is clearly illustrated by
Figs.~\ref{fig:rtimg_press} and \ref{fig:rtimg_vort} for Models W00,
W00F, and W12F.

The pattern of the pressure perturbations in Fig.~\ref{fig:rtimg_press}
reveals the presence
of a particular radius  $R_\varphi$ where a phase shift occurs.
The dash-dotted line in these figures is defined
as the radius $R_\nabla$ where the velocity gradient of the
unperturbed flow has a local extremum. This particular radius seems
to have an important influence on the properties and behavior of
pressure perturbations; in all studied cases the two radii coincide:
\begin{equation}
R_\nabla\sim R_\varphi\,.
\end{equation}
This striking coincidence might be interpreted as the consequence of a
particularly efficient coupling between advected and acoustic perturbations
in layers where the accretion flow is strongly decelerated.
In order to test this hypothesis, we have compared the
wavelength $2\pi v/\omega_r$ ($\omega_r$ was defined in the context
of Eq.~\ref{eq:dispersion}) of advected perturbations at radius
$R_\nabla$ to the length scale $|{\rm d}\ln v/{\rm d}r|^{-1}$ of
this deceleration zone. An efficient coupling is expected if the flow
velocity varies on scales shorter than the wavelength of advected
perturbations:
\begin{equation}
\left|\frac{{\rm d}\ln v}{{\rm d} r}\right|^{-1} \lesssim
\frac{2\pi v}{\omega_r}
\quad \longleftrightarrow \quad
\tau_{\rm osc} \gtrsim \left|\frac{{\rm d} v}{{\rm d} r}\right|^{-1}\ .
\end{equation}
This hypothesis is confirmed by our simulations for Models W12F and W00F
in Fig.~\ref{fig:dvelx_dr_w00fa}. In the case of the neutrinoless
Model W00FA, where only a very weak SASI mode develops, a prominent
deceleration peak is indeed absent. Therefore the interpretation
of $R_\nabla$ within the framework of the AAC
as an effective coupling radius between advected 
perturbations and acoustic feedback, i.e.
\begin{equation}
\Rc\equiv R_\nabla\sim R_\varphi \,,
\end{equation}
is consistent with the results of our simulations: efficient coupling
of advective and acoustic perturbations requires a well localized 
deceleration peak, in which case a strong SASI mode can be expected 
to develop.
We would like to mention that the extrema of the flow deceleration that
are present in our simulations with approximative neutrino transport
are also found in simulations with more sophisticated energy-dependent
neutrino transport and therefore seem to be generic features of the
neutrino-cooled accretion flow.

\begin{figure}[tbph!]
\centering
\includegraphics[angle=0,width=8.5cm]{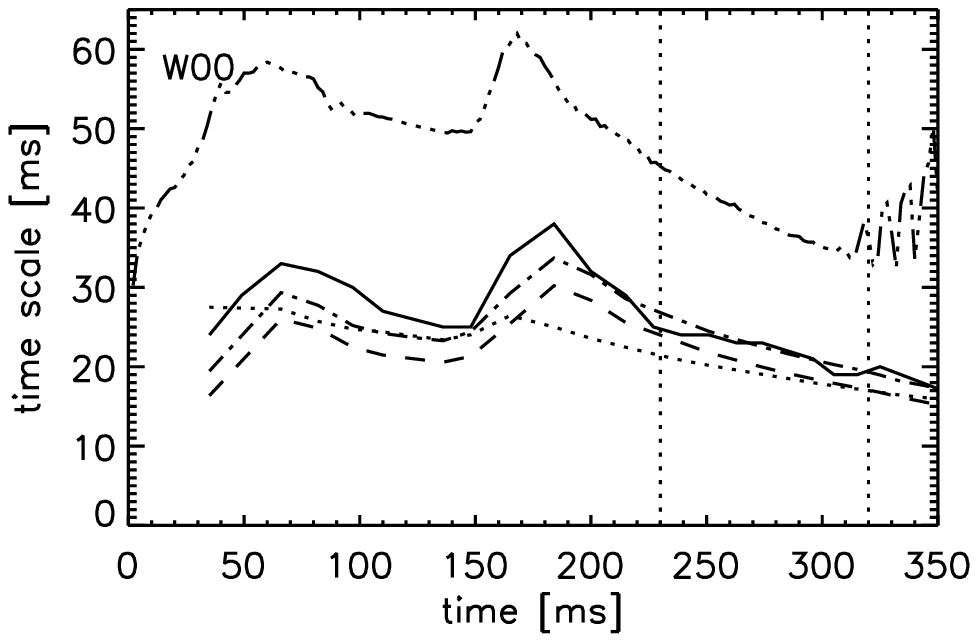}
\includegraphics[angle=0,width=8.5cm]{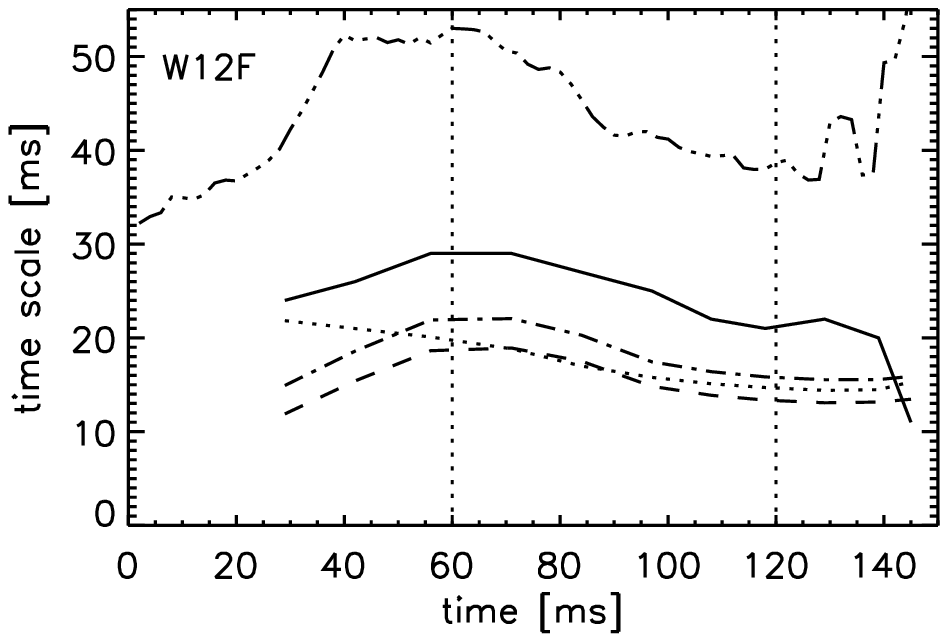}
\includegraphics[angle=0,width=8.5cm]{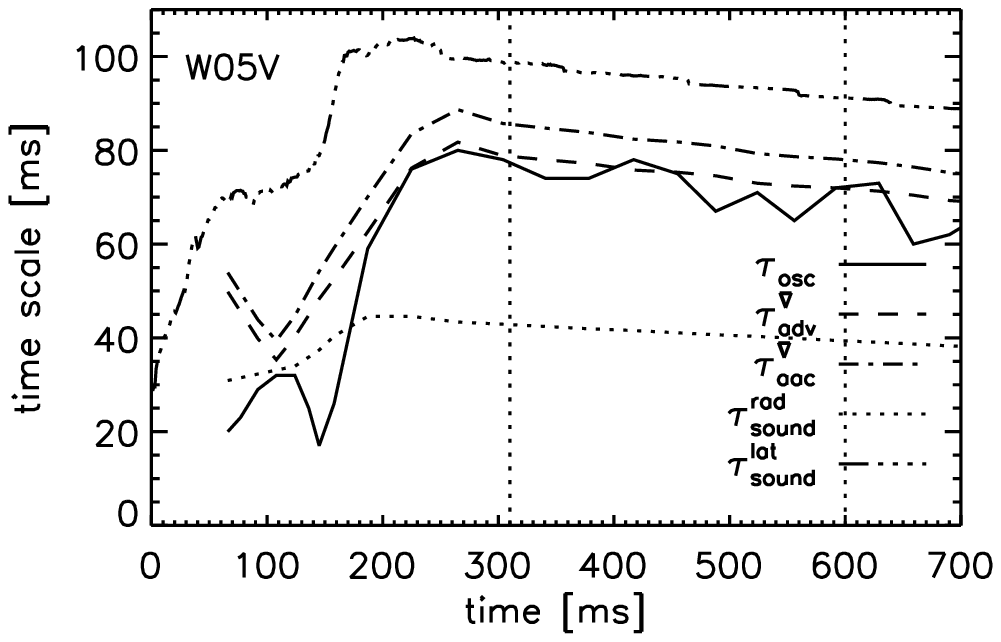}
\caption{Evolution of the $l = 1$ mode oscillation period,
  $\tau_{\rm osc}$, the
  advection time $\tau_{\rm adv}^\nabla$ from the shock to the radius $R_
\nabla$ of strongest
  deceleration, and the time $\tau_{\rm aac}^\nabla$ for a radial 
advective-acoustic cycle, for Models W00, W12F and W05V.
The advection time $\tau_{\rm adv}^\nabla$ agrees with the oscillation time for
Models W00 and W05V. In Model W12F, the oscillation period is longer than
both $\tau_{\rm adv}^\nabla$ and $\tau_{\rm aac}^\nabla$, which can be 
explained by the consequences of strong neutrino heating (see text).
 The oscillation period is also compared to the radial sound travel time
 $\tau_{\rm sound}^{\rm rad}$ through the shock cavity and back, and
  the maximum lateral sound travel time $\tau_{\rm sound}^{\rm lat}$ just
behind the shock. $\tau_{\rm osc}$ is close to $\tau_{\rm sound}^{\rm rad}$
for Model W00, but is closer to
  $\tau_{\rm sound}^{\rm lat}$ for Model W05V. The vertical dotted lines
enclose the time intervals
  considered for the evaluations of Figs.~\ref{fig:tscaling} and
\ref{fig:Q_aac}. }
\label{fig:toscevo}
\end{figure}

\begin{figure}[tbph!]
\centering
\resizebox{\hsize}{!}{\includegraphics[angle=0,width=8.5cm]{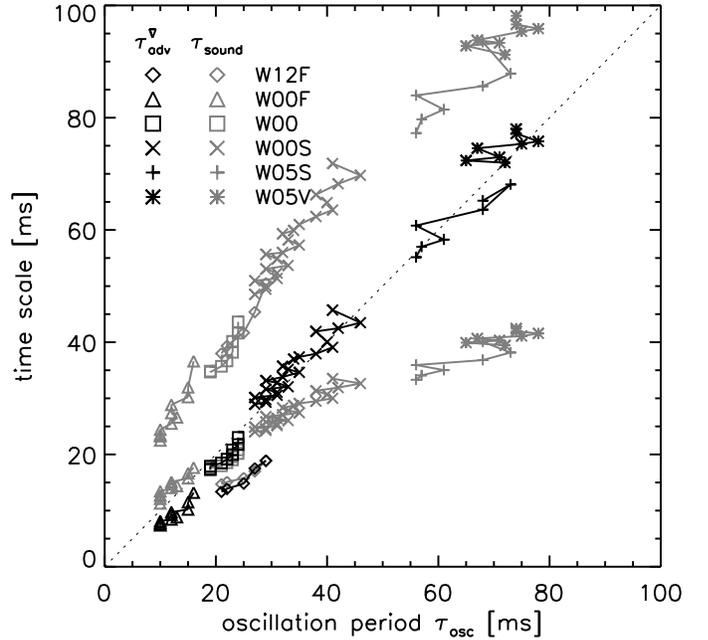}}
\caption{Advection time $\tau_{\rm adv}^{\nabla}$ of the fluid moving
  from the shock to the radius of maximum deceleration, $R_{\nabla}$,
 and the two acoustic times $\tau_{\rm sound}^{\rm rad}$ (lower grey
 symbols) and $\tau_{\rm sound}^{\rm lat}$ (upper grey symbols),
  versus the oscillation period, $\tau_{\rm osc}$, for
  Models W00F, W00, W00S, W05S, W05V and W12F. The data used in
  this figure are selected from phases in which the oscillations can be clearly
  identified and in which the flow is quasi-stationary (see
  Fig.~\ref{fig:toscevo}). While $\tau_{\rm sound}^{\rm lat}$ is clearly
  too long in all cases and $\tau_{\rm sound}^{\rm rad}$ is too short
  for all models except W00 and W00F, the oscillation period is well 
  approximated by $\tau_{\rm adv}^{\nabla}$ for all models except
  W12F. The special role of Model~W12F can be explained by the effects
  of strong neutrino heating (see text).}
\label{fig:tscaling}
\end{figure}

It is interesting to compare the oscillation period $\tau_{\rm osc}$
measured for our models with the timescale $\tau_{\rm aac}^{\rm f}$ of
the fundamental AAC mode, approximated
by the advection time $\tau_{\rm adv}^\nabla$ of the fluid moving
from the shock to $R_\nabla$ ($\tau_{\rm aac}^{\rm f} \approx
\tau_{\mathrm{aac}}^\nabla \approx \tau_{\mathrm{adv}}^\nabla$;
Eqs.~\ref{eq:tadv}, \ref{eq:taac}, and Fig.~\ref{fig:toscevo}). 
A systematic comparison
between the measured oscillation timescale, the advection timescale,
and the acoustic timescales $\tau_{\rm sound}^{\rm rad}$ and
$\tau_{\rm sound}^{\rm lat}$ is shown in Fig.~\ref{fig:tscaling}
for six of our eight models.
In all models except W12F, the advection time is very
close to the oscillation period, whereas in Model~W12F we find
$\tau_{\mathrm{adv}}^\nabla < \tau_{\mathrm{osc}}$.
In the light of the perturbative analysis of \cite{Yamasaki_Yamada+07},
this finding can be interpreted as a consequence of the strong neutrino
heating in Model W12F\footnote{Note that in Model W12F a larger core
neutrino luminosity was assumed than in the other models
(see Table~\ref{tab:restab_limcas}). Moreover, the
prescribed rapid contraction of the nascent neutron star leads to 
a large accretion luminosity. Both contributions to the neutrino
emission cause a particularly strong neutrino energy
deposition in the gain layer.}. \cite{Yamasaki_Yamada+07}
measured the continuous transition of the
eigenfrequency from the oscillatory SASI to the purely growing
($\omega_r=0$) convective instability when neutrino heating is increased.
According to their
work, the oscillation frequency $\omega_r$ of the
SASI is sensitively decreased by the effect
of buoyancy in the gain region, resulting in a significantly 
longer oscillation timescale (see Fig.~3 in \citealt{Yamasaki_Yamada+07}). 
This agrees well with our results, comparing in particular Models W00F
and W12F, whose prescribed contraction of the lower radial grid
boundary is similar, but the latter model has a much larger core
(and higher total) neutrino luminosity 
(see Table~\ref{tab:restab_limcas}), much stronger neutrino
heating, stronger buoyancy, and therefore a larger value of 
$\tau_{\mathrm{osc}}$. In contrast, the advection timescale is 
increased by convection to a lesser extent (see Fig.~4 in 
\citealt{Yamasaki_Yamada+07}), consistent with our finding of the
data points $(\tau_{\mathrm{osc}},\tau_{\mathrm{adv}}^\nabla)$
for Model~W12F lying below the diagonal, dotted line in 
Fig.~\ref{fig:tscaling}. 
The effect of
buoyancy can be seen in the pressure evolution of Model W12F, shown in
the lower plot of Fig.~\ref{fig:rtimg_press},
where a phase shift $\varphi$
takes place in the vicinity of the gain radius (cf.\ Eq.~\ref{phaseQ}).

\begin{figure}[tbph!]
\centering
\resizebox{\hsize}{!}{\includegraphics[angle=0,width=8.5cm]{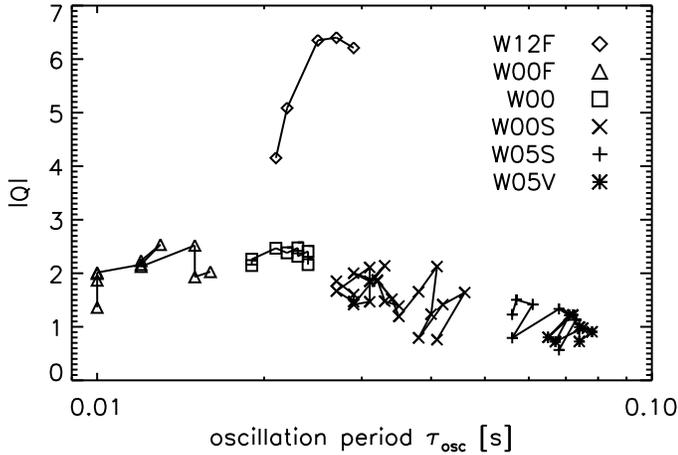}}
\caption{Cycle efficiency, $|Q|$, as a function of the
  oscillation period, $\tau_{\rm osc}$, for Models W00F, W00, W00S,
  W05S, W05V and W12F. The particularly high values for Model W12F
  can be explained as a consequence of strong neutrino heating in
  the gain layer (see text).}
\label{fig:Q_aac}
\end{figure}

The effect of buoyancy in Model W12F is also visible in Fig.~\ref{fig:Q_aac}
showing the amplification factor $Q$ for six of
our eight simulated models . The amplification
factor has modest values between 1 and 3 in most cases, whereas it is
as high as $Q\ga 6$ in Model W12F. This high value of $Q$ can be
understood as a direct
consequence of the small value of the oscillation frequency $\omega_r$,
see Eq.~(\ref{eq:def_Qaac_meas}), because according to \cite{Yamasaki_Yamada+07}
stronger neutrino heating sensitively increases $\tau_{\mathrm{osc}}$,
i.e. reduces $\omega_r = 2\pi/\tau_\mathrm{osc}$, 
but hardly affects the growth rate $\omega_i$
(cf.\ Fig.~2 in \citealt{Yamasaki_Yamada+07}), which appears in the 
numerator of the exponent in Eq.~(\ref{eq:def_Qaac_meas}).
Model W00F exhibits similar trends of 
$\tau_{\mathrm{adv}}^\nabla < \tau_{\mathrm{osc}}$ and $Q$-enhancement
as Model W12F,
however much less strongly. Although in this model the rapid 
contraction of the inner grid boundary leads to a significant 
accretion luminosity, only a very small neutrino flux
from the excised inner core was assumed and therefore the 
neutrino heating in the gain layer is less strong than in
Model W12F.

We wish to point out that our calculation of the amplification factor $Q$ does
not rely on any interpretation of the underlying mechanism. Interestingly, 
however, the
values between 1 and 3 are consistent with those measured by
\citeauthor{Foglizzo+06b} (\citeyear{Foglizzo+06b}, Fig.~17) 
for a shock radius $R_{\rm sh} \sim 2R_\nabla$ 
in a much simpler context.
From the point of view of the underlying mechanism, these values
for $Q$ are consistent with numbers obtained by downward extrapolation
of the efficiency $|Q|_{\rm WKB}$ of the advective-acoustic cycle 
from the region of its validity at larger shock radii (also shown
in Fig.~17 of \citealt{Foglizzo+06b}). For each of the models depicted in
Fig.~\ref{fig:Q_aac},
the amplitude of the spread of amplification factors can receive a
natural explanation in the context of the advective-acoustic mechansim: the
contribution of the acoustic cycle can be either constructive or destructive,
depending on the relative phase of the two cycles, 
which varies with time as the
size of the cavity evolves. This dispersion can be interpreted as a measure of
the efficiency of the acoustic cycle, which is consistently smaller than unity.

The comparison of the oscillation periods with the acoustic timescales shows
that
$\tau_{\rm osc}$ is similar to the radial acoustic timescale
$\tau_{\rm sound}^{\rm rad}$ only in Models W00F and
W00. It is longer by up to 30\% in the case of Model W00S
and by up to a factor of about two in the case of Models W05S and W05V
(Figs.~\ref{fig:toscevo} and \ref{fig:tscaling}). For all models,
the upper bound of
the acoustic time, $\tau_{\rm sound}^{\rm lat}$, is always larger than
$\tau_{\rm osc}$ by 20--50\%.
Note that the setup of Models W05S and W05V (with slow contraction of the
inner grid boundary and non-negligible core neutrino luminosities and thus
significant neutrino heating) was chosen such that the radius of the
standing accretion shock in these models is larger
than in the other cases and therefore the accretion velocities in the
postshock layer are smaller. This enhances the discrepancy
between the advection time and the radial sound crossing time in these
models. Given the lack of any better suggestions for a unique
definition of the timescale of the acoustic cycle than
the lower and upper bounds considered here, and because of
the remarkable correlation
between the oscillation time and $\tau_{\rm adv}^\nabla$, we interpret
Fig.~\ref{fig:tscaling} as a clear support of our hypothesis that the
SASI oscillations are a consequence of the AAC and not of a purely acoustic
amplification process as suggested by \cite{Blondin_Mezzacappa06}.

\subsection{Conclusions about the instability mechanism}

The flow properties that are consistent with an advective-acoustic cycle
as the physical mechanism for the SASI are summarized as follows:
\begin{itemize}
\item[(i)] The acoustic structure of the unstable modes is strongly correlated
with the structure of the velocity gradients (Fig.~\ref{fig:rtimg_press}),
\item[(ii)] the deceleration region is more localized in unstable flows, while
smoothly decelerated flows are more stable (Fig.\ref{fig:dvelx_dr_w00fa}),
\item[(iii)] the advection time $\tau_{\rm adv}^\nabla$ is in good agreement
with the oscillation period of the instability (Fig.~\ref{fig:tscaling}),
\item[(iv)] the typical efficiencies $Q\sim 1$--3 computed in
Fig.~\ref{fig:Q_aac}
are consistent with those extrapolated from the WKB analysis of the
advective-acoustic cycle in Fig.~17 of \cite{Foglizzo+06b}.
Their dispersion smaller
than unity is consistent with the expected marginal effect of the
purely acoustic cycle.
\end{itemize}

In contrast, a purely acoustic interpretation would have to consider the
properties (i), (ii), and (iii) as coincidences,
and the distribution of efficiencies $Q$ remain uninterpreted.
Although the oscillation period is consistent with
an acoustic timescale along a carefully chosen acoustic path,
Fig.~\ref{fig:tscaling} indicates that this acoustic path should be
close to radial for some models, and much more lateral in others.
A theory of the purely acoustic
instability would have to explain this behavior.

Without claiming that our present knowledge of the advective-acoustic theory
is fully satisfactory in the complex core-collapse context, its mechanism is
understandable from the physics point of view and allows us to explain several
features of the simulations, which would not be understood otherwise.

%=====================================================================
\section{Interpretation of the nonlinear phase}
\label{sec:nonlinear}

In the following we will discuss our simulations
during the nonlinear phase of the evolution in which the SASI
cannot be considered as a small perturbation. In particular,
we will analyse the relation between the SASI and convective 
instability, as well as the role these instabilities play for 
the explosion mechanism and the resulting energy of the explosion.

\subsection{The SASI as trigger of convective overturn}
\label{sec:trigger_conv}

In models with low core neutrino luminosity convective 
activity does initially not occur because the corresponding
instability is suppressed in the accretion flow of the 
neutrino-heating layer according to Eqs.~\eqref{eq:cond_chi} and
\eqref{eq:delta_crit}. 
The first large-scale non-radial perturbations in the postshock
flow of such models are therefore caused by SASI oscillations. 
Once large average lateral velocities around $10^9\,$cm$\,$s$^{-1}$
or more are reached in the gain layer at $t > t_{\mathrm{nl}}$ 
(cf.\ Table~\ref{tab:restab_limcas}), however, also the 
smaller-scale mushroom-like structures that are typical of the 
onset of Rayleigh-Taylor instability start to grow.
Within only a few more oscillation
cycles, plumes of neutrino-heated matter and supersonic downdrafts 
of low-entropy matter develop and violent convective overturn 
sets in very similar to what we found in the case of
the models described in Paper~I. There are two effects that are 
mainly responsible for the corresponding change of the flow
character; these are linked to the unsteady motion and the growing
deformation of the shock, respectively.

Firstly, in course of radial expansion and contraction phases 
the shock reaches velocities of ${\cal O}(10^9 \mathrm{cm/s})$,
which is a significant fraction of the preshock velocity.
Since the postshock entropy depends on the preshock
velocity in the frame of the shock, such fast shock oscillations
cause strong variations of the entropy in the downstream region. 
Rapid outward motion of the shock produces high entropies in
the postshock flow, whereas phases in which the shock retreats
lead to lower postshock entropies. Periodic shock expansion and
contraction thus results in alternating layers with high and low
entropies, which are compressed as the accreted matter is 
advected towards the neutron star. With increasing amplitude of the
shock oscillations the convectively unstable entropy gradients 
between these layers eventually become so steep that the growth
timescale of Rayleigh-Taylor instabilities shrinks to about
$1\,$ms, which is much shorter than the advection timescale. 
Therefore non-radial perturbations are able to grow quickly at the
entropy interfaces and vortices and mushroom-like
structures begin to form (see Fig.~\ref{fig:w00_evo_stot}).

\begin{figure}[tbph!]
\centering
\includegraphics[angle=0,width=8.5cm]{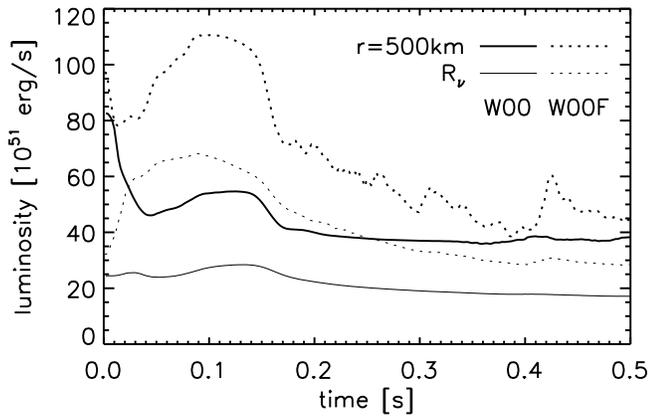}
\caption{Evolution of the sum of the $\nue$ and
  $\nuebar$ luminosities at the neutrinosphere and at $r=500$\,km for
  Models W00 and W00F. The luminosities decay with time because
  the largest mass accretion rates are present at early
  times ($t<0.2\,$s). 
  Model W00F has the higher luminosity, because the neutron star
  contracts faster, setting free more gravitational energy.}
\label{fig:limcas_w00f_evo_lum}
\end{figure}

\begin{figure*}[tbph!]
\sidecaption
\mbox{
\includegraphics[angle=0,width=6cm]{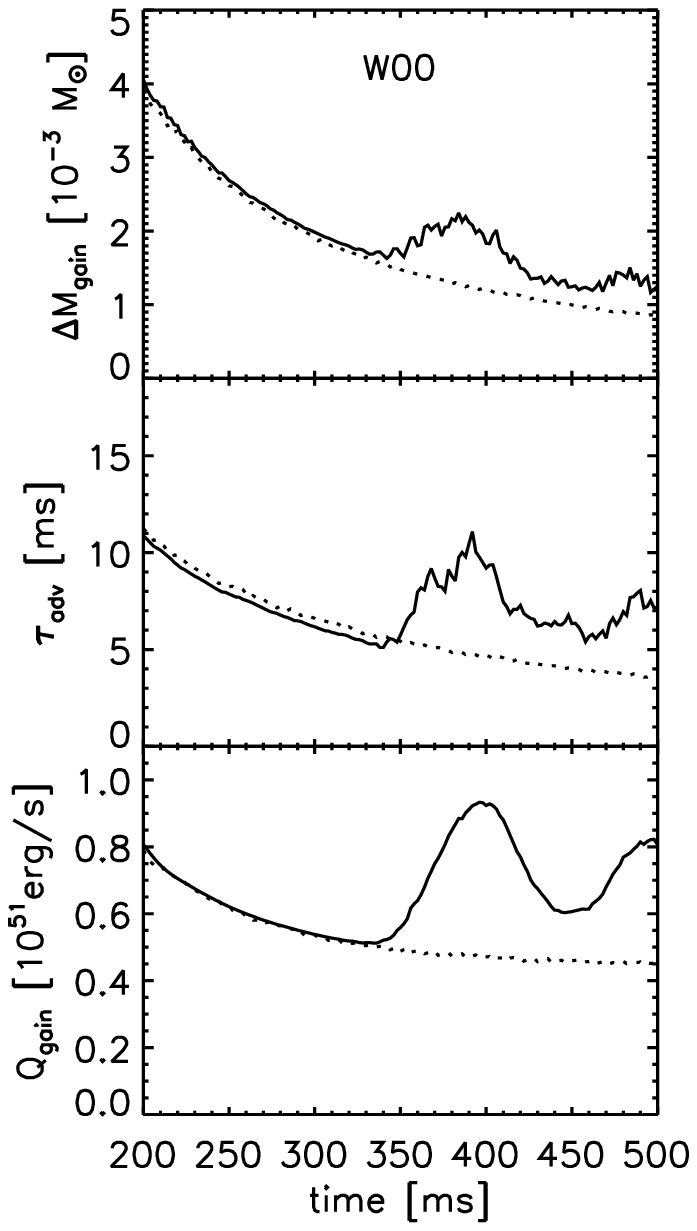}
\includegraphics[angle=0,width=6cm]{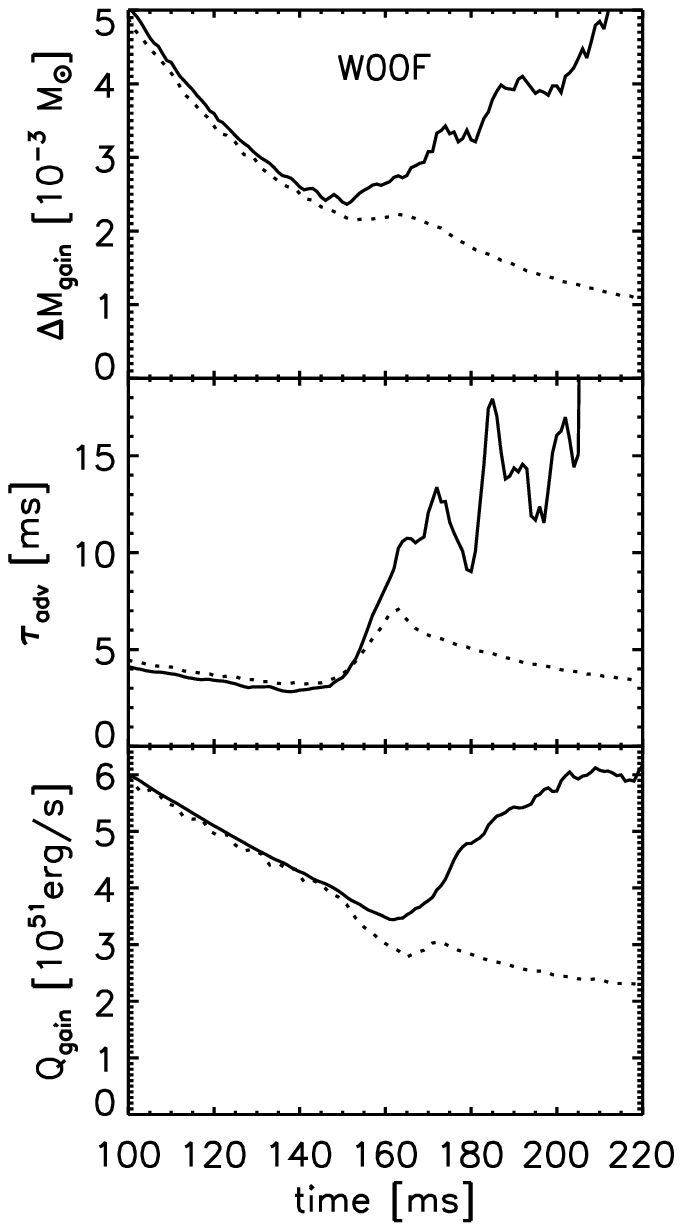}
}
\caption{Evolution of mass in the gain layer, $\Delta M_{\mathrm{gain}}$,
  advection timescale $\tau_{\mathrm{adv}}$ from the shock to the gain radius,
  and neutrino heating rate integrated over the whole gain layer, 
  $Q_{\mathrm{gain}}$, for Models W00 and W00F (solid lines). The dotted
  lines are results from corresponding 1D simulations.
  When the SASI becomes nonlinear ($t
  \approx 350\,$ms for W00, $t \approx 150\,$ms for W00F) and triggers
  convection ($\sim$30$\,$ms later), $\tau_{\mathrm{adv}}$ and 
  $\Delta M_{\mathrm{gain}}$ begin to grow. As a consequence, 
  $Q_{\mathrm{gain}}$ increases significantly compared to the 1D
  simulations. In the case of Model 
  W00F the enhanced heating is strong enough to gravitationally
  unbind the matter in the gain layer and to drive an explosion.
  The advection timescale shows an increase also in the 1D counterpart
  of Model W00F at $t \approx 150\,$ms, because 
  a composition interface of the progenitor star falls through the
  shock at this time and the strong decrease of the mass accretion 
  rate leads to a transient expansion of the shock and of the gain 
  layer. Nevertheless the 1D model does not develop an explosion 
  because without the aid of multi-dimensional effects neutrino heating
  cannot become powerful enough. In Model W00 the neutrino 
  energy deposition rate is so low that even an increase by almost
  a factor of two between 350$\,$ms and 400$\,$ms is not sufficient
  for an explosion.}
\label{fig:limcas_w00f_evo_qmgain}
\end{figure*}

Secondly, also the off-center displacement of the accretion shock
by the $l=1$ SASI mode and the shock deformation caused by $l \ge 2$
modes play an important role when the amplitudes become large enough.
The radial preshock flow hits the deformed or displaced shock 
at an oblique angle. Since the velocity component
tangential to the shock is not changed when the gas passes through the
shock, in contrast to the normal component, which is strongly reduced,
the flow is deflected and attains a substantial lateral velocity
component, whose size and sign changes during the cycle period, see
Fig.~\ref{fig:w00_evo_vely}. As long as the cycle amplitude is small,
the lateral velocity components are also small, and the postshock flow 
remains approximately radial.
In the case of a strongly deformed shock, however, the postshock flow 
becomes mainly non-radial because the lateral velocity reaches a
significant fraction of the preshock velocity (up to several
$10^9\,$cm/s, i.e. the lateral flow becomes supersonic).  For an $l=1$
mode, the highest negative lateral velocities are obtained when the
shock has its maximum displacement in the negative $z$-direction, see
Fig.~\ref{fig:w00_evo_vely}, upper left panel. A shell of matter with
high negative lateral velocity formed in this phase is advected
towards the neutron star, and half an oscillation period later the
highest positive lateral velocities are generated right behind the shock
when the shock expands into positive $z$-direction 
(Fig.~\ref{fig:w00_evo_vely}, middle left panel).  With increasing
oscillation amplitude the shock radius -- and consequently also the
advection timescale and the cycle period -- begin to vary so strongly
during one cycle that the northern and southern hemispheres run ``out
of phase'' so that the shock radii at the north pole
($\theta=0^{\circ}$) and at the south pole ($\theta=180^{\circ}$) reach
their maximum values not alternatingly any more, but at almost the same
time.  In this case streams of matter with high positive and high
negative lateral velocities emerge simultaneously near the north
and south pole, respectively.  These streams collide and one of them
is deflected upwards, producing a bump bounded by two ``kinks'' in the 
shock surface, while the other one is directed downwards, forming a 
supersonic downflow (see Fig.~\ref{fig:w00_evo_vely}, lower left panel 
and Fig.~\ref{fig:w12f_evo_stot}, middle right panel), a phenomenon
that we have also observed in the simulations of Paper~I and that
was also reported by \cite{Burrows06,Burrows07}.

Large-amplitude SASI oscillations are thus able to trigger 
nonlinear convective overturn even in models in which the growth
of buoyancy instabilities is initially suppressed because of 
unfavorable conditions in the accretion flow as discussed in the
context of Eqs.~\eqref{eq:cond_chi} and \eqref{eq:delta_crit}.

\subsection{From SASI oscillations to explosions}
\label{sec:explosion}

Why is Model W00F able to develop an explosion while Model W00 and 
the models with even slower boundary contraction (W00S, W05S, W05V)
do not explode? Models W00 and W00F differ in the assumed contraction
of the nascent neutron star, i.e. in the parameters describing their
final inner boundary radius, $R_{\mathrm{ib}}^{\mathrm{f}}$, and the 
contraction timescale, $\tib$. A smaller value of
$R_{\mathrm{ib}}^{\mathrm{f}}$ implies that the matter accreted on the
forming neutron star sinks deeper into the gravitational potential and thus 
more gravitational energy is released. The smaller value of $\tib$ causes
this release of energy to happen earlier. Most of the potential energy
that is converted to internal energy by $p$d$V$-work is radiated away
in the form of neutrinos. Consequently, the neutrino luminosity that
leads to heating in the gain layer is much
higher at early times in the case of Model W00F 
(Fig.~\ref{fig:limcas_w00f_evo_lum}).

\begin{figure}[tbph!]
\centering
\includegraphics[width=8.5cm]{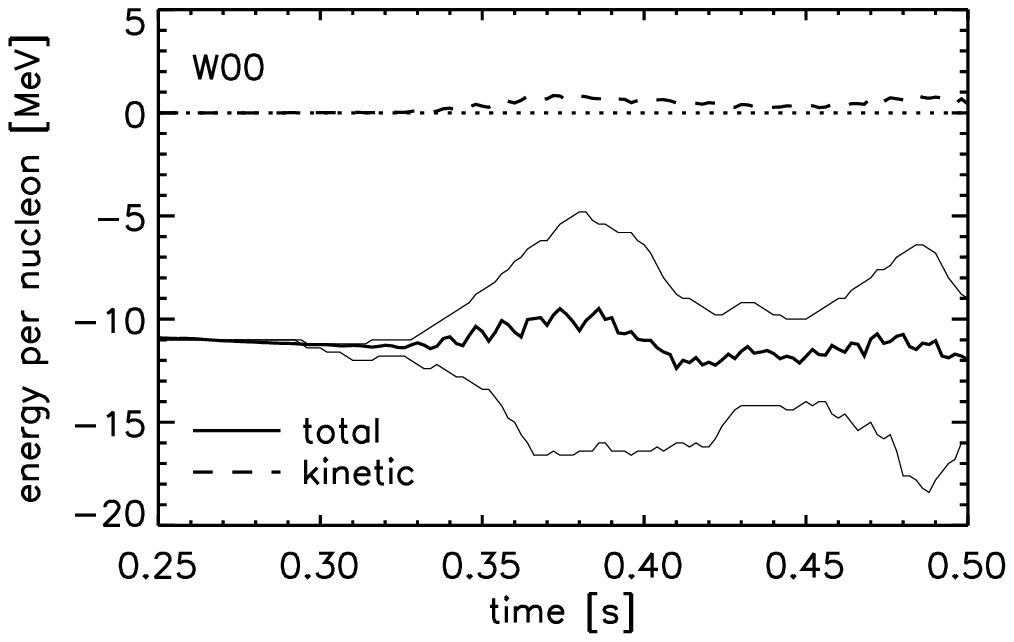}
\includegraphics[width=8.5cm]{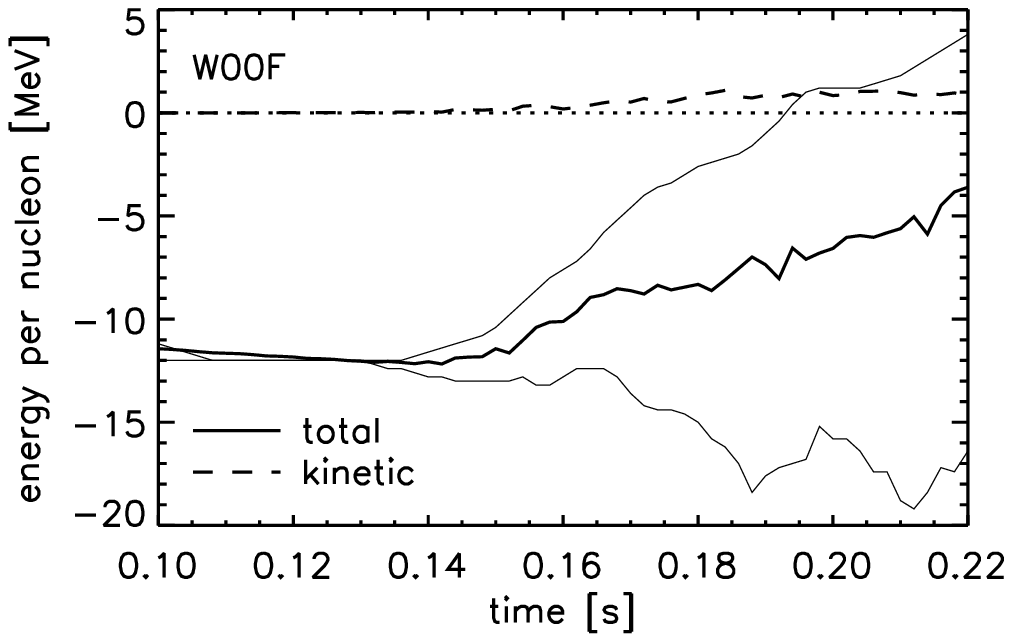}
\caption{Average total (kinetic plus internal plus 
  gravitational) energy per baryon (thick solid line) and kinetic
  energy per baryon (dashed line) versus time in the gain layer 
  of Models W00 and W00F. The thin solid lines correspond to the
  energy interval that contains 90\% of the mass of the gain layer.}
\label{fig:limcas_w00f_egh_mass}
\end{figure}

Yet, these high luminosities alone are not sufficient to start an
explosion. This is demonstrated by a one-dimensional simulation with
the same boundary parameters as Model W00F, which fails to explode.  
It is well known that 
in the multi-dimensional case convection leads to an enhancement of
the efficiency of neutrino energy deposition in the gain layer, on
the one hand because non-radial convective motions stretch the time 
fluid elements can stay in the gain layer and are thus exposed to
efficient neutrino heating in the vicinity of the gain radius,
on the other hand because high-entropy, neutrino-heated matter
becomes buoyant, expands quickly, and thus cools, which
reduces the energy loss by the reemission of neutrinos.
The former of these two effects effectively leads to an increase of
the advection timescale of accreted matter from the shock to the gain
radius (see also \citealt{Buras+06b}), as a consequence of which the 
mass in the gain layer becomes larger. The same effect can also be
produced by large-amplitude SASI oscillations, because such non-radial
motions expand the average shock radius, thus leading to smaller 
postshock velocities, and deflect the postshock flow in non-radial
direction, also leading to a longer advection time of accreted matter
through the gain layer.

In Model W00F we observe such a rise of the advection timescale
starting at $t \approx 150\,$ms (Fig.~\ref{fig:limcas_w00f_evo_qmgain})
when the postshock flow becomes strongly non-radial, 
but violent convective overturn has not yet set in 
(Fig.~\ref{fig:w00_evo_vely}, right middle panel). This increase of the 
advection time leads to a significant growth of the integrated neutrino heating
rate in the gain layer, an effect that becomes even more pronounced when 
the convective activity gains strength ($t \gtrsim 170\,$ms).
Initially the total specific energy of most of the matter in the gain 
region is in a narrow range around $-11\,$MeV per nucleon, but the
distribution of specific particle energies becomes broader
by the influence of the large-amplitude SASI and of convective
overturn (Fig.~\ref{fig:limcas_w00f_egh_mass}). Due to the large
energy deposition by neutrinos the mean value of the total energy rises 
and ultimately some fraction of the matter in the gain layer acquires 
positive total energy and the explosion sets in.
Also in Model W00 we see enhanced neutrino
heating (up to two times higher than in the corresponding
one-dimensional simulation) from $t \approx 350\,$ms on, caused by a
combination of nonlinear SASI motions and convective activity
(Fig.~\ref{fig:limcas_w00f_evo_qmgain}). However, due to the low
accretion rate at this late time the neutrino luminosity
and thus the neutrino heating rate are much lower than in Model
W00F at $t \approx 200\,$ms. The total energy in the gain layer 
of Model W00 increases only temporarily by about $1\,$MeV per
nucleon but then drops again soon and continues to decrease
slowly later on (Fig.~\ref{fig:limcas_w00f_egh_mass}). The distribution
of specific energies of matter in the gain layer does not become very
broad and none of the matter gets unbound. In both the Models W00 and
W00F the specific kinetic energy in the gain layer remains
relatively small (only about $1\,$MeV/nucleon, see
Fig.~\ref{fig:limcas_w00f_egh_mass}).

% (1) redistribution of energy is not sufficient for explosion

Different from \cite{Blondin+03} we do not observe a continuous 
increase of the kinetic energy associated with lateral (turbulent)
motion of the matter behind the shock. In their simulations without
neutrino effects, \cite{Blondin+03} observed that the SASI 
oscillation can redistribute some of the gravitational binding 
energy of the accreted matter from coherent fluid motion to 
turbulent energy, in fact with sufficient efficiency to drive an
expansion of the accretion shock. Since some of the turbulent 
material had obtained positive total energy at the end of their
simulations, \cite{Blondin+03} concluded that the SASI in their
calculations was able to lead to an explosion. We do not see this
kind of process going on in our simulations (in agreement with
the results of \citealt{Burrows06,Burrows07}). The reason for this
discrepancy may be the inclusion of neutrino heating and cooling
in our models. It is possible that the energy loss by neutrino 
emission below the gain radius prevents the efficient conversion
of gravitational binding energy to turbulent energy. Another
reason may be the 
use of different conditions at the outer radial grid boundary 
in our models; while \cite{Blondin+03} assumed steady-state accretion
and thus held the mass accretion rate fixed with time, the 
stellar progenitor structure employed in our work leads to a 
continuous decrease of the mass accretion rate at the shock.
Therefore less total kinetic energy is available that can be 
converted to turbulent motions by the distorted accretion shock.

In our simulations a growth of the turbulent kinetic energy of
the matter in the gain layer is definitely not the reason for 
starting the explosions. The corresponding lateral kinetic energy 
never exceeds a few $10^{49}\,$erg in any of our models.
This is well
below the size of neutrino energy deposition and of the energy needed
for unbinding matter and triggering an explosion. 
Nevertheless, the non-radial flow associated with the SASI
is certainly helpful, in combination with convection actually crucial  
for making the neutrino-heating mechanism work. The failure of 
one-dimensional simulations with the same treatment of the neutrino
physics clearly demonstrates the importance of non-radial fluid
instabilities, convection and the SASI, for a success of the neutrino-driven
explosion mechanism. These hydrodynamic instabilities affect the 
gas motion in the gain layer such
that the advection timescale from the shock to the gain radius is 
effectively increased. This enhances
the efficiency of neutrino energy deposition by allowing more matter to
be exposed to the intense neutrino flux near the gain radius for a longer
time. Thus both convection and the SASI can be considered as ``catalysts''
that facilitate neutrino-driven explosions rather than being direct drivers 
or energy sources of the explosion. As a consequence, explosions in 
multi-dimensional simulations, i.e. even with the support by convective
overturn and the SASI, still require the presence of strong enough
neutrino heating. Our set of simulations clearly demonstrates that only in 
the case of a sufficiently large neutrino luminosity and thus only for 
sufficiently
powerful neutrino heating behind the shock, the models are able to 
develop an explosion. This finding is in agreement with the results of
\cite{Ohnishi+06}, who also obtained an explosion by neutrino heating
only in a simulation with high neutrino luminosity, while lower-luminosity 
cases failed to explode.

\subsection{The importance of the seed perturbations}
\label{sec:w12f_interpretation}

% (4) nonlinear phase must be reached soon enough
The comparison of our results shows that explosions during the first
second after core bounce do not only require the neutrino
luminosities to be large enough but also that the SASI and
convection are able to reach the nonlinear phase sufficiently quickly.
Whether this is the case or not depends on their growth rates, which
in turn depend on the properties of the postshock flow. The latter
are a complex function of the progenitor structure, the neutron
star contraction, and neutrino cooling and heating in the layer
between neutron star and shock. Last but not least, also the size
of the seed perturbations, i.e. the inhomogeneities present in the
collapsing star, can play a role for the development and growth of 
non-radial hydrodynamic instabilities after core bounce.

In our Models W12F and W12F-c (as well as in most of the recent 
simulations with multi-energy group neutrino transport by
\citealt{Buras+06,Buras+06b}) the advection time through the
gain layer is so short and the growth rate of convective
instabilities (Eq.~\ref{omegabuoy}) in that region so small that the
timescale ratio $\chi$ of Eq.~(\ref{eq:def_chi}) remains below the 
critical threshold $\chi_0$ for a linear growth of globally 
unstable modes, i.e., $\chi < \chi_0 \approx 3$ according to 
\cite{Foglizzo+06}, see Eq.~(\ref{eq:cond_chi}) and also
Fig.~3 in \cite{Buras+06b}. This means that the fast advection
of the flow from the shock to the gain radius suppresses the 
growth of convective modes according to linear stability analysis.
However, as explained in Sect.~\ref{sec:convection}, in this case
buoyancy can nevertheless drive bubble rise and convective
instability  
in a nonlinear way if the initial density perturbations $\delta$ 
in matter falling through the shock are large enough, i.e.
$\delta > \delta_{\mathrm{crit}}\exp(-\chi)$ according to
Eq.~(\ref{eq:cond_deltas}), with $\delta_{\mathrm{crit}}$ being 
typically of the order of some percent (see Eq.~\ref{eq:delta_crit}).

The inhomogeneities in the matter upstream of the shock originate
from seed perturbations in the progenitor star, whose size and
amplitude are not well known because three-dimensional, long-time 
stellar evolution simulations for full-sphere models until the
onset of core collapse have not been possible so far (see, e.g.,
\citealt{Bazan_Arnett98,Murphy+04,Young+05,Meakin_Arnett06,Meakin_Arnett07a,Meakin_Arnett07b}).
With our assumed initial inhomogeneities in the case of Models W12F 
and W12F-c (see Table~\ref{tab:restab_limcas} and 
Sect.~\ref{sec:modelpars}), the perturbation amplitudes remain
below the critical value $\delta_{\mathrm{crit}}$ in the former
case, whereas they become larger than this threshold value in the
latter case (see Fig.~\ref{fig:chi_denspert}). Therefore, as
visible in Fig.~\ref{fig:w12f_evo_stot}, the 
fastest growing non-radial instability on large scales is the
SASI in Model W12F, whereas it is convective overturn in Model
W12F-c. Only because of the growth of SASI modes does Model W12F
also develop convective activity in the gain layer, which enhances
the efficiency of neutrino energy deposition and finally leads to
an explosion also in this case. The crucial role of these
non-radial instabilities is demonstrated again by
a corresponding one-dimensional simulation that does not develop
an explosion. In Model W12F the SASI is a key feature in the
multi-dimensional evolution, because the development of convective
modes is not possible in the first place due to the low initial 
amplitude of perturbations and the insufficient growth of these
perturbations in the advection flow from the shock to the gain 
radius.

Considering Models W12F and W12F-c, however, no noticeable memory 
of the initial source of the low-mode asymmetries is retained 
during the long-time evolution.
Although the early postbounce evolution of
these two models is clearly different and the times of the onset
of the explosion differ, the global parameters of the explosion
become very similar (see Table~\ref{tab:restab_limcas}).
Neither the explosion energy nor the neutron
star mass and kick velocity are strongly affected by the different 
explosion times, because the conditions in the infalling stellar 
core change only on longer timescales and the ejecta energy and
neutron star recoil build up over a much more extended period of
time after the launch of the explosion (see Paper~I).
Since the anisotropies develop chaotically and in a very irregular 
way during the nonlinear phase, the final ejecta morphology is the 
result of a stochastic process and does not depend in a
deterministic and characteristic manner on the type of non-radial 
instability that has grown fastest after core bounce. It therefore
seems unlikely that observational parameters of 
supernova explosions are able to provide evidence of the
initial trigger of the large-scale anisotropies that develop in
the early stages of the explosion. Future simulations with
a more detailed treatment of the neutrino transport (instead 
of our approximative description) and without the use of the
inner boundary condition of the present models will have to show 
whether the gravitational-wave and neutrino signals carry 
identifiable fingerprints of this important aspect of supernova
dynamics.

%=====================================================================
\section{Summary and conclusions}
\label{sec:conclusions}

We performed a set of two-dimensional hydrodynamic simulations with
approximative neutrino transport to investigate the role of
non-convective instabilities in supernova explosions. As initial data
we used a postbounce model of a 15$\,M_\odot$ progenitor star, 
which had been 
evolved through core collapse and bounce in a computation with detailed,
energy-dependent neutrino transport. For following the subsequent,
long-time evolution, the neutron star core (above a neutrino optical 
depth of about 100) was excised and replaced by a contracting 
Lagrangian inner boundary that was intended to mimic the behavior of
the shrinking, nascent neutron star. The models of our set differed
in the choice of the neutrino luminosities assumed to be radiated by 
the excised core, in the prescribed speed and final radius of the 
contraction of the neutron star, and in the initial velocity 
perturbations imposed on the 1D collapse model after bounce.

Our hydrodynamic simulations indeed provide evidence --- supporting
previous linear analysis \citep{Foglizzo+06,Foglizzo+06b,Yamasaki_Yamada+07}
--- that two different hydrodynamic instabilities, convection and the 
SASI \citep{Blondin+03}, occur at conditions present during the 
accretion phase of the stalled shock in collapsing stellar cores and 
lead to large-scale, low-mode asymmetries. These non-radial
instabilities
can clearly be distinguished in the simulations by their growth 
behavior, location of development, and spatial structure. While
convective activity grows in a non-oscillatory way and its onset can
be recognized from characteristic mushroom-type structures appearing
first in regions with steep negative entropy gradients, the SASI 
starts in an oscillatory manner, encompasses the whole postshock
layer, and leads to low-mode shock deformation and displacement.

As discussed by \cite{Foglizzo+06}, the growth of convection is 
suppressed in the accretion flow because of the rapid infall 
of the matter from the shock to the gain radius, unless either
neutrino heating is so strong and therefore the entropy gradient
becomes so steep that the advection-to-growth
timescale ratio ($\chi$ of Eq.~\ref{eq:def_chi}) exceeds the
critical value $\chi_0 \approx 3$, or, alternatively, 
sufficiently large density perturbations in the accretion flow
(cf. Eq.~\ref{eq:cond_deltas}) 
cause buoyant bubbles rising in the infalling matter.
While convective instability is damped by faster infall of
accreted matter, the growth rate of the SASI {\em increases} when the 
advection timescale is shorter. 
Moreover, the quasi-periodic shock expansion
and contraction with growing amplitude due to the SASI
produce strong entropy variations in the postshock flow, which
can then drive convective instability. Even when the neutrino-heated
layer is not unstable to convection in the first place, the 
perturbations caused by the SASI oscillations can thus be the trigger
of ``secondary'' convection.

Our detailed analysis of the evolution of the SASI modes in our
simulations, of their dependence on the model parameters, and of
the cooperation between convection and the SASI in the nonlinear
regime revealed the following facts:
 
\begin{enumerate}

\item
  When the SASI reaches large amplitudes and supersonic lateral
  velocities occur in the postshock flow, sheets with very steep
  unstable entropy gradients are formed. As a consequence,
  low-mode convective overturn grows in a highly nonlinear way.
  Since the SASI and strong convective activity push the accretion
  shock to larger radii, they reduce the infall velocity in the
  postshock layer. Moreover, the flow in the neutrino-heating 
  region develops large non-radial velocity components and 
  therefore the accreted matter stays 
  in the gain layer for a longer time. This increases the
  energy deposition by neutrinos in this region and facilitates
  the explosion. However, like convection the SASI does not
  guarantee an explosion on the timescale considered in our
  simulations (in agreement with the findings of 
  \citealt{Burrows06,Burrows07}): the kinetic energy associated with the
  SASI remains negligible for the explosion energetics. Therefore
  sufficiently strong neutrino heating and consequently a sufficiently
  large neutrino luminosity are still necessary to obtain explosions.

\item
  The growth rate (and amplification) and oscillation frequency of the 
  SASI depend sensitively on the advection time
  from the shock to the coupling region and the structure of the
  flow in this region. The latter, in turn, depends on the neutron star
  contraction (which has a strong influence on the shock radius), 
  on the neutrino luminosities and the corresponding
  heating and cooling, and on the mass accretion rate of the stalled
  shock in the collapsing star. For a wide range of investigated
  conditions (changing the parameters of our models), we found the
  SASI being able to develop large amplitudes on timescales relevant
  for the explosion. Therefore the SASI turned out to
  create large-scale anisotropy also in cases where convective
  activity was not able to set in in the first place.

\item
  In our simulations we could clearly identify a faster contraction and
  a smaller radius of the excised core of the nascent neutron star
  as helpful for an explosion. We therefore conclude that a softer
  high-density equation of state and general relativity, which both
  lead to a more compact neutron star, are favorable for an explosion.
  This is so because on the one hand the
  accretion luminosities of neutrinos become higher, correlated with
  stronger neutrino heating behind the stalled shock, and on the
  other hand the amplification of the SASI increases with enhanced
  neutrino heating. Since the initial
  growth of convection is damped or suppressed by more rapid infall,
  the presence of the SASI instability and its ability to trigger
  convection as a secondary phenomenon, play a crucial role for the
  final success of the delayed neutrino-heating mechanism.

\item
  The amplitude of the initial seed perturbations in the collapsing
  core of the progenitor star, which is not well constrained due to the
  nonexistence of fully consistent and long-evolved three-dimensional
  stellar evolution models, has an influence on the question
  whether convection or the SASI develop more rapidly after core bounce.
  While this can determine how fast the explosion sets in,
  we found that once the two non-radial instabilities are simultaneously
  present and cooperate in the nonlinear regime, the global properties
  of the explosions are essentially insensitive to the initial phase.
  Since the final anisotropic distribution of the ejecta is the result
  of a very stochastic and chaotic process, it has also lost the memory
  of the early evolution. Neither the explosion energy nor the neutron
  star kick velocity are therefore good indicators of the initial seed
  perturbations that existed in the progenitor star and of the type
  of the fastest growing non-radial instability. Future
  supernova models without the approximations used in our simulations
  will have to show whether the neutrino and
  gravitational-wave signals, which will be measurable in detail from a
  galactic event, carry any useful information about this crucial
  aspect of the postbounce explosion dynamics.

\item
  While most of the above conclusions and the corresponding 
  evaluation of our simulations are independent of the
  exact physical mechanism that is responsible for
  the growth of the SASI, we nevertheless 
  tried to explore this important question,
  which is still controversially discussed in the literature. To this
  end we analysed our models in the linear regime of the SASI
  and compared the results on the one hand to predictions based on the 
  hypothesis that the SASI growth is due to an advective-acoustic cycle
  (AAC), and on the other hand to the possibility that the SASI is 
  driven by a purely acoustic mechanism.
  Our analysis shows that many of the observed SASI properties are
  consistent and can be well understood with the AAC hypothesis. This
  is the case for: (1) The oscillation period 
  of the SASI, which agrees well with the advection time of perturbations
  from the shock to a radius $R_\nabla$ where the deceleration of the
  accretion flow is strongest. This radius is located in the 
  neutrino-cooling layer somewhat outside of the neutrinosphere.
  (2) The acoustic structure of the unstable modes is strongly 
  correlated with the velocity gradient in the postshock layer
  and more SASI-unstable flows are obtained in more abruptly
  decelerated accretion flows.
  (3) The amplification factors found for the SASI agree with
  extrapolated WKB results for the AAC. Moreover, the effect that
  stronger neutrino heating causes a larger SASI amplification
  efficiency can be explained on grounds of an assumed AAC.
  In contrast, our measured oscillation timescales for the SASI 
  are not consistent with a uniquely chosen path for the sound wave 
  propagation through the shock cavity in all models, but would
  require that the acoustic waves travel more radially in some
  cases and more in angular direction in other models. This as
  well as the other mentioned features are not satisfactorily 
  accounted for by the existing theory of a purely acoustic 
  instability. We therefore think that our analysis provides
  significant support for the suggestion that the amplification of the 
  SASI happens through an AAC rather than a purely acoustic mechanism.

\end{enumerate}

While the presented simulations employed a number of approximations
like the neutrino transport scheme and the use of an inner boundary
condition instead of following the evolution of the neutron star 
core, we are confident that our main findings do not depend on these
simplifications. In fact, recent long-time stellar core-collapse 
simulations with detailed, multi-energy-group neutrino transport
and fully consistent consideration of the central part of the 
nascent neutron star basically confirm the importance of the SASI
and its nonlinear interaction with convective instability for the
viability of the delayed neutrino-driven explosion mechanism
\citep{Marek_Janka07}.

%%%%%%%%%%%%%%%%%%%%%%%%%%%%%%%%%%%%%%%%%%%%%%%%%%%%%%%%%%%%%%%%%%%%%%

\begin{acknowledgements}
  We thank R.~Buras and M.~Rampp for providing us with post-bounce
  models and S.~Woosley and A.~Heger for their progenitor models. 
  Support by the Sonderforschungsbereich~375
  on ``Astroparticle Physics'' of the Deutsche Forschungsgemeinschaft
  in Garching and funding by DAAD (Germany) and Egide (France) through
  their ``procope'' exchange program are acknowledged. The 
  computations were performed on the IBM p655 of the Max-Planck-Institut 
  f\"ur Astrophysik and on the IBM p690 clusters of the Rechenzentrum
  Garching and of the John-von-Neumann Institute for Computing in
  J{\"u}lich. 
\end{acknowledgements}

%%%%%%%%%%%%%%%%%%%%%%%%%%%%%%%%%%%%%%%%%%%%%%%%%%%%%%%%%%%%%%%%%%%%%%

\bibliography{paper}

\begin{thebibliography}{44}
\expandafter\ifx\csname natexlab\endcsname\relax\def\natexlab#1{#1}\fi

\bibitem[{{Arcones} {et~al.}(2006){Arcones}, {Janka}, \&
  {Scheck}}]{Arcones+06}
{Arcones}, A., {Janka}, H.-Th., \& {Scheck}, L. 2006, \aap, 467, 1227

\bibitem[{{Arzoumanian} {et~al.}(2002){Arzoumanian}, {Chernoff}, \& 
{Cordes}}]{Arzoumanian+02}
{Arzoumanian}, Z., {Chernoff}, D.~F., \& {Cordes}, J.~M. 2002, \apj, 568, 289

\bibitem[{{Bazan} \& {Arnett}(1998)}]{Bazan_Arnett98}
{Bazan}, G. \& {Arnett}, W.~D. 1998, \apj, 496, 316

\bibitem[{{Blondin} \& {Mezzacappa}(2006)}]{Blondin_Mezzacappa06}
{Blondin}, J.~M. \& {Mezzacappa}, A. 2006, \apj, 642, 401

\bibitem[{{Blondin} \& {Shaw}(2007)}]{Blondin_Shaw07}
{Blondin}, J.~M. \& {Shaw}, S. 2007, \apj, 656, 366

\bibitem[{{Blondin} {et~al.}(2003){Blondin}, {Mezzacappa}, \&
  {DeMarino}}]{Blondin+03}
{Blondin}, J.~M., {Mezzacappa}, A., \& {DeMarino}, C. 2003, \apj, 584, 971

\bibitem[{{Buras} {et~al.}(2003){Buras}, {Rampp}, {Janka}, \&
  {Kifonidis}}]{Buras+03}
{Buras}, R., {Rampp}, M., {Janka}, H.-T., \& {Kifonidis}, K. 2003, \prl, 90,
  241101

\bibitem[{{Buras} {et~al.}(2006{\natexlab{a}}){Buras}, {Rampp}, {Janka}, \&
  {Kifonidis}}]{Buras+06}
{Buras}, R., {Rampp}, M., {Janka}, H.-T., \& {Kifonidis}, K.
  2006{\natexlab{a}}, \aap, 447, 1049

\bibitem[{{Buras} {et~al.}(2006{\natexlab{b}}){Buras}, {Janka}, {Rampp}, \&
  {Kifonidis}}]{Buras+06b}
{Buras}, R., {Janka}, H.-T., {Rampp}, M., \& {Kifonidis}, K.
  2006{\natexlab{b}}, \aap, 457, 281

\bibitem[{{Burrows} {et~al.}(1995){Burrows}, {Hayes}, \& {Fryxell}}]{BHF95}
{Burrows}, A., {Hayes}, J., \& {Fryxell}, B.~A. 1995, \apj, 450, 830

\bibitem[{{Burrows} {et~al.}(2006){Burrows}, {Livne}, {Dessart}, {Ott},
 \& {Murphy}}]{Burrows06}
{Burrows}, A., {Livne}, E., {Dessart}, L., {Ott}, C.~D., \& {Murphy}, J. 
 2006, \apj, 640, 878

\bibitem[{{Burrows} {et~al.}(2007){Burrows}, {Livne}, {Dessart}, {Ott},
 \& {Murphy}}]{Burrows07}
{Burrows}, A., {Livne}, E., {Dessart}, L., {Ott}, C.~D., \& {Murphy}, J. 
 2007, \apj, 655, 416

\bibitem[{{Chandrasekhar}(1961)}]{Chandra61}
{Chandrasekhar}, S. 1961, {Hydrodynamic and hydromagnetic stability} (New York:
  Dover)

\bibitem[{{Chatterjee} {et~al.}({2005}){Chatterjee}, {Vlemmings}, {Brisken},
  {Lazio}, {Cordes}, {Goss}, {Thorsett}, {Fomalont}, {Lyne}, \&
  {Kramer}}]{Chatterjee+05}
{Chatterjee}, S., {Vlemmings}, W.~H.~T., {Brisken}, W.~F., {et~al.} {2005},
  \apjl, {630}, L61

\bibitem[{Colella \& Woodward(1984)}]{CW84}
Colella, P. \& Woodward, P.~R. 1984, J. Comput. Phys., 54, 174

\bibitem[{{Cordes} {et~al.}(1993){Cordes}, {Romani}, \& {Lundgren}}]{Cordes+93}
{Cordes}, J.~M., {Romani}, R.~W., \& {Lundgren}, S.~C. 1993, \nat, 362, 133

\bibitem[{{Foglizzo}(2001)}]{Foglizzo01}
{Foglizzo}, T. 2001, \aap, 368, 311

\bibitem[{{Foglizzo}(2002)}]{Foglizzo02}
{Foglizzo}, T. 2002, \aap, 392, 353

\bibitem[{{Foglizzo} {et~al.}(2006){Foglizzo}, {Scheck}, \&
  {Janka}}]{Foglizzo+06}
{Foglizzo}, T., {Scheck}, L., \& {Janka}, H.-T. 2006, \apj, 652, 1436

\bibitem[{{Foglizzo} {et~al.}(2007){Foglizzo}, {Galletti},
  {Scheck}, \& {Janka}}]{Foglizzo+06b}
{Foglizzo}, T., {Galletti}, P., {Scheck}, L., \& {Janka}, H.-T.
  2007, \apj, 654, 1006

\bibitem[{{Foglizzo} \& {Tagger}(2000)}]{Foglizzo_Tagger00}
{Foglizzo}, T. \& {Tagger}, M. 2000, \aap, 363, 174

\bibitem[{{Galletti} \& {Foglizzo}(2005)}]{Galletti_Foglizzo05}
{Galletti}, P. \& {Foglizzo}, T. 2005, in Proc. SF2A-2005: 
Semaine de l'Astrophysique Francaise, Strasbourg, France, June 27--July 1, 2005, 
eds. F.~Casoli, T.~Contini, J.~M. Hameury, \& L.~Pagani, 
EdP-Sciences, Conference Series, 2005, p. 487

% \bibitem[{{Goldreich} {et~al.}(1996){Goldreich}, {Lai}, \&
%   {Sahrling}}]{Goldreich+96}
% {Goldreich}, P., {Lai}, D., \& {Sahrling}, M. 1996, in {Unsolved Problems in
%   Astrophysics}, ed. J.~N. {Bahcall} \& J.~P. {Ostriker} (Princeton: Princeton
%  Univ. Press)

\bibitem[{{Hansen} \& {Phinney}(1997)}]{HP97}
{Hansen}, B.~M.~S. \& {Phinney}, E.~S. 1997, \mnras, 291, 569

\bibitem[{{Herant}(1995)}]{Herant95}
{Herant}, M. 1995, \physrep, 256, 117

\bibitem[{Herant {et~al.}(1994)Herant, Benz, Hix, Fryer, \& Colgate}]{HBFC94}
Herant, M., Benz, W., Hix, W.~R., Fryer, C.~L., \& Colgate, S.~A. 1994, \apj,
  435, 339

\bibitem[{{Hobbs} {et~al.}(2005){Hobbs}, {Lorimer}, {Lyne}, \&
{Kramer}}]{Hobbs+05}
{Hobbs}, G., {Lorimer}, D.~R., {Lyne}, A.G., \& {Kramer}, M. 2005, \mnras, 360, 963

\bibitem[{{Houck} \& {Chevalier}(1992)}]{Houck_Chevalier92}
{Houck}, J.~C. \& {Chevalier}, R.~A. 1992, \apj, 395, 592

\bibitem[{{Janka} \& {M{\"u}ller}(1994)}]{JM94}
{Janka}, H.-T. \& {M{\"u}ller}, E. 1994, \aap, 290, 496

\bibitem[{{Janka} \& {M{\"u}ller}(1995)}]{JM95}
{Janka}, H.-T. \& {M{\"u}ller}, E. 1995, \apj, 448, L109

\bibitem[{{Janka} \& {M{\"u}ller}(1996)}]{JM96}
{Janka}, H.-T. \& {M{\"u}ller}, E. 1996, \aap, 306, 167

\bibitem[{{Kifonidis} {et~al.}(2003){Kifonidis}, {Plewa}, {Janka}, \&
  {M\"uller}}]{Kifonidis+03}
{Kifonidis}, K., {Plewa}, T., {Janka}, H.-T., \& {M\"uller}, E. 2003, \aap,
  408, 621

\bibitem[{{Kuhlen} {et~al.}(2003){Kuhlen}, {Woosley}, \&
  {Glatzmaier}}]{Kuhlen+03}
{Kuhlen}, M., {Woosley}, W.~E., \& {Glatzmaier}, G.~A. 2003, in ASP Conf. Ser.
  Vol. 293: 3D Stellar Evolution, eds. S.~Turcotte, S.~Keller, \& R.~Cavallo
  (San Francisco: Astron. Society of the Pacific), 147

% \bibitem[{{Lai} \& {Goldreich}(2000)}]{Lai+00}
% {Lai}, D. \& {Goldreich}, P. 2000, \apj, 535, 402

\bibitem[{{Laming}(2007)}]{Laming+07}
{Laming}, J.~M. 2007, \apj, 659, 1449

\bibitem[{{Leonard} {et~al.}(2006){Leonard}, {Filippenko}, {Ganeshalingam},
  {Serduke}, {Li}, {Swift}, {Gal-Yam}, {Foley}, {Fox}, {Park}, {Hoffman}, \&
  {Wong}}]{Leonard+06}
{Leonard}, D.~C., {Filippenko}, A.~V., {Ganeshalingam}, M., {et~al.} 2006,
  \nat, 440, 505

\bibitem[{{Liebend{\" o}rfer} {et~al.}(2001){Liebend{\" o}rfer}, {Mezzacappa},
  {Thielemann}, {Messer}, {Hix}, \& {Bruenn}}]{Liebendoerfer+01}
{Liebend{\" o}rfer}, M., {Mezzacappa}, A., {Thielemann}, F., {et~al.} 2001,
  \prd, 63, 103004

\bibitem[{{Liebend{\"o}rfer} {et~al.}(2005){Liebend{\"o}rfer}, {Rampp},
  {Janka}, \& {Mezzacappa}}]{Liebendoerfer+05}
{Liebend{\"o}rfer}, M., {Rampp}, M., {Janka}, H.-T., \& {Mezzacappa}, A. 2005,
  \apj, 620, 840

\bibitem[{{Lyne} \& {Lorimer}(1994)}]{LL94}
{Lyne}, A.~G. \& {Lorimer}, D.~R. 1994, \nat, 369, 127

\bibitem[{{Marek} \& {Janka}(2007)}]{Marek_Janka07}
{Marek}, A. \& {Janka}, H.-Th. 2007, submitted to \apj, arXiv:0708.3372

\bibitem[{{Meakin} \& {Arnett}(2006)}]{Meakin_Arnett06}
{Meakin}, C.~A. \& {Arnett}, D. 2006, \apj, 637, L53

\bibitem[{{Meakin} \& {Arnett}(2007a)}]{Meakin_Arnett07a}
{Meakin}, C.~A. \& {Arnett}, D. 2007a, \apj, 665, 690

\bibitem[{{Meakin} \& {Arnett}(2007b)}]{Meakin_Arnett07b}
{Meakin}, C.~A. \& {Arnett}, D. 2007b, \apj, 667, 448

\bibitem[{{Murphy} {et~al.}(2004){Murphy}, {Burrows}, \&
{Heger}}]{Murphy+04}
{Murphy}, J.~W., {Burrows}, A., \& {Heger}, A. 2004, \apj, 615, 460

\bibitem[{{Nobuta} \& {Hanawa}(1994)}]{Nobuta+94}
{Nobuta}, K. \& {Hanawa}, T. 1994, PASJ, 46, 257

\bibitem[{{Ohnishi} {et~al.}(2006){Ohnishi}, {Kotake}, \&
  {Yamada}}]{Ohnishi+06}
{Ohnishi}, N., {Kotake}, K., \& {Yamada}, S. 2006, \apj, 641, 1018

\bibitem[{{Rampp} \& {Janka}(2002)}]{RJ02}
{Rampp}, M. \& {Janka}, H.-T. 2002, \aap, 396, 361

\bibitem[{{Scheck} {et~al.}(2006){Scheck}, {Kifonidis}, {Janka}, \& 
 {M{\"u}ller}}]{Scheck+06}
{Scheck}, L., {Kifonidis}, K., {Janka}, H.-T., \& {M{\" u}ller}, E. 2006, \aap,
  457, 963 (Paper~I)

\bibitem[{{Scheck} {et~al.}(2004){Scheck}, {Plewa}, {Janka}, {Kifonidis}, \&
  {M{\" u}ller}}]{Scheck+04}
{Scheck}, L., {Plewa}, T., {Janka}, H.-T., {Kifonidis}, K., \& {M{\" u}ller},
  E. 2004, \prl, 92, 011103

\bibitem[{Shen {et~al.}(1998)Shen, Toki, Oyamatsu, \& Sumiyoshi}]{Shen+98}
Shen, H., Toki, H., Oyamatsu, K., \& Sumiyoshi, K. 1998, Nucl. Phys. A, 637,
  435

\bibitem[{{Thompson}(2000)}]{Thompson00}
{Thompson}, C. 2000, \apj, 534, 915

\bibitem[{{Thompson} {et~al.}(2003){Thompson}, {Burrows}, \&
  {Pinto}}]{Thompson+03}
{Thompson}, T.~A., {Burrows}, A., \& {Pinto}, P.~A. 2003, \apj, 592, 434

\bibitem[{{Wang} {et~al.}(2003){Wang}, {Baade}, {H{\" o}flich}, \&
  {Wheeler}}]{Wang+03}
{Wang}, L., {Baade}, D., {H{\" o}flich}, P., \& {Wheeler}, J.~C. 2003, \apj,
  592, 457

\bibitem[{{Wang} {et~al.}(2001){Wang}, {Howell}, {H{\" o}flich}, \&
  {Wheeler}}]{Wang+01}
{Wang}, L., {Howell}, D.~A., {H{\" o}flich}, P., \& {Wheeler}, J.~C. 2001,
  \apj, 550, 1030

\bibitem[{{Wang} {et~al.}(2002){Wang}, {Wheeler}, {H{\" o}flich}, {Khokhlov},
  {Baade}, {Branch}, {Challis}, {Filippenko}, {Fransson}, {Garnavich},
  {Kirshner}, {Lundqvist}, {McCray}, {Panagia}, {Pun}, {Phillips}, {Sonneborn},
  \& {Suntzeff}}]{Wang+02}
{Wang}, L., {Wheeler}, J.~C., {H{\" o}flich}, P., {et~al.} 2002, \apj, 579, 671

\bibitem[{{Woodward} {et~al.}(2003){Woodward}, {Porter}, \&
  {Jacobs}}]{Woodward+03}
{Woodward}, P.~R., {Porter}, D.~H., \& {Jacobs}, M. 2003, in ASP Conf. Ser.
  Vol. 293: 3D Stellar Evolution, eds. S.~Turcotte, S.~Keller, \& R.~Cavallo
  (San Francisco: Astron. Society of the Pacific), 45

\bibitem[{{Woosley} \& {Weaver}(1995)}]{WW95}
{Woosley}, S.~E. \& {Weaver}, T.~A. 1995, \apjs, 101, 181

\bibitem[{{Yamasaki} \& {Yamada}(2007)}]{Yamasaki_Yamada+07}
{Yamasaki}, T. \& {Yamada}, S. 2007, \apj, 656, 1019

\bibitem[{{Young} {et~al.}(2005){Young}, {Meakin}, {Arnett}, \&
{Fryer}}]{Young+05}
{Young}, P.~A., {Meakin}, C., {Arnett}, D., \& {Fryer}, C.~L. 2005, \apj, 629, L101

\bibitem[{{Zou} {et~al.}(2005){Zou}, {Hobbs}, {Wang}, {Manchester}, {Wu}, \&
  {Wang}}]{Zou+05}
{Zou}, W.~Z., {Hobbs}, G., {Wang}, N., {et~al.} 2005, \mnras, 362, 1189

\end{thebibliography}

%%%%%%%%%%%%%%%%%%%%%%%%%%%%%%%%%%%%%%%%%%%%%%%%%%%%%%%%%%%%%%%%%%%%%%

%\online
%\onecolumn
%\appendix

\end{document}